\documentclass[aps,twocolumn,prd,showpacs,showkeys,preprintnumbers,superscriptaddress,nobibnotes,floatfix,longbibliography,nofootinbib]{revtex4-1}

\pdfoutput=1

\usepackage[english]{babel}
\usepackage{amsmath,amssymb,amsbsy,amstext,amsthm,amsfonts,array,booktabs,exscale,relsize,slashed,graphicx,upgreek,xcolor,ulem}
\usepackage{multirow,dcolumn,bm,enumerate,url,hyperref}
\usepackage[capitalise]{cleveref}
\usepackage{fontawesome}
\usepackage{mathrsfs}
\usepackage{longtable}
\usepackage{comment}
\definecolor{lightsabergreen}{rgb}{.14,.64,.14}
\definecolor{lightgreen}{rgb}{.14,.44,.14}

\hypersetup{
  colorlinks   = true, 
  urlcolor     = lightsabergreen, 
  linkcolor    = lightsabergreen, 
  citecolor   = lightsabergreen 
}

\begin{document}

\title{Direct Detection of Mechanism-Agnostic Fast-Moving Dark Matter}

\author{Haider Alhazmi}
\thanks{{\scriptsize Email}: \href{hmalhazmi@jazanu.edu.sa}{hmalhazmi@jazanu.edu.sa}}
\affiliation{Department of Physical Sciences, Jazan University, Jazan 45142, Saudi Arabia}

\author{Doojin Kim}
\thanks{{\scriptsize Email}: \href{doojin.kim@tamu.edu}{doojin.kim@usd.edu}}
\affiliation{Department of Physics, University of South Dakota, Vermillion, SD 57069, USA}
\affiliation{Mitchell Institute for Fundamental Physics and Astronomy, Department of Physics and Astronomy, Texas A\&M University, College Station, TX 77845, USA}

\author{Kyoungchul Kong}
\thanks{{\scriptsize Email}: \href{kckong@ku.edu}{kckong@ku.edu}}
\affiliation{Department of Physics and Astronomy, University of Kansas, Lawrence, KS 66045, USA}

\author{Aishah Sumayli}
\affiliation{Department of Physical Sciences, Jazan University, Jazan 45142, Saudi Arabia}


\begin{abstract}
We present a comprehensive framework for interpreting electron recoil signals induced by fast-moving dark matter (DM), applicable across a wide range of theoretically motivated models. 
Amid both null results in conventional weakly interacting massive particle searches and growing interest in alternative DM scenarios, we focus on (semi-)relativistic DM components that can arise from mechanisms such as DM annihilation, decay, or cosmic-ray acceleration. These boosted DM candidates produce distinct experimental signatures that differ qualitatively from non-relativistic DM, necessitating a dedicated treatment. Our framework incorporates relativistic kinematics and atomic effects through ionization form factors, enabling accurate predictions of differential cross sections in both low- and high-energy regimes. We demonstrate how atomic effects become negligible at high recoil energies, validating the free-electron approximation in specific parameter regions. Furthermore, we highlight the complementarity between low-threshold direct detection experiments and high-threshold neutrino observatories in probing fast-moving DM across broad kinematic domains. This formalism provides a robust and model-independent foundation for interpreting current and future searches for relativistic DM.

\end{abstract}

\maketitle


\section{Introduction}
\label{sec:intro}

The existence of dark matter (DM) in the universe provides compelling observational evidence---all gravity-based---for physics beyond the Standard Model (SM), as no SM particle has yet been shown to consistently account for DM-related observations. One of the earliest and most well-motivated candidate frameworks involves weakly interacting massive particles (WIMPs), which are thermally produced in the early universe and remain consistent with the observed cosmological evolution. The WIMP framework predicts weakly interacting DM candidates with weak-scale masses to be consistent with the observed DM relic abundance, $\Omega_C h^2 \simeq 0.12$~\cite{Planck:2018vyg}. Interestingly, other fundamental issues, such as the gauge hierarchy problem, also motivate new physics at the weak scale (often referred to as the ``WIMP miracle''), further energizing the community and driving experimental efforts, including DM direct and indirect detection as well as collider searches. However, despite these extensive efforts over the past few decades, no conclusive evidence of DM has been observed through non-gravitational interactions (see, e.g., Refs.~\cite{Schumann:2019eaa,Cooley:2022ufh}), gradually motivating the community to reconsider and broaden the theoretical framework. 

While many alternative scenarios continue to envision non-relativistic DM with different mass scales, a series of ideas predicting fast-moving or even relativistic DM in the present universe have emerged over the past decade. Typical scenarios assume particular mechanisms through which a certain fraction of ambient DM acquires nontrivial boost factors in the present universe, hence boosted, while remaining consistent with the overall cosmological history. Moreover, many of these models offer new realizations of the WIMP miracle or alternative ways for preserving the thermal DM framework, while largely evading the strong experimental constraints placed on conventional WIMP DM~\cite{Agashe:2014yua,Belanger:2011ww,Berger:2014sqa,Kong:2014mia}. Earlier ideas include two-component boosted DM~\cite{Agashe:2014yua} with the assisted freeze-out mechanism~\cite{Belanger:2011ww}, semi-annihilation~\cite{DEramo:2010keq}, self-annihilation~\cite{Carlson:1992fn,Hochberg:2014dra}, DM-induced nucleon decay~\cite{Davoudiasl:2010am,Huang:2013xfa}, and decaying DM~\cite{Bhattacharya:2014yha,Kopp:2015bfa,Bhattacharya:2016tma,Cui:2017ytb}. More recent proposals include charged cosmic-ray acceleration~\cite{Cappiello:2018hsu,Bringmann:2018cvk,Ema:2018bih,Dent:2019krz}, cosmic-ray neutrino acceleration~\cite{Jho:2021rmn,Das:2021lcr,Chao:2021orr}, various astrophysical processes~\cite{Kouvaris:2015nsa,Hu:2016xas,An:2017ojc,Emken:2017hnp,DeRocco:2019jti,Calabrese:2021src,Wang:2021jic}, and inelastic collision of cosmic rays with the atmosphere~\cite{Alvey:2019zaa,Su:2020zny}.

Unlike the conventional DM candidates, the expected phenomenology of fast-moving DM is qualitatively different. As mentioned above, since only a certain fraction of the relic DM becomes boosted, large-volume detectors---including ton-scale DM direct detection experiments~\cite{Cherry:2015oca,Giudice:2017zke,McKeen:2018pbb,DeRocco:2019jti,Wang:2019jtk,Kannike:2020agf,Su:2020zny,Fornal:2020npv,Cao:2020bwd,Alhazmi:2020fju,Chigusa:2020bgq,Guo:2020oum,Jia:2020omh,Borah:2021jzu,Borah:2021yek,Bardhan:2022bdg,Maity:2022exk,Bell:2023sdq,Guha:2024mjr,Alhazmi:2025nvt} and large-volume neutrino detectors such as DUNE~\cite{Necib:2016aez,Alhazmi:2016qcs,Kim:2016zjx,Kim:2019had,Berger:2019ttc,Kim:2020ipj,Dent:2020syp,Dent:2020syp,Dutta:2024kuj}, Super-/Hyper-Kamiokande~\cite{Agashe:2014yua,Berger:2014sqa,Kong:2014mia,Necib:2016aez,Alhazmi:2016qcs,Kim:2016zjx,Aoki:2018gjf,Kim:2020ipj,Dent:2020syp,Bardhan:2022bdg,Bell:2023sdq,Dutta:2024kuj,Choi:2024ism}, IceCube~\cite{Agashe:2014yua,Kong:2014mia,Bhattacharya:2014yha,Kopp:2015bfa,Aoki:2018gjf,Kim:2020ipj,Cappiello:2024acu}, prototype detectors for DUNE~\cite{Chatterjee:2018mej,Kim:2018veo}, and detectors in the Short Baseline Neutrino Program~\cite{Kim:2018veo}--are preferred to achieve signal sensitivity in regions of parameter space beyond existing constraints. Moreover, by construction, the incoming DM particles carry sufficient energy to produce a variety of relativistic experimental signatures, ranging from simple electron or nucleon scattering~\cite{Agashe:2014yua,Kong:2014mia,Kim:2016zjx,Giudice:2017zke,Kim:2020ipj,Necib:2016aez,Alhazmi:2016qcs,Chatterjee:2018mej,Kim:2018veo,Wang:2019jtk,DeRoeck:2020ntj,Kannike:2020agf,Su:2020zny,Fornal:2020npv,Cao:2020bwd,Alhazmi:2020fju,Chigusa:2020bgq,Jia:2020omh,Guo:2020oum,Dent:2020syp,Borah:2021jzu,Borah:2021yek,Alhazmi:2025nvt} to deep inelastic scattering~\cite{Berger:2019ttc,Kim:2020ipj}, nuclear inelastic scattering~\cite{Dutta:2024kuj}, and even DM upscattering followed by cascade decays~\cite{Kim:2016zjx,Giudice:2017zke,Chatterjee:2018mej,Kim:2019had,Heurtier:2019rkz,Kim:2020ipj,DeRoeck:2020ntj,Bell:2021xff}. Motivated by these intriguing proposals, searches for fast-moving or boosted DM have been conducted in various experiments---including Super-Kamiokande~\cite{Super-Kamiokande:2017dch,Super-Kamiokande:2022ncz}, COSINE-100~\cite{COSINE-100:2018ged,COSINE-100:2023tcq}, PandaX~\cite{PandaX-II:2021kai,PandaX:2024pme}, CDEX~\cite{CDEX:2022fig,CDEX:2022dda,CDEX:2022dda}, NEWSdm~\cite{NEWSdm:2023qyb}, and ICARUS-Gran Sasso~~\cite{ICARUS:2024lew}---across both elastic and inelastic scattering channels. Beyond these experimental search efforts, fast-moving DM may also have intriguing astrophysical and cosmological implications \cite{Kim:2024ltz,Kim:2023onk} such as the core-cusp problem~\cite{Liu:2019bqw} and DM self-heating~\cite{Kamada:2021muh}.

Independent of theoretical developments and experimental efforts along this line, the phenomenology of fast-moving DM received significant attention in connection with the now-subsided Xenon1T electron recoil excess~\cite{XENON:2020rca}. Since the typical energy scale of the reported electron signals is $\mathcal{O}$(keV), the electron binding energy is no longer negligible, and proposals to explain the excess (e.g., Refs.~\cite{Kannike:2020agf,Su:2020zny,Fornal:2020npv,Cao:2020bwd,Alhazmi:2020fju,Chigusa:2020bgq,Jia:2020omh,Borah:2021jzu,Borah:2021yek}) incorporated the effects from ionization form factors (also often referred to as atomic excitation form factors) into their calculations. Indeed, the importance of the ionization form factors was already highlighted in Refs.~\cite{Essig:2011nj,Essig:2012yx,Roberts:2015lga,Roberts:2016xfw,Essig:2017kqs,Catena:2019gfa,Bloch:2020uzh} in the context of electron scattering induced by non-relativistic DM, as they affect the electron-DM scattering rates. Several of these studies further emphasized how different ionization form factor formalisms influence the recoil momentum spectrum, especially at high momentum transfers. 
Many studies implementing the ionization form factors have employed the plane-wave approximation~\cite{Essig:2011nj,Lee:2015qva}, even in the analysis of fast-moving DM. However, this method has limitations compared to more sophisticated approaches that compute outgoing electron wave functions by solving the Schr\"{o}dinger equation~\cite{Essig:2012yx,Essig:2017kqs,Catena:2019gfa,Bloch:2020uzh} or using the relativistic Dirac-Hartree-Fock method~\cite{Roberts:2015lga,Roberts:2016xfw}. These differences become particularly significant at high momentum transfers, where the effects of fast-moving DM are most pronounced. A recent study of ours~\cite{Alhazmi:2025nvt} revisited and extended the atomic ionization formalism, systematically comparing different methods for formulating the ionization form factor, specifically in the context of fast-moving DM interactions mediated by either a vector or a pseudoscalar mediator.

Building on the context described above, in this work, we fully generalize the program initiated in Ref.~\cite{Alhazmi:2025nvt} by undertaking the following tasks:
\begin{itemize}
    \item Establishing the formalism for electron scattering with any fast-moving DM across a broad range of well-motivated DM scenarios,
    \item Identifying the effective regime in which atomic effects can be neglected, validating the use of the free-electron approximation,
    \item Demonstrating the complementarity between low- and high-energy-threshold experiments in the search for (fast-moving) DM.
\end{itemize}

First, in our formalism, the incoming (fast-moving) DM is treated as relativistic, rather than assuming the non-relativistic limit (see, e.g., Ref.~\cite{Roberts:2016xfw}), and the wave functions for both the bound electrons (initial state) and free electrons (final state) are taken as solutions to the Dirac equation instead of the Schr\"{o}dinger equation. The resulting ionization form factors depend on both the momentum transfer and the recoil electron energy. This relativistic treatment provides the flexibility to accommodate various Lorentz interaction structures through the ionization form factors. We will consider scenarios where interactions between the SM particles and scalar or fermionic DM are mediated by scalar, pseudoscalar, vector, and axial vector mediators through renormalizable operators.  

Obviously, the formalism should reduce to that of free electrons in the limit of large recoil energy, which also serves as a consistency check of the formalism. For a given recoil energy, the ionization form factor determines the range of momentum transfers capable of producing that recoil. In the low recoil energy regime, the form factor is spread over a broad range of allowed momenta. However, we will demonstrate that in the high recoil energy limit, the ionization form factor becomes sharply peaked around a single momentum transfer value---effectively approaching a Dirac delta function---thereby recovering the result for free electrons. The boundary between the low and high recoil energy regimes---and thus the boundary for the effective regime where the free-electron approximation is valid---depends on the details of the underlying model. We identify the parameter regions where incorporating the ionization form factor is essential for accurate signal rate estimates and quantify the resulting differences across all DM scenarios considered. We further find that, contrary to the common expectation that the ionization form factor suppresses event rates, its inclusion can enhance the signal, depending on the model specifics---particularly in the case of a pseudoscalar mediator. 

Finally, we emphasize that the search for fast-moving DM benefits significantly from the complementarity between low- and high-energy-threshold experiments. Low-threshold detectors, such as DM direct detection experiments, are particularly sensitive to low-energy electron recoils and can probe lighter fast-moving DM candidates or heavier DM with a mild boost. In contrast, high-threshold detectors, including large-volume neutrino observatories, excel at identifying higher-energy interactions associated with more massive or strongly boosted DM particles. Together, these experimental approaches cover a broad range of recoil energies and DM parameter space, enabling a more comprehensive and robust exploration of fast-moving DM models than would be possible with either class of detector alone. For illustration, we consider the XENONnT~\cite{XENON:2017lvq} and Super-Kamiokande~\cite{Super-Kamiokande:2002weg} experiments, which respectively employ the ionization form factor and free-electron models to estimate signal sensitivity, in the contexts of two-component boosted DM and cosmic-ray-induced boosted DM scenarios.

This paper is organized as follows.
In Section \ref{sec:electron_recoils} we provide the notation and the general formulation of the DM interaction with the bound electrons by providing the necessary formulas for the differential cross section and the elementary interactions of interest. Section~\ref{sec:limits} presents a quantitative numerical analysis that identifies the kinematic regions where the ionization form factor becomes significant and, consequently, where atomic-physics effects materially impact the limits. Section~\ref{sec:prospects} presents sensitivity estimates both with the ionization form factor included and under the free-electron approximation, and compares the two, assuming a fixed reference DM flux.
We then apply these results in Section~\ref{sec:interpretation} to two benchmark fast-moving DM scenarios with realistic fluxes: two-component boosted DM and cosmic-ray–boosted DM. Section~\ref{sec:conclusion} provides the summary and outlook. Results for various DM–mediator spin combinations are presented in Appendix~\ref{appendix:A}.

\section{Setup}
\label{sec:electron_recoils}

\subsection{Electron Recoils}

\subsubsection{Kinematics}
\label{subsec:NotationandKinematics}

\begin{figure}[t]
    \centering
    \includegraphics[width=0.85\linewidth]{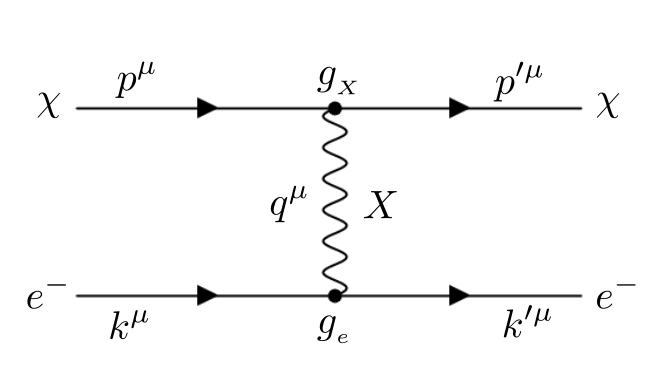}
    \caption{Feynman diagram depicting the interaction between a DM particle and a target electron.}
    \label{fig:feyn_diag}
\end{figure}

In this section, we discuss the scattering between an electron and a DM particle---denoted by $\chi$ of mass $m$---incoming with a speed $v$ and an energy $E$. 
We assume that the DM particle $\chi$ is either a fermion or a scalar that interacts elastically with an electron $e$:
\begin{equation}
    \chi(p^{\mu}) \,e(k^{\mu}) \rightarrow \chi(p^{\prime \mu}) \,e(k^{\prime \mu}),
\end{equation} 
where $p^\mu$ and $k^\mu$ denote the four-momenta of the DM particle and the electron, respectively, with unprimed and primed quantities referring to the initial and final states. 
As shown in the Feynman diagram in Figure~\ref{fig:feyn_diag}, the interaction is mediated by a $t$-channel exchange of a particle $X$ with mass $m_X$, which couples to $\chi$ and $e$ via the couplings $g_X$ and $g_e$, respectively. 
As mentioned earlier, the mediator can be a vector, scalar, axial vector, or pseudoscalar, provided the interaction remains renormalizable. Further details on the interaction structures will be presented later.

For clarity and future reference, we adopt the following labeling for the four-momenta: 
\begin{equation} 
\begin{split}
p^{\mu} & =(E, \mathbf{p}), 
\quad \,\, p^{\prime \mu} =\left(E^{\prime}, \mathbf{p}^{\prime}\right), \\
k^{\mu}  & =\left(E_{e} , \mathbf{k}\right),  \quad k^{\prime \mu}=\left(E_{e}^{\prime} , \mathbf{k}^{\prime}\right).
\end{split}
\end{equation}
The deposited energy is defined as the energy difference between the outgoing electron and the incoming electron, $\Delta E \equiv E_{e}^{\prime}-E_{e}$, while the momentum transfer is given by $\mathbf{q} \equiv \mathbf{p}-\mathbf{p}^{\prime}$. 
Applying energy conservation and the on-shell condition, the deposited energy is given by
\begin{equation}
\Delta E=E-\sqrt{E^{2}+q^{2}-2 p q \cos \theta_{q}},
\end{equation}
where $\theta_{q}$ is the angle between $\mathbf{p}$ and $\mathbf{q}$. Note that we use the notation $p = |\mathbf{p}|$ and $q = |\mathbf{q}|$ to denote the magnitudes of the momenta---a convention we will maintain throughout the paper.
Solving the above relation for $q$, we find
\begin{equation}
q=p \cos \theta_{q} \pm \sqrt{p^{2} \cos ^{2} \theta_{q}+\Delta E(\Delta E-2 E)} .
\end{equation}

Let us analyze the above equation in the limiting cases $\cos \theta_{q}=\pm 1$. 
First, for nonzero deposited energy, energy conservation implies that the quantity $\Delta E(\Delta E-2 E)$ is strictly negative, as it equals $E^{\prime 2}-E^{2}$. 
With this in mind, the expression under the square root becomes $E^{\prime 2}-m^{2} \geq 0$, ensuring that the square root is real. 
Consequently, the deposited energy $\Delta E$, must be chosen carefully to guarantee that the momentum transfer $q$ remains physical. The condition ensuring physical values of $q$, consistent with energy conservation, is given by\footnote{$\Delta E \geq E+m$ is also allowed, but it violates energy conservation.}
\begin{equation}
\Delta E \leq E-m.
\label{eq:DE_cond}
\end{equation}

Second, we note that the case when $\cos \theta_{q}=-1$ is physically not allowed since it leads to negative values as $q=-p \pm p^{\prime}$, where we have used $p^{\prime 2}=E^{\prime 2}-m^{2} \geq 0$. The reason that $q$ is negative here is because $p>p^{\prime}$ for non-zero $\Delta E$. We are now left with the condition $\cos \theta_q = 1$, which corresponds to the maximum and minimum values of $q$. We denote these values by $q^\pm$, which are determined by
\begin{equation}
q^{\pm}=p \pm \sqrt{p^{2}+\Delta E(\Delta E-2 E)}.
\label{eq:qpm}
\end{equation}
Although the condition defined in \eqref{eq:DE_cond} is required to ensure real values of $q$, energy conservation imposes an upper kinematic bound on $\Delta E$. For a free, stationary electron, the maximum scattered electron energy occurs at $\cos \theta_q = 1$ and is given by
\begin{equation}
    E_{e}^{\prime {\rm max}} = m_e \frac{(E - m_e)^2 + E^2 - m^2}{(E - m_e)^2 - E^2 + m^2}.
\end{equation}
This leads to the maximum energy deposition
\begin{equation}
    \Delta E^{\rm max} = E_{e}^{\prime {\rm max}} - E_e \le E - m,
    \label{eq:DEmaxcond}
\end{equation}
where the equality holds when $m = E_e$.

Finally, we comment on the non-relativistic limit of the above equation. Using the approximation $E \approx m + \frac{1}{2} m v^2$, where $v$ is the velocity of the DM particle, and assuming $m \gg \Delta E$, we obtain
\begin{equation}
q_{\mathrm{non-rel}}^{\pm} \approx p \pm \sqrt{p^{2}-2 m \Delta E}.
\label{eq:qpmNR}
\end{equation}
Here, we have neglected the $\frac{1}{2} m \Delta E v^{2}$ term, as it is subleading compared to the $p^2$ term. Equation \eqref{eq:qpmNR} agrees with the result in Ref.~\cite{Roberts:2016xfw}. For the non-relativistic case, one uses Eq.~\eqref{eq:qpmNR} with $p \approx m v$, while in the relativistic case, one uses Eq.~\eqref{eq:qpm} with $p = \gamma m v$ and $E = \gamma m$, where $\gamma$ is the Lorentz factor.

\subsubsection{Generalized Differential Cross Section}
\label{subsec:GDCS}
We now discuss the full formulation of the cross section for the $\chi$-$e$ scattering, taking into account that electrons are bound to atoms.
For concreteness, we use the fully relativistic, spin-averaged matrix element squared, denoted by $\overline{|\mathcal{M}|^2}$, throughout this paper.
In the DM literature---particularly in the context of WIMP physics---a non-relativistic treatment of the matrix element squared is often preferred. This is typically done using the Born approximation, where the cross section is parametrized as
\begin{equation}
{\sigma}_{\rm B } = \frac{\mu^2}{16 \pi m^2 m_e^2} \, \overline{|\mathcal{M}|^2_{\rm B}},
\label{eq:Borncs}
\end{equation}
where $\mu$ is the $\chi$-$e$ reduced mass, and $\overline{|\mathcal{M}|^2_{\rm B}}$ is given by
\begin{equation}
\overline{|\mathcal{M}|^2_{\rm B}} = \frac{16 g_e^2 g_X^2 m_e^2 m^2}{(q^2 + m_X^2)^2}.
\label{eq:MESB}
\end{equation}

For later uses, we define the fully relativistic dark matter form factor $|F_{\rm DM}|^2$ and its Born approximation counterpart $|F_{\rm DM}|_{\rm B}^2$ as follows:
\begin{eqnarray}
    |F_{\rm DM}|^2 &\equiv& \frac{\overline{|\mathcal{M}|^2}}{\overline{|\mathcal{M}|^2_{\rm B}}(q=\alpha_{\rm em} m_e)},\\
    |F_{\rm DM}|_{\rm B}^2 &\equiv& \frac{\overline{|\mathcal{M}|_{\rm B}^2}}{\overline{|\mathcal{M}|^2_{\rm B}}(q=\alpha_{\rm em} m_e)},
\end{eqnarray}
where we have chosen a reference momentum transfer of $\alpha_{\rm em} m_e$, with $\alpha_{\rm em}$ denoting the fine-structure constant. 
Defining the parameter $\overline{\sigma} \equiv \sigma_{\rm B}(q = \alpha_{\rm em} m_e)$, which serves as a reference cross section and encapsulates model-dependent parameters such as couplings, we can express the squared matrix element as
\begin{equation}
\overline{|\mathcal{M}|^2} = \frac{16 \pi m^2 m_e^2 \overline{\sigma}}{\mu^2} \times |F_{\rm DM}|^2.
\end{equation} 
Hence, the differential cross section can be generally written as \cite{Cao:2020bwd}
\begin{eqnarray}
d\sigma &=& \frac{m^2 m_e^2 \overline{\sigma}}{4 \pi E  E^{\prime}  E_e  E^{\prime}_e \mu^2 v} \, |F_{\rm DM}|^2 \, \, \times \label{eq:dsigma_1} \\ &&  \delta(E^{\prime} + E^{\prime}_e - E - E_e) \frac{|f_{e_{i} \rightarrow e_{f}}|^2}{(2 \pi)^3}  \, d^3q d^3k^{\prime}, \nonumber
\end{eqnarray}
where $|f_{e_{i} \rightarrow e_{f}}|$ is the transition factor from a bound state electron $e_i$ to a free electron $e_f$ and the volume element $d^3 p'$ is replaced with $d^3q$.
The Dirac delta function here can be re-expressed in terms of the angle, $\theta_q$, as
\begin{equation}
\delta\left(E^{\prime} - (E + E_e - E^{\prime}_e)\right) = \frac{E^{\prime}}{p q \sin \theta_q} \delta(\theta_q).
\end{equation}

Since the transition factor does not depend on the angular component of the vector, $\mathbf{q}$, we can integrate over the solid angle element $d \hat{\Omega}_{q}$, simplifying the differential cross section to
\begin{equation}
d\sigma = \frac{m^2 m_e^2 \overline{\sigma}}{2 \mu^2}\frac{qdq}{p E E_e  E^{\prime}_e v} \, |F_{\rm DM}|^2 \, \frac{|f_{e_{i} \rightarrow e_{f}}|^2}{(2 \pi)^3} \, \, d^3k^{\prime}.
\label{eq:dsigma_2}
\end{equation}
We now express the differential cross section in terms of the recoil energy $E_r$, defined via $E^{\prime}_e = m_e + E_r$. Using the relation $d^3k^{\prime} = E^{\prime}_e k^{\prime} d\Omega_{k^{\prime}} dE_r$, we obtain
\begin{equation}
\frac{d\sigma}{d E_r} = \frac{m^2 m_e^2 \overline{\sigma}}{4 \mu^2 p {k^{\prime}}^2 E E_e v} \int^{q^{+}}_{q^{-}} qdq \, |F_{\rm DM}|^2 \, |f_{\rm ion}|^2,
\label{eq:dsigma_3}
\end{equation}
where $|f_{\rm ion}|^2$ is the dimensionless ionization form factor, defined as 
\begin{equation}
    |f_{\rm ion}|^2 \equiv \frac{2 {k^{\prime}}^3}{(2 \pi)^{3}} \int d {\Omega}_{k^{\prime}} |f_{e_{i} \rightarrow e_{f}}|^2.
\end{equation}
For a spherically symmetric transition factor squared, we have
\begin{equation}
|f_{\rm ion}|^2 = \frac{4 {k^{\prime}}^3}{(2 \pi)^{2}} \, |f_{e_{i} \rightarrow e_{f}}|^2.
\label{eq:trans}
\end{equation}
Finally, the differential cross section above becomes
\label{eq:ionf}
\begin{equation}
\frac{d\sigma}{d E_r} = \frac{m^2 m_e^2 \overline{\sigma}}{4 \mu^2 p {k^{\prime}}^2 E E_e v} \int^{q^{+}}_{q^{-}} qdq \, |F_{\rm DM}|^2 \, |f_{\rm ion}|^2.
\label{eq:dsigma_4}
\end{equation}

It is straightforward to see that the fully relativistic treatment of the differential cross section smoothly reduces to the standard WIMP case in the non-relativistic limit, under the following substitutions:
\begin{equation}
    p \sim m v, \quad E E_e \sim m m_e, \quad {\rm and} \quad k^{\prime} \sim \sqrt{2 m_e E_r}.
\end{equation}
This yields the standard WIMP formula
\begin{equation}
\frac{d\sigma^{\rm WIMP}}{d E_r} = \frac{\overline{\sigma}_{\rm B}}{8 \mu^2 E_r v^2} \int^{q_{\mathrm{NR}}^{+}}_{q_{\mathrm{NR}}^{-}} qdq  \, |F_{\rm DM}|_{\rm B}^2 \, |f_{\rm ion}|^2.
\label{eq:dsigma_NR}
\end{equation}
Note that we have used the approximate form factor $|F_{\rm DM}|_{\rm B}^2$ in place of the full relativistic form $|F_{\rm DM}|^2$, and the non-relativistic limits $q_{\mathrm{NR}}^{\pm}$ instead of $q^{\pm}$. In this context, the form factor typically takes one of two limiting forms:
\begin{equation}
|F_{\rm DM}|_{\rm B}^2=\begin{cases}
          \,\,\,\,\, 1 \quad &\text{if} \quad m_X \gg q,  \\[6pt]
          \dfrac{\alpha_{\rm em}^2 m_e^2}{q^2} \quad &\text{if} \quad m_X \ll q. \\
     \end{cases}
\end{equation}

Equation \eqref{eq:dsigma_4} provides the full treatment of the $\chi$–$e$ interaction, incorporating atomic physics effects through the ionization form factor.
It is instructive to show that the simpler case of a free, stationary electron---neglecting atomic effects---can be easily recovered.
This is accomplished by making the following replacement in Eq.~\eqref{eq:dsigma_1}:
\begin{equation}
    |f_{e_{i} \rightarrow e_{f}}|^2 \rightarrow (2 \pi)^3 \delta^3(\mathbf{q} - \mathbf{k}^{\prime}),
    \label{eq:free_limit1}
\end{equation}
which corresponds to treating the electron as free and at rest. For a spherically symmetric ionization form factor, this replacement implies
\begin{equation}
    |f_{\rm ion}|^2 \rightarrow 2 \frac{{k^{\prime}}^3}{q^2}\delta(q -k^{\prime}).
    \label{eq:free_limit2}
\end{equation}
Substituting this into the cross section expression yields the simplified differential cross section for a free electron:
\begin{equation}
\frac{d\sigma^{\rm free}}{d E_r} = \frac{m^2 m_e \overline{\sigma}}{2 \mu^2 p^2}  \, |F_{\rm DM}|^2_{q = {k^{\prime}}},
\label{eq:dsigma_5}
\end{equation}
where we have used $E_e = m_e$ for an electron at rest.
In the following section, we examine the effects of atomic physics by analyzing the impact of the ionization form factor $|f_{\rm ion}|^2$ on the kinematically allowed region of the parameter space. Before doing so, however, we first describe the class of interactions under consideration.

\subsection{Dark Matter Models}
\label{subsec:inter}

Here, we outline the types of interactions we aim to study, classifying them based on the spin of the DM particle and the structure of the Lorentz interaction. In this work, we consider a generic DM particle $\chi$ in two forms: a spin-1/2 fermion denoted by $\psi$ and a spin-0 scalar denoted by $\phi$.
The generic mediator particle $X$ responsible for the interaction is classified as one of the following types: vector ($V^{\mu}$), scalar ($S$), axial vector ($A^{\mu}$), or pseudoscalar ($P$).
Given the set of dark matter particles
\begin{eqnarray}
    \chi = \{\psi, \phi \},
\end{eqnarray}
and the set of the mediators
\begin{equation}
    X = \{V^{\mu}, S, A^{\mu}, P \},
\end{equation}
there are seven distinct combinations that can be described by renormalizable interaction Lagrangians.
To clarify our notation, each interaction type is abbreviated using two characters: the first denotes the DM particle type, and the second denotes the mediator type.
For example, FV refers to fermionic dark matter with a vector mediator, while SP represents scalar dark matter with a pseudoscalar mediator.
The coupling between a particle and a mediator is denoted by superscripting the particle symbol and subscripting the mediator symbol. For example, the coupling between fermionic dark matter and an axial vector mediator is labeled as $g^{\psi}_{A}$, while the coupling of the electron to a scalar mediator is labeled as $g^{e}_{S}$.

In all cases, the spin-averaged squared matrix element shares a common denominator of the form $(m_X^2 - t)^2$, where $t$ is the Mandelstam variable associated with the $t$-channel exchange diagram, as shown in Figure~\ref{fig:feyn_diag}. 
Furthermore, as evident from Eqs.~\eqref{eq:dsigma_4} and \eqref{eq:dsigma_5}, the cross section appears as an overall factor in the reference cross section $\overline{\sigma}$.  
Consequently, all interaction scenarios include this common prefactor, defined by Eqs.~\eqref{eq:Borncs} and \eqref{eq:MESB}. 
For clarity, we denote it as
\begin{equation}
    \overline{\sigma_{\chi X}} = \frac{\mu^2}{\pi} \frac{({g^e_X} {g^{\chi}_X})^2}{(m_X^2 + \alpha_{\rm em}^2 m_e^2)^2}.
    \label{eq:refcsmod}
\end{equation}
This notation is important, as we will present exclusion plots based on this cross section for various interaction types.
For example, in the case of fermionic DM with a vector mediator, limits on the cross section are expressed as bounds on $\overline{\sigma_{\rm FV}}$, from which constraints on the couplings and other model parameters can be derived accordingly.

It is also of interest to examine the behavior of the differential cross section for the various interactions considered.
For practical purposes, and in terms of the squared matrix element, the differential cross section for scattering off free electrons can be derived from Eq.~\eqref{eq:dsigma_5} as
\begin{equation}
   \dfrac{d \sigma^{\rm free}_{\chi _X}}{d E_r}=\frac{({g^e_X} {g^{\chi}_X})^2}{32 {\pi} m_e  m_{\chi}^2\gamma^2} \,\,|\mathcal{A}|_{\chi X}^2,
\end{equation}
where the condition $\gamma \gg 1$ is assumed, and the coupling-normalized matrix element squared is defined as
\begin{equation}
    \overline{|\mathcal{A}|_{\chi X}^2} \equiv \frac{\overline{|\mathcal{M}|_{\chi X}^2}}{({g^e_X} {g^{\chi}_X})^2}.
\end{equation}
For simplicity, we express the cross section parameters in terms of the four dimensionless quantities defined in Table~\ref{tab:parameters}. In this case, the differential cross section becomes 
\begin{equation}
   \dfrac{d \sigma^{\rm free}_{\chi X}}{d x}=\frac{({g^e_X} {g^{\chi}_X})^2}{32 {\pi} m_e^2} \,\, \frac{\overline{|\mathcal{A}|_{\chi X}^2}(x, \alpha,\gamma,\beta)}{\alpha^2\gamma^2},
   \label{eq:dcsdx}
\end{equation}
where we explicitly express that $\overline{|\mathcal{A}|_{\chi X}^2}$ is a function of the four dimensionless parameters. 

\begin{table}[t]
    \centering
    \def\arraystretch{1.5}
    \scalebox{1}{
    \begin{tabular}{|cc||c|}
    \hline 
        \multicolumn{2}{|c||}{Parameter}  & ~~Definition~~ \\
        \hline \hline
        Boost factor            & $\gamma$~~  & $E/m$ \\
        Dark matter mass        & $\alpha$~~  & $m/m_e$ \\
        Mediator mass           & $\beta$~~   & $m_X/m_e$ \\
        ~~Electron recoil energy  & $x$~~       & $E_r/m_e$ \\
        \hline
    \end{tabular}}
    \caption{Dimensionless parameters used in the shape analysis of the coupling-normalized matrix element squared, $\overline{|\mathcal{A}|^2}$. Except for the boost factor, all quantities are normalized by the electron mass.}
    \label{tab:parameters}
\end{table}

After expressing $\overline{|\mathcal{A}|^2}$ in terms of the dimensionless parameters, we analyze its behavior in two limiting cases: i) the heavy mediator limit ($\beta \gg 1$) where the expressions are expanded and truncated at terms of order $\mathcal{O}(1/\beta^6)$ and ii) the light mediator limit ($\beta \ll 1$) where the expressions are expanded and truncated at terms of order $\mathcal{O}(\beta^2)$.
Table~\ref{tab:RMES} summarizes the leading-order behavior of the cross section in the two limiting cases.
\begin{table*}[t!]
    \centering
    \def\arraystretch{2.8}
    \scalebox{1.1}{
    \begin{tabular}{|c||c||c|}
    \hline
     $\chi X$  & Heavy mediators: $\overline{|\mathcal{A}|_{\chi X}^2}(x, \alpha,\gamma,\beta\gg1)$ & Light mediators: $\overline{|\mathcal{A}|_{\chi X}^2}(x, \alpha,\gamma,\beta\ll1)$ \\ \hline \hline
       FV  & $8 \, \dfrac{ 2 \alpha ^2 \gamma ^2 - x \left(\alpha ^2+2 \alpha  \gamma +1\right) + x^2 }{\beta^4}$ & $2 -\dfrac{2 \left(\alpha ^2+2 \alpha  \gamma +1\right)}{x} + \dfrac{4 \alpha ^2 \gamma ^2}{x^2}$\\
       FS  & $4 \, \dfrac{4 \alpha ^2 + 2 x\left(1 + \alpha ^2\right) + x^2}{\beta ^4}$ & $1 +\dfrac{2 \left(\alpha ^2+1\right)}{x}+ \dfrac{4 \alpha ^2}{x^2}$\\
       FA  & $\,\,\,\,8 \, \dfrac{2 \alpha ^2 \left(2 + \gamma ^2\right) + x \left(1 - 2 \alpha  \gamma + \alpha ^2 \right) + x^2}{\beta ^4}\,\,\,\,$ & $\,\,\,\,2 + \dfrac{16 \alpha ^2}{\beta ^4}+\dfrac{2
   \left(1 - 2 \alpha  \gamma + \alpha ^2 \right)}{x}+\dfrac{4 \alpha ^2 \left(1 + \gamma ^2\right)}{x^2}\,\,\,\,$\\
       FP  & $4 \, \dfrac{x^2}{\beta ^4}$ & $1  - \dfrac{\beta^2}{x}$\\ \hline\hline
       SV  & $8 \alpha \,   \dfrac{2 \alpha  \gamma ^2 - x (\alpha +2 \gamma )}{\beta^4}$ & $2 \alpha \,   \dfrac{2 \alpha  \gamma ^2 - x (\alpha +2 \gamma )}{x^2}$\\
       SS  & $8 \alpha^2 \, \dfrac{2 + x}{\beta^4}$ & $2 \alpha^2 \, \dfrac{2 + x}{x^2}$\\
       SP  & $\dfrac{8 x \alpha ^2}{\beta^4}$ & $\dfrac{2 \alpha ^2}{x}$\\ \hline
    \end{tabular}}
    \vspace{0.2cm}
    \caption{Summary table of the coupling-normalized matrix element squared, $\overline{|\mathcal{A}|^2}$ in terms of the dimensionless parameters defined in Table~\ref{tab:parameters} at two mediator mass limits: heavy and light mediators.}
    \label{tab:RMES}
\end{table*}

Instead of describing each limiting case for every interaction individually---and to avoid repetition---we highlight the generic features that emerge across interaction types as follows:
\begin{itemize}
    \item \textit{Heavy mediators:} In the heavy mediator regime ($\beta \gg 1$), the differential cross sections generally do not exhibit a decreasing trend with recoil energy $E_r$, denoted here as $x$. Most distributions start out flat in $x$, although some cases show a rising behavior depending on specific parameter choices. A notable exception occurs in pseudoscalar-mediated interactions: the FP case displays a quadratic rise, while the SP case exhibits a linear increase with $x$. The flat behavior is particularly pronounced when the squared matrix element includes constant terms (independent of $x$) proportional to $\gamma^2$, which is especially evident in the FV, FA, and SV cases. For example, in the FV case, the distribution transitions from flat to rising when $x \gg \sqrt{2} \alpha \gamma$, a condition satisfied only within a restricted region of parameter space where $\sqrt{2} \alpha \gamma < x^{\rm max}$. In contrast, scalar-mediated interactions (FS and SS) exhibit weaker flatness. The FS case transitions from flat to rising when $x \gg 2\alpha$, while for the SS case, this transition occurs at $x > 2$.
    
    \item \textit{Light mediators:} In the light mediator regime ($\beta \ll 1$), the differential cross section distributions are generally falling or flat. With the exception of the FP case, all interactions exhibit a decreasing trend at small values of $x$, and in some cases, this trend continues up to $x^{\rm max}$. This falling behavior is particularly characteristic of the three scalar dark matter interactions: SV, SS, and SP. In other cases, the extent of the decrease depends on the value of the $\gamma$ parameter. For instance, in the FV case, the distribution transitions from falling to flat when $x \gg \sqrt{2} \alpha \gamma$, and a similar transition occurs in the FS case when $x \gg 2\alpha$.
    
    \item \textit{General observations:} The analysis reveals several distinctive features associated with specific interaction types. Scalar and pseudoscalar mediators show no dependence on the $\gamma$ parameter, a property that holds beyond leading-order approximations. Additionally, the FP interaction is entirely independent of $\alpha$, whereas all scalar dark matter interactions (SV, SS, and SP) exhibit at least a linear dependence on $\alpha$.
\end{itemize}
Overall, these observations illustrate how the mediator type, kinematic variables, and interaction structure collectively shape the behavior of differential cross sections, offering valuable insights into expected DM detection rates and exclusion limits, which will be elaborated upon shortly.
 
\subsection{Ionization Form Factor}
\label{sec:free_electron_recoil_rate}
Here we aim to provide a quick overview of the ionization form factor.
The ionization form factor $|f_{\rm ion}|^2$ relies on the estimation of the transition factor squared $|f_{e_{i} \rightarrow e_{f}}|^2$ which is defined as the overlap of the initial bound electron state $|e_i\rangle$ and the final free electron state $|e_f\rangle$ as
\begin{equation}
f_{e_{i} \rightarrow e_{f}} = \langle e_f | e^{i {\bf{q}}\cdot{\bf{r}}}| e_i \rangle = \int {d^{3}{\bf{r}}} \, \psi^{*}_{e_f}({\bf{r}}) \, e^{i {\bf{q}}\cdot{\bf{r}}} \,\psi_{e_i}({\bf{r}}),
\label{eq:transf}
\end{equation}
where $\psi_{e_i}$ is the initial state electron wave function representing the original bound electron and $\psi_{e_f}$ is the final state electron wave function representing the outgoing electron. 
In this work, we focus our attention on the relativistic ionization form factor while different treatments have been considered in various studies; see, for example, Refs.~\cite{Kopp:2009et, Essig:2011nj, Essig:2012yx, Essig:2015cda, Lee:2015qva, Roberts:2015lga, Roberts:2016xfw, Essig:2017kqs, Catena:2019gfa, Cao:2020bwd, Alhazmi:2021qgd, Hamaide:2021hlp, Emken:2024nox}.

The use of relativistic treatment allows for simple natural generalization for the various Lorentz interaction as has been pointed out in Ref. \cite{Roberts:2016xfw}.
The transition factor in this case is defined through wave functions that are solutions to the Dirac equation as
\begin{equation}
f^{\rm{DCE}}_{e_{i} \rightarrow e_{f}} = \int {d^{3}{\bf{r}}} \, \psi^{\dagger}_{{E_r \kappa' m'}}({\bf{r}}) \, e^{i {\bf{q}}\cdot{\bf{r}}} \,\psi_{n \kappa m}({\bf{r}}),
\label{eq:transfDCE}
\end{equation}
where the superscript DCE explicitly indicates the fact that the ionization form factor is related to the Dirac Continuous Energy case. Here $\psi_{n\kappa m}$ is the general bound-state solution to the eigenvalue problem $\hat{H}\psi_{n\kappa m}=E_{n\kappa}\psi_{n\kappa m}$ where $\hat{H}$ is the relativistic Dirac Hamiltonian,
\begin{equation}
\hat{H}=\vec{\alpha}\cdot \hat{\vec{p}}+m_e(\beta-1)+V_{\rm eff}(r)
\end{equation}
where $\vec{\alpha}$ and $\beta$ denote the usual Dirac matrices in the standard representation. The effective potential $V_{\rm eff}$ depends solely on the radial coordinate, assuming that the electrons are subject to a spherically symmetric potential. 
We employ the Dirac-Hatree-Fock-Slater form for $V_{\rm eff}$ that implements the electron-nucleus and electron-electron interactions~\cite{Alhazmi:2021qgd}.\footnote{We assume that the interactions potentially arising via new mediator exchanges are negligible.} Therefore, the solution $\psi_{n\kappa m}$ depends on radius $r=|{\bf r}|$ only:
\begin{equation}
    \psi_{n\kappa m}({\bf r})=\frac{1}{r}
    \begin{pmatrix}
        P_{n\kappa}(r)\Omega_{\kappa m}(\hat{r}) \\
        iQ_{n\kappa}(r)\Omega_{-\kappa m}(\hat{r})
    \end{pmatrix}\,,
\end{equation}
where $\Omega_{\kappa m}$ is the two-component spherical spinor. 
Note that the above full Dirac wave function is a four-component spinor, with the upper (lower) two components corresponding to the large (small) component radial wave functions, denoted by $P$ ($Q$).
For the free-state solution $\psi_{E_r \kappa m}$, the principal quantum number $n$ is replaced with the recoil energy $E_r$. 
Finally, the Dirac quantum number $\kappa$ is determined by the orbital and total angular momentum quantum numbers, $\ell$ and $j$, respectively, 
\begin{equation}
    \kappa = (\ell -j)(2j+1)\,.
\end{equation}

The operator between the two wave functions in Eq.~\eqref{eq:transfDCE} is governed by the form of the interaction. In other words, this is where the model dependence appears in the ionization form factor. 
For example, the above equation assumes that the dark matter mediator is a vector boson that interacts with electrons through a vector coupling.
In this case, the operator is simply $e^{i {\bf{q}}\cdot{\bf{r}}}$ times a 4-by-4 identity matrix $I$, which leaves the large and small components unmixed when performing the dot product. 
We can readily consider the situation where the mediator interacts with electrons through scalar ($\gamma^{0}$), axial vector ($\gamma_5$), or pseudoscalar ($\gamma^{0} \gamma_{5}$) with the identity matrix in Eq.~\eqref{eq:transfDCE} replaced by the gamma matrices in the respective brackets as per Ref. \cite{Roberts:2016xfw}. 
Denoting the vector, scalar, axial vector, and pseudoscalar mediators by V, S, A, and P, respectively, we express the squared transition factor for each mediator as follows:
\begin{eqnarray}
\left|f^{\rm{DCE}}_{e_{i} \rightarrow e_{f}} \right|_{\rm{V}}^{2} &=& \sum_{\kappa'} \,\sum_{L} \, C_{\kappa \kappa'}^{L} \, (R_{P,P} + R_{Q,Q})^2, \nonumber \\
\left|f^{\rm{DCE}}_{e_{i} \rightarrow e_{f}} \right|_{\rm{S}}^{2} &=& \sum_{\kappa'} \,\sum_{L} \, C_{\kappa \kappa'}^{L} \, (R_{P,P} - R_{Q,Q})^2, \nonumber\\
\left|f^{\rm{DCE}}_{e_{i} \rightarrow e_{f}} \right|_{\rm{A}}^{2} &=& \sum_{\kappa'} \,\sum_{L} \, D_{\kappa \kappa'}^{L} \, (R_{P,Q} - R_{Q,P})^2, \nonumber\\
\left|f^{\rm{DCE}}_{e_{i} \rightarrow e_{f}} \right|_{\rm{P}}^{2} &=& \sum_{\kappa'} \,\sum_{L} \, D_{\kappa \kappa'}^{L} \, (R_{P,Q} + R_{Q,P})^2,
\label{eq:relFFdiffmed}
\end{eqnarray}
where $L=|\ell - \ell'|$ with $\ell$ and $\ell'$ encapsulated in $\kappa$.
Here, $R$ is the radial integral, and the two subscripts, either $P$ or $Q$, indicate which components contribute to the integral:
\begin{equation}
R_{X,Y} = \int_0^{\infty} \, X_{E_r \kappa'} \,Y_{n \kappa} j_L(q r) dr,
\end{equation}
where $j_L(x)$ is the usual spherical Bessel function.
Finally, The overall coefficient $C_{\kappa \kappa^{\prime}}^L$ is written in terms of the Wigner 3j-symbol as
\begin{widetext}
    \begin{equation}
        \begin{aligned} & C_{\kappa \kappa^{\prime}}^L=\frac{1}{4}(-1)^{j+j^{\prime}-{\ell}-l^{\prime}}(2 L+1)\left(\begin{array}{ccc}{\ell}^{\prime} & {\ell} & L \\ 0 & 0 & 0\end{array}\right)^2\left(\begin{array}{ccc}j^{\prime} & L & j \\ -\frac{1}{2} & 0 & \frac{1}{2}\end{array}\right)^{-2}\left[(-1)^{j+j^{\prime}-{\ell}-{\ell}^{\prime}}(2 j+1)\left(2 j^{\prime}+1\right)\left(\begin{array}{ccc}l^{\prime} & {\ell} & L \\ 0 & 0 & 0\end{array}\right)^2\right. \\ &\left.+8 \sqrt{{\ell}^{\prime}\left({\ell}^{\prime}+1\right) {\ell}({\ell}+1)}\left(\begin{array}{ccc}{\ell}^{\prime} & {\ell} & L \\ 0 & 0 & 0\end{array}\right)\left(\begin{array}{ccc}{\ell}^{\prime} & {\ell} & L \\ -1 & 1 & 0\end{array}\right)-4\left(\kappa^{\prime}+1\right)(\kappa+1)\left(\begin{array}{ccc}{\ell}^{\prime} & {\ell} & L \\ -1 & 1 & 0\end{array}\right)^2\right].\end{aligned}
    \end{equation}
\end{widetext}

The coefficient $D_{\kappa \kappa^{\prime}}^L$ has a similar form structure and can be obtained from the expression for $C_{\kappa\kappa'}^L$ by making simple index substitutions: specifically, replacing $\kappa$ with $-\kappa$ and $\ell$ with $|1/2 - \kappa| - 1/2$. 
For a detailed discussion of the numerical stability of the above integrals and the methods used to compute the relativistic wave functions, see Refs.~\cite{Roberts:2015lga,Roberts:2016xfw, Alhazmi:2021qgd, Alhazmi:2025nvt}. Ref.~\cite{Caddell:2023zsw} highlighted the importance of accurately calculating the ionization form factor, which requires precise wave–function modeling to preserve the orthogonality of free–electron states. Our results show excellent agreement with Ref.~\cite{Caddell:2023zsw}, as well as with Ref.~\cite{Roberts:2016xfw}.

\section{Effective Limit of Ionization Form Factor}
\label{sec:limits}
Equipped with the full relativistic treatment, we return to one of our original questions: ``how does the ionization form factor deform the cross section?'' 
In particular, we are interested in whether the full treatment of the cross section incorporating the ionization form factor in Eq.~\eqref{eq:dsigma_4} can be approximated to the simpler free electron formalism in Eq.~\eqref{eq:dsigma_5}.
To this end, we investigate two aspects: we study (i) how the ionization form factor behaves at increasing electron recoil energies and whether it converges to a desired asymptotic form and (ii) the impact of the ionization form factor on the cross section for a wide range of the kinematic parameter space, focusing on low electron recoil energies only.

\subsection{Asymptotic behavior at high energies}
\begin{figure*}[t!]
\centering

\includegraphics[width=.955\textwidth]{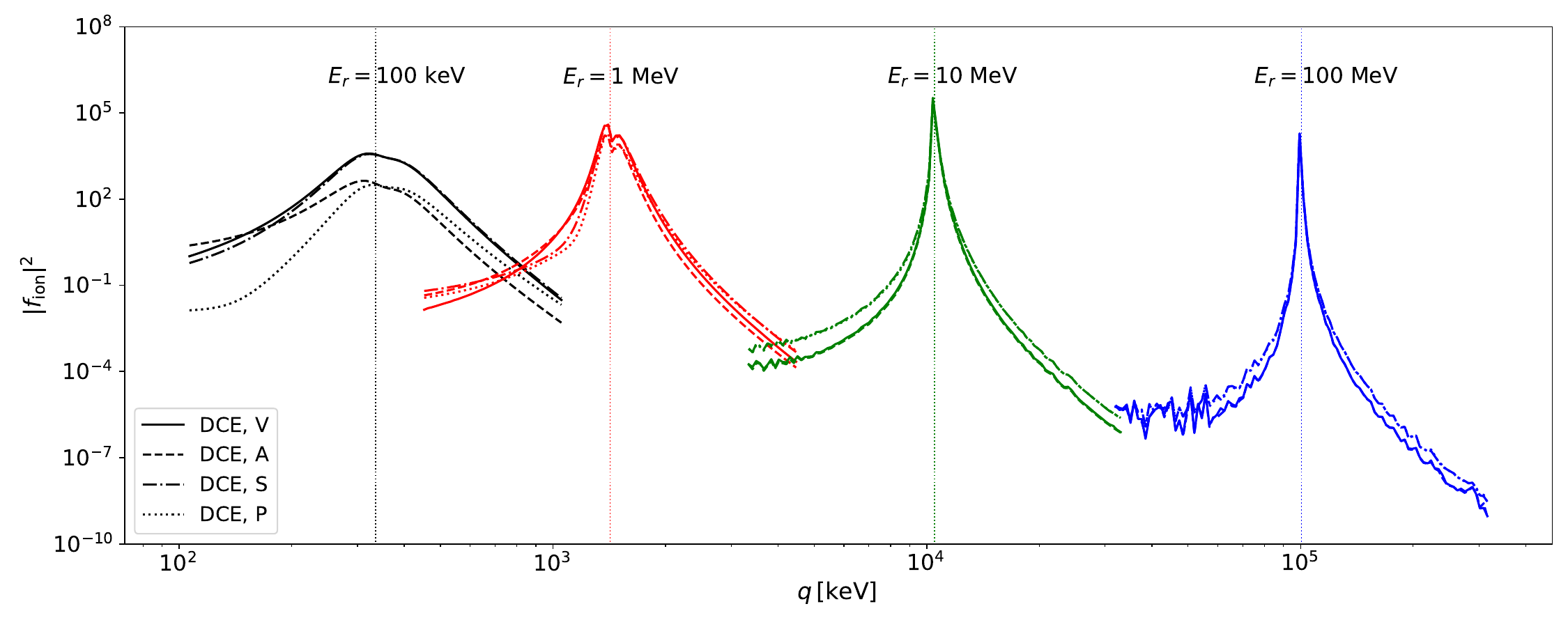}\\
\includegraphics[width=.955\textwidth]{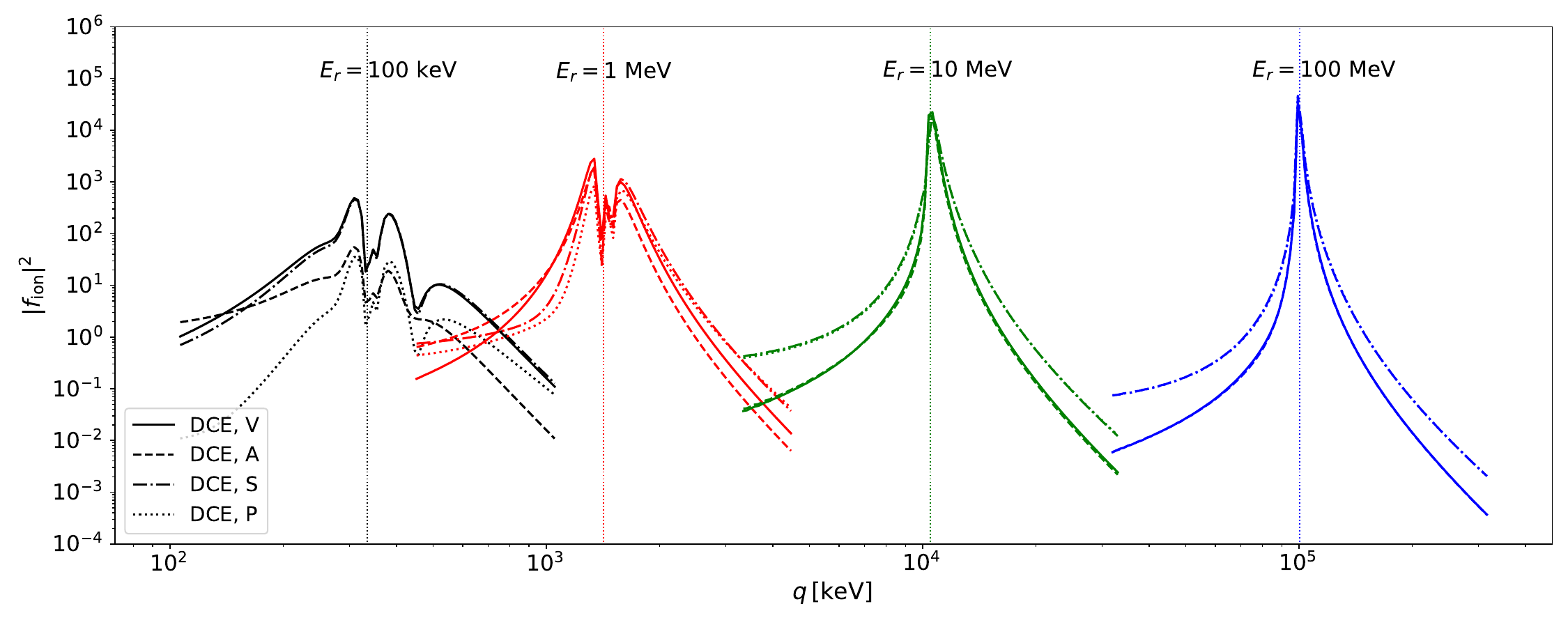}\\

\caption{The asymptotic behavior of the ionization form factor of the xenon atom for the 2p (top) and 3s (bottom) shells. All ionization form factor curves are estimated at various electron recoil energies: 100 keV (black), 1 MeV (red), 10 MeV (green), and 100 MeV (blue). The four mediator cases are represented by solid (vector), dashed (axial vector), dot-dashed (scalar), and dotted (pseudoscalar) curves. The vertical dotted lines represent the corresponding $k'$ value for each of the given $E_r$ values.}
\label{fig:free_limit}
\end{figure*}

As we mentioned earlier in Eqs. (\ref{eq:free_limit1}-\ref{eq:free_limit2}), at some limit, the ionization form factor should be able to retrieve the free electron case. 
To this end, we take the ionization form factor, varying the electron recoil energy as in Figure \ref{fig:free_limit}. 
For illustration purposes, we chose two example energy shells of the xenon atom: 2p (top) and 3s (bottom). 
We study the shape of the ionization form factor as the electron recoil energy increases: 100 keV (black), 1 MeV (red), 10 MeV (green), and 100 MeV (blue) for all of the mediator types. At each energy, we show the ionization form factor curves: vector (solid), axial vector (dashed), scalar (dot-dashed), and pseudoscalar (dotted).
We observe that, at each energy $E_r$, the curves are centralized at momentum transfer values of $q = k'$ as required by the momentum conservation.
At low electron recoil, the curves are broadly distributed, showing clear model dependence between various mediator types.
However, as the recoil energy increases, the curves exhibit a sharper peak centered at $q = k'$.
We clearly see that as the energy increases, the shape of the ionization form factor behaves like a Dirac delta function, conforming to the free electron treatment.
In other words, if the electron recoil energy is high ($E_r \gtrsim$ 10 MeV), the ionization form factor is unlikely to deform the cross section, validating the free initial electron assumption.

\subsection{Effective behavior at low energies}
Our results in Figure~\ref{fig:free_limit} clearly suggest that the effects due to the ionization form factor are significant at low electron recoil energies.
In fact, many DM detectors operate at electron recoil energy near the keV scale, making them sensitive to the ionization form factor. We now aim to quantitatively determine the validity of the free-electron approximation by numerically analyzing the relevant regions of parameter space. 
A straightforward approach to assess this is by comparing the total cross sections---for example, through the quantity $|1-\sigma/\sigma^{\rm free}|$ where $\sigma$ denotes the total cross section computed from Eq.~\eqref{eq:dsigma_4} and $\sigma^{\rm free}$ corresponds to that from Eq.~\eqref{eq:dsigma_5}.
While this method can provide a useful estimate of the approximation's validity in many cases, it may be less reliable in others. This limitation arises because integrating to obtain the total cross section inherently discards valuable information about the shape of the differential cross section.
To retain the shape information, we define the following measure\footnote{This is an example measure. We find that similar conclusions can be drawn using alternative measures, such as mean absolute-difference, maximum absolute-difference, and maximum square-difference.}
\begin{equation}
    R_{\sigma} \equiv \sqrt{ \frac{\sum_{i=1}^{N}\left( \dfrac{d\sigma^{\rm free}}{d E^{i}_r} - \dfrac{d\sigma}{d E^{i}_r} \right)^2}{\sum_{i=1}^{N}\left( \dfrac{d\sigma}{d E^{i}_r} \right)^2}}.   
\end{equation}
One may argue that the free-electron assumption is valid in regions where $R_{\sigma}$ approaches zero, but becomes unreliable when $R_{\sigma}$ deviates significantly from zero. 
For illustration, we consider the recoil energy range from 0 to 200 keV with a fixed interval of 0.5 keV. Consequently, the index 
$i$ runs from 1 to $N=400$.
Although this choice is primarily illustrative, the selected energy range is relevant to certain DM detectors, such as the XENONnT experiment.
We choose energy levels from 3s to 5p which account for effectively 44 electrons of the xenon atom, as we find that the contributions from the remaining innermost electrons are negligible.
\begin{figure*}[t!]
\centering

\includegraphics[width=.325\textwidth]{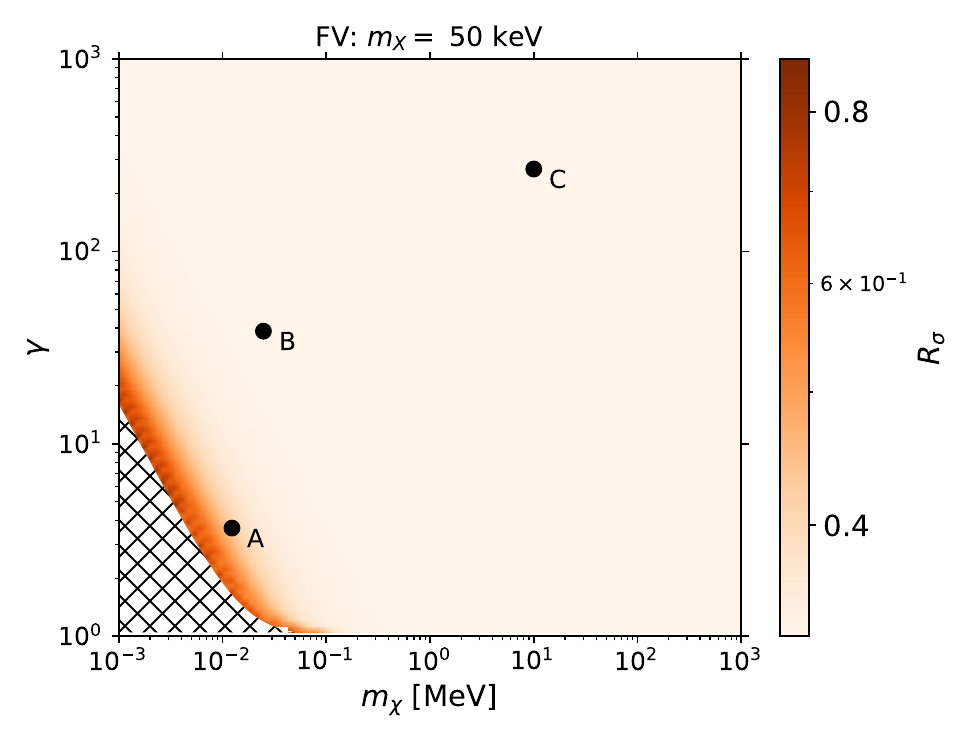}
\includegraphics[width=.325\textwidth]{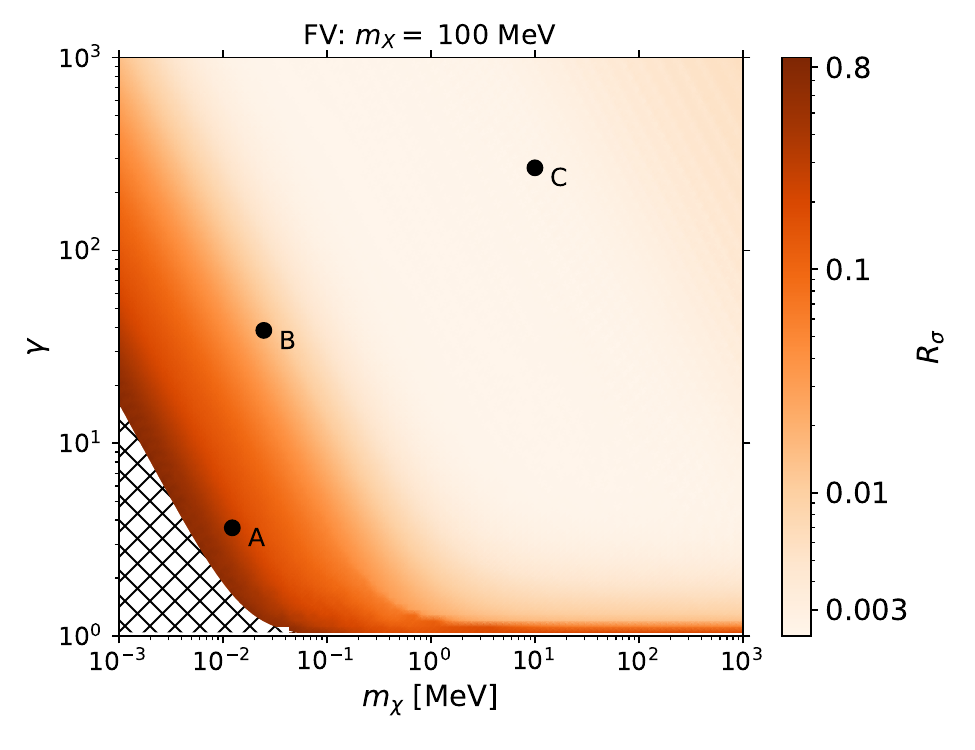}
\includegraphics[width=.325\textwidth]{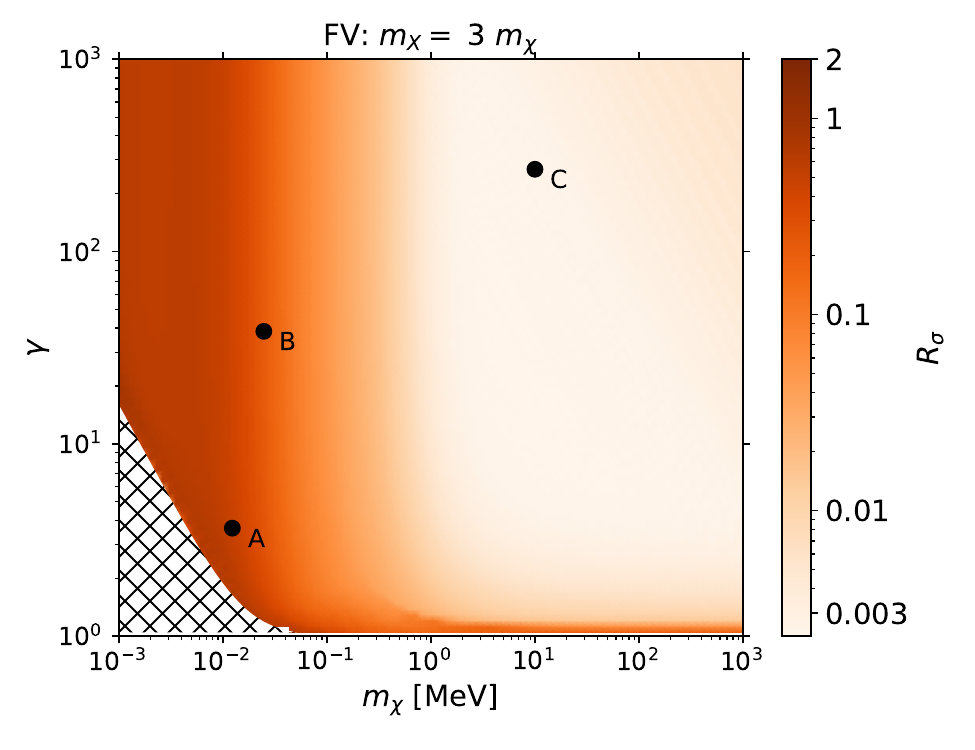}\\

\caption{$R_{\sigma}$ for the FV case representing different mediator mass regimes of $m_X =$ 50 keV (left), $m_X =$ 100 MeV (center), and $m_X = 3 \, m_{\chi}$ (right). The color scale in each panel is normalized to the maximum $R_\sigma$ value within that panel.}
\label{fig:csvar_ABC_FV}
\end{figure*}
\begin{figure*}[t!]
\centering

\includegraphics[width=.32\textwidth]{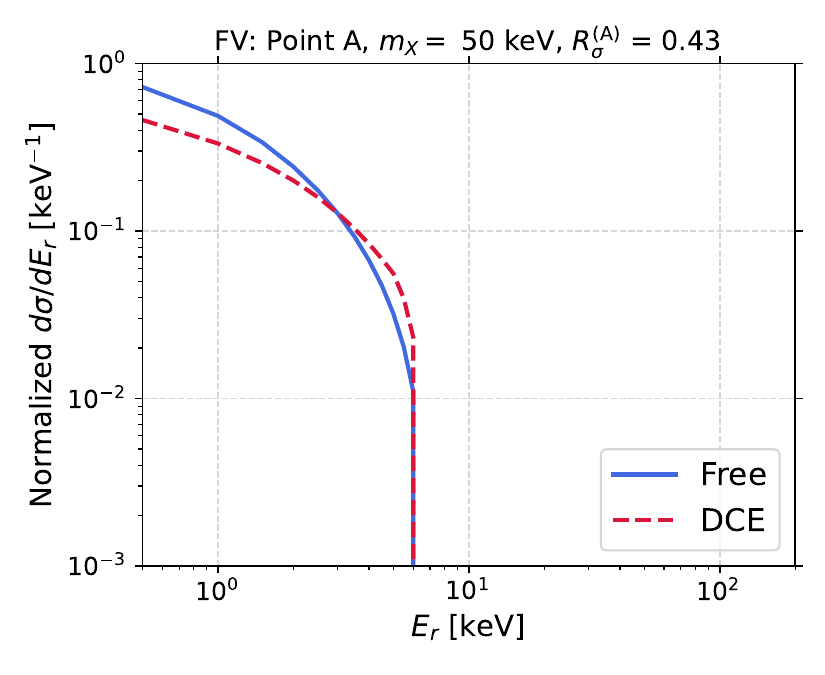}
\includegraphics[width=.32\textwidth]{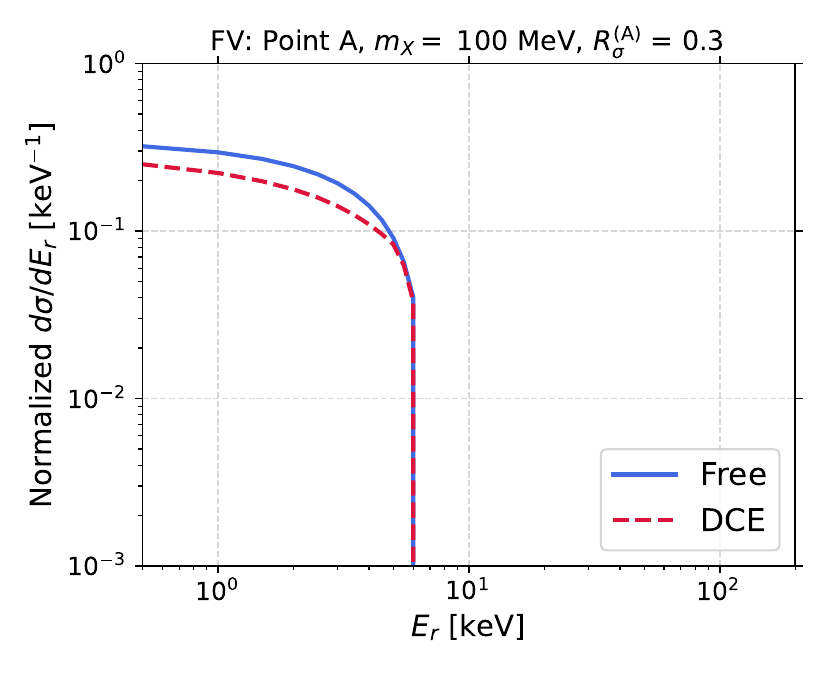}
\includegraphics[width=.32\textwidth]{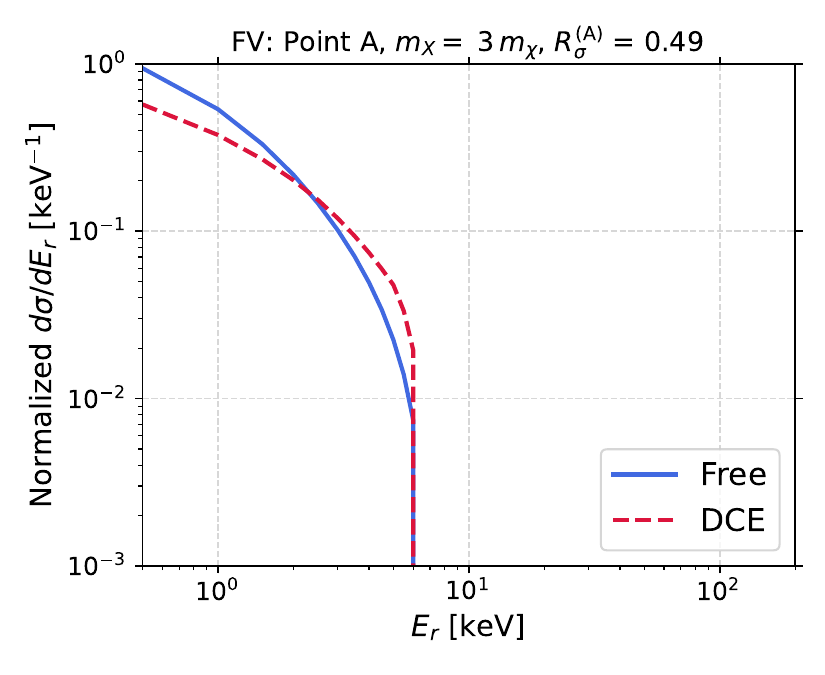}\\

\includegraphics[width=.32\textwidth]{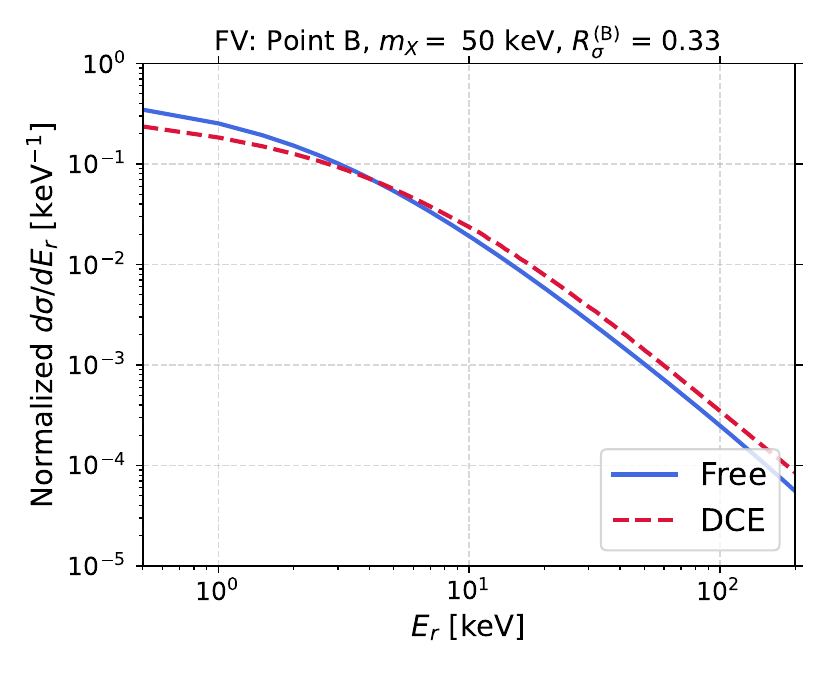}
\includegraphics[width=.32\textwidth]{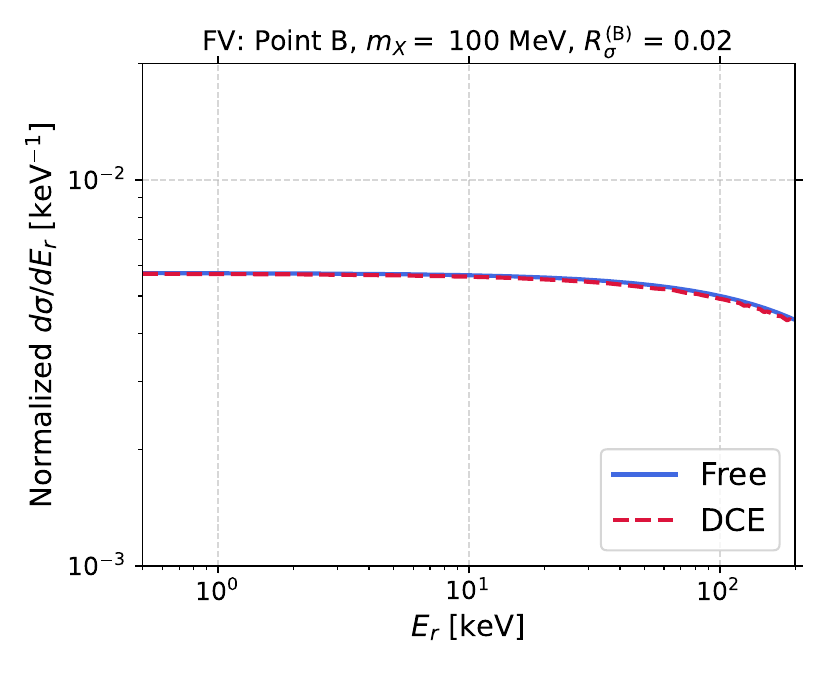}
\includegraphics[width=.32\textwidth]{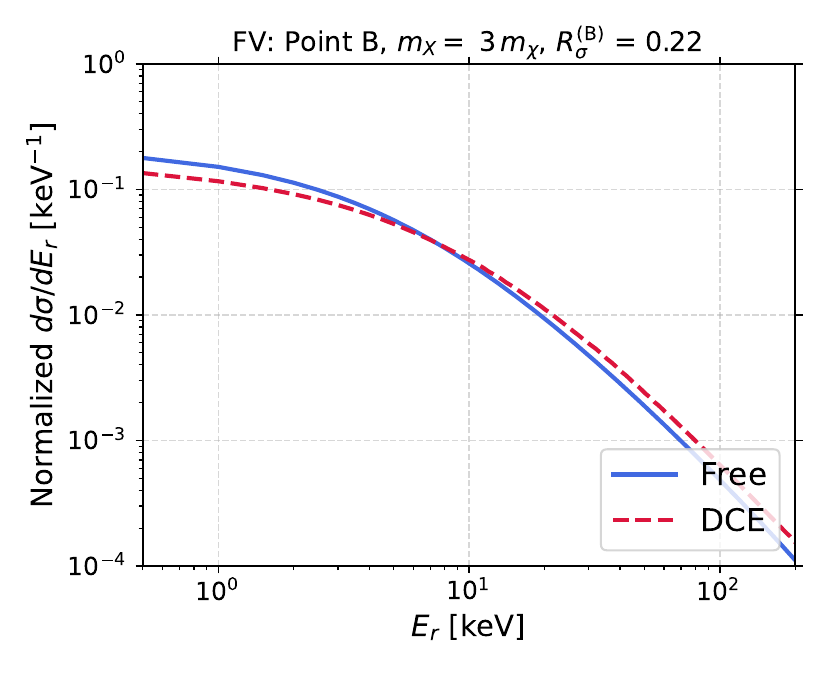}\\

\includegraphics[width=.32\textwidth]{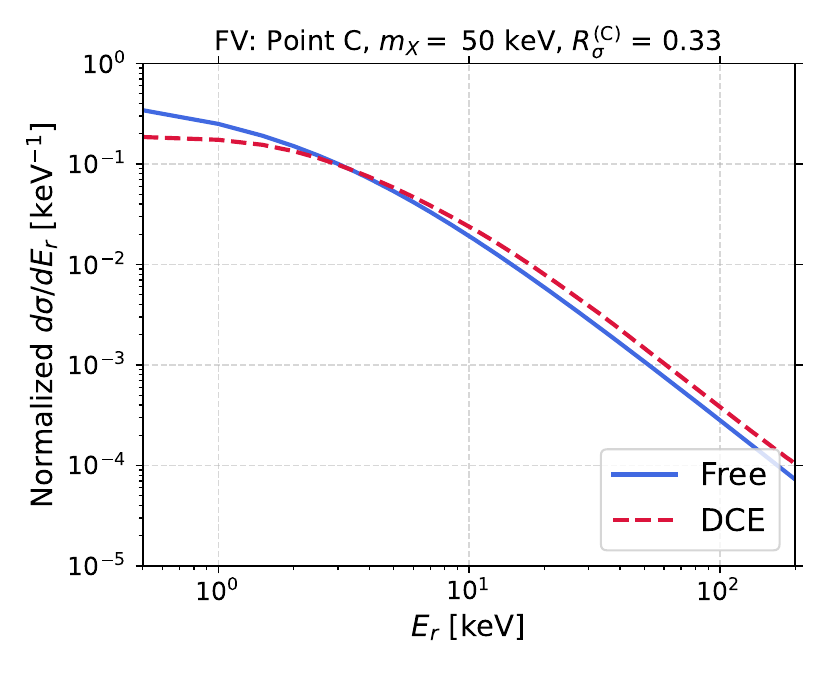}
\includegraphics[width=.32\textwidth]{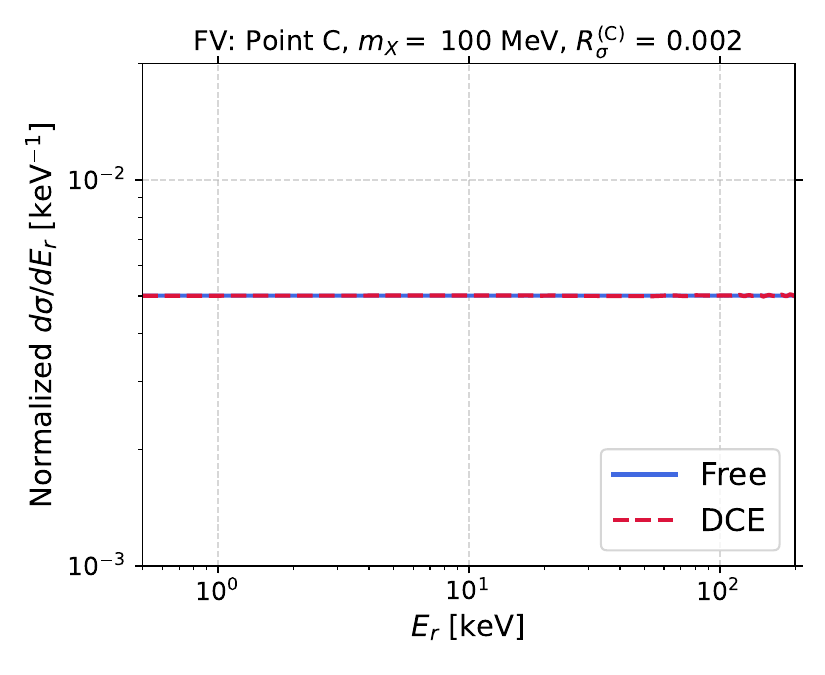}
\includegraphics[width=.32\textwidth]{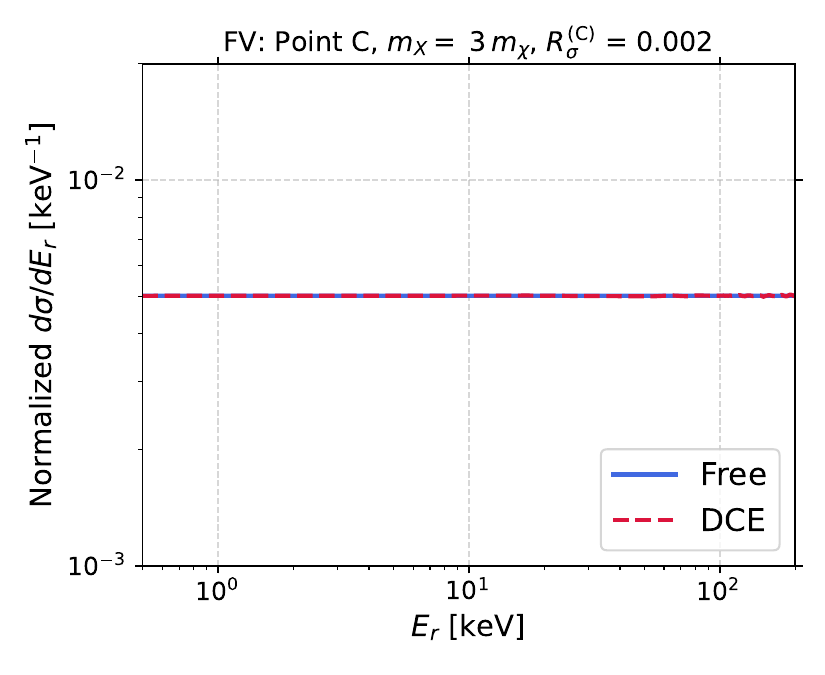}\\

\caption{Differential cross sections for the FV case in terms of the electron recoil energy for the free-electron assumption in solid blue and for the relativistic ionization form factor in dashed red. Left, center, and right panels respectively correspond to mediator masses $m_X =$ 50 keV, $m_X =$ 100 MeV, and $m_X = 3 \, m_{\chi}$. Top, middle, and bottom panels respectively correspond to points A, B, and C as indicated on Figure~\ref{fig:csvar_ABC_FV}.}
\label{fig:comparison_FV}
\end{figure*}

In Figure~\ref{fig:csvar_ABC_FV} we show color maps of $R_{\sigma}$ in the kinematic parameter space $(m_{\chi}, \,\gamma)$ for the vector mediator, shown at three representative mediator masses: $m_X = 50$ keV (left), $m_X = 100$ MeV (center), and $m_X = 3 \, m_{\chi}$ (right).
We henceforth adopt these mass choices for the rest of the paper. Several noteworthy observations follow.
Notice that the range of the color map is different based on the case. First and foremost, the validity of the free-electron assumption---and consequently, the necessity of incorporating the ionization form factor---depends critically on the specifics of the underlying model.
For example, when the mediator mass is $m_X = 100$ MeV, $R_\sigma$ ranges from 0.003 to 0.8, whereas for $m_X = 3\,m_{\chi}$, it spans from 0.003 to 4.
Nevertheless, a common feature in all panels is the hatched region in the bottom-left corner, corresponding to portions of the parameter space that cannot yield electron recoil energies above 1 keV.

At a fixed mediator mass, as illustrated in the left and center panels, we observe a gradual decrease in the value of $R_\sigma$ from the bottom-left to the top-right corner. This trend indicates that atomic effects on the cross section diminish with increasing energy $\gamma m_\chi$. This observation is consistent with the common expectation that electrons can be approximated as free particles when the incoming DM particle possesses sufficiently high energy. While this holds true in many cases, the precise behavior depends on the specific form of the squared matrix element and the dynamics of its convolution with the ionization form factor.
For example, the descent in the $R_\sigma$ value to the top-right corner is more pronounced for a mediator mass of 100 MeV than for 50 keV, as heavier mediators are more likely to facilitate larger momentum transfer to the target electron.
In contrast, the value of $R_{\sigma}$ for the dynamic choice $m_X = 3 \, m_{\chi}$ (right panel) appears more sensitive to the DM mass than to the boost factor. 

These observations indicate that the impact of the ionization form factor is highly dependent on the mediator mass choice, making it to difficult to draw a generalized conclusion.
To develop our intuition on the atomic effects in the FV case, we choose three representative benchmark points A, B, and C (marked in Figure~\ref{fig:csvar_ABC_FV}) as follows:
\begin{eqnarray}
    \text{Point A} &=& \{m_{\chi} = 15 \text{ keV}, \quad \gamma = 4\}, \nonumber \\
    \text{Point B} &=& \{m_{\chi} = 25 \text{ keV}, \quad \gamma = 40\}, \label{eq:ABCpoints} \\ 
    \text{Point C} &=& \{m_{\chi} = 10 \text{ MeV}, \quad \gamma = 300\}. \nonumber
\end{eqnarray}
For these benchmark points, the energy $\gamma m_{\chi}$ ranges from well below the electron mass (A), to near the electron mass (B), and to significantly above the electron mass (C).
The full differential cross section given by Eq.~\eqref{eq:dsigma_4} (dashed red) and that of the free-electron assumption in Eq.~\eqref{eq:dsigma_5} (solid blue) at the three points are contrasted in Figure~\ref{fig:comparison_FV}. 
Here, the points A, B, and C are arranged from top to bottom, while the mediator mass scenarios $m_X =$ 50 keV, $m_X =$ 100 MeV, and $m_X = 3 \, m_{\chi}$ are arranged from left to right.

In the left panel, the mediator mass is chosen to be much smaller than that of the electron mass, corresponding to the FV case in the $\beta \ll 1$ limit as defined in Table~\ref{tab:RMES}.
For small values of $x$, i.e., electron recoil energies much smaller than the electron mass, the dominant contribution in the matrix element squared is due to $\alpha^2 \gamma^2/x^2$ term.
This consequently makes the differential cross section in Eq.~\eqref{eq:dcsdx} dominated by a $1/E_r^2$ dependence, leading to the falling behavior observed in all distributions in the left panel.
Moreover, the distributions incorporating atomic effects closely resemble those based on the free-electron treatment, with only mild deviations. 
This is reflected in the estimated $R_\sigma$ measure, which decreases from 0.43 to 0.33 as we move from point A to point C. 
Neglecting atomic physics effects in this regime can lead to errors in the cross section at the level of several tens of percent.

In the center panel, the mediator mass lies in the $\beta\gg1$ regime (see the left column of Table~\ref{tab:RMES} for the FV case).
In this limit, the matrix element squared is dominated by the constant term, making the leading contribution independent of the electron recoil energy. As a result, the differential cross section becomes approximately flat over a broad range of electron recoil energy values. 
We observe that the cross section obtained using the atomic effects treatment closely matches that of the free-electron approximation, exhibiting minimal deviations. Indeed, the associated error remains well below the percent level across most of the parameter space, as illustrated in the center panel of Figure~\ref{fig:csvar_ABC_FV}.

In the right panel, the mediator mass is chosen as 
$m_X=3m_\chi$, making it dependent on the DM mass. 
Consequently, the parameter $\beta$ spans a wide range, varying from values much smaller than 1 to those much greater than 1.
As the DM mass increases, the form of the matrix element squared transitions from that corresponding to the right column to the left column of Table~\ref{tab:RMES}.
First, this explains why the right panel of Figure~\ref{fig:csvar_ABC_FV} exhibits growing sensitivity to the DM mass. 
Second, it accounts for the gradual shift in the distributions shown in the right panel of Figure~\ref{fig:comparison_FV}---from steeply falling to nearly flat---as one moves from the top to the bottom panel, i.e., as the dark matter mass increases.
In effect, this choice of mediator mass encapsulates and reinforces the observations discussed above. The discrepancy between the atomic effects treatment and the free-electron approximation ranges from substantial at low dark matter masses to negligible at high masses. This variation reflects the underlying transition from a light mediator regime (on the order of a few keV) to a heavy mediator regime (on the order of a GeV).

A key observation that encapsulates the FV case is that the differential cross section exhibits either a flat or falling behavior, depending on the mediator mass scenario and the specific kinematic parameters. Notably, the value of the measure $R_\sigma$ becomes negligible when the distribution is flat. In such cases, the free-electron approximation can be reliably employed without significantly underestimating or overestimating the cross section.

\begin{figure*}[t!]
\centering

\includegraphics[width=.325\textwidth]{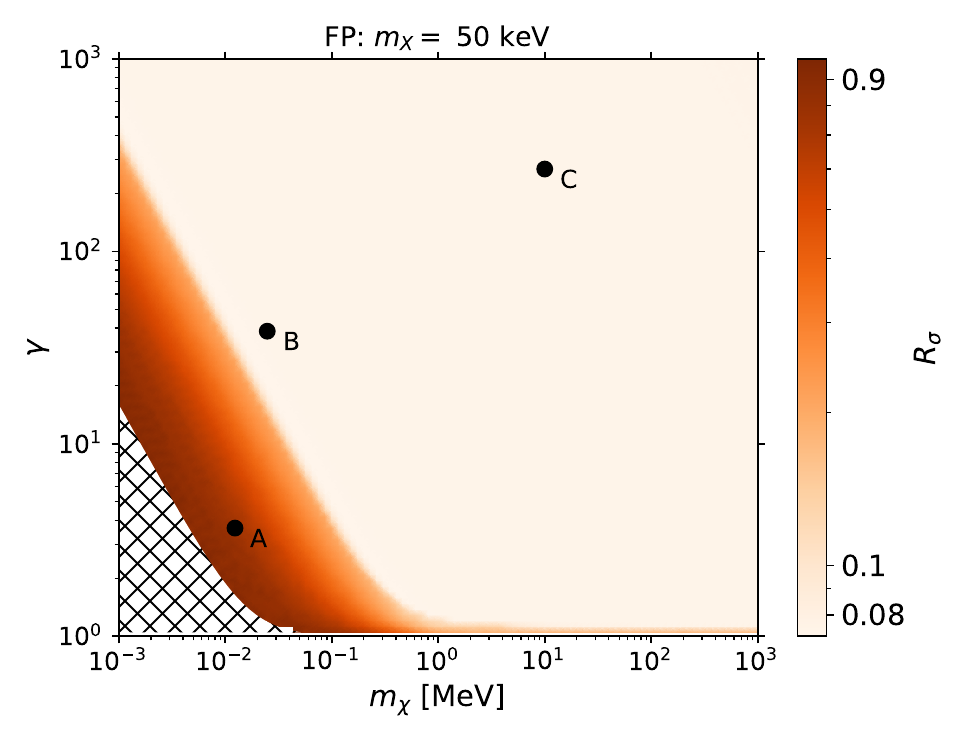}
\includegraphics[width=.325\textwidth]{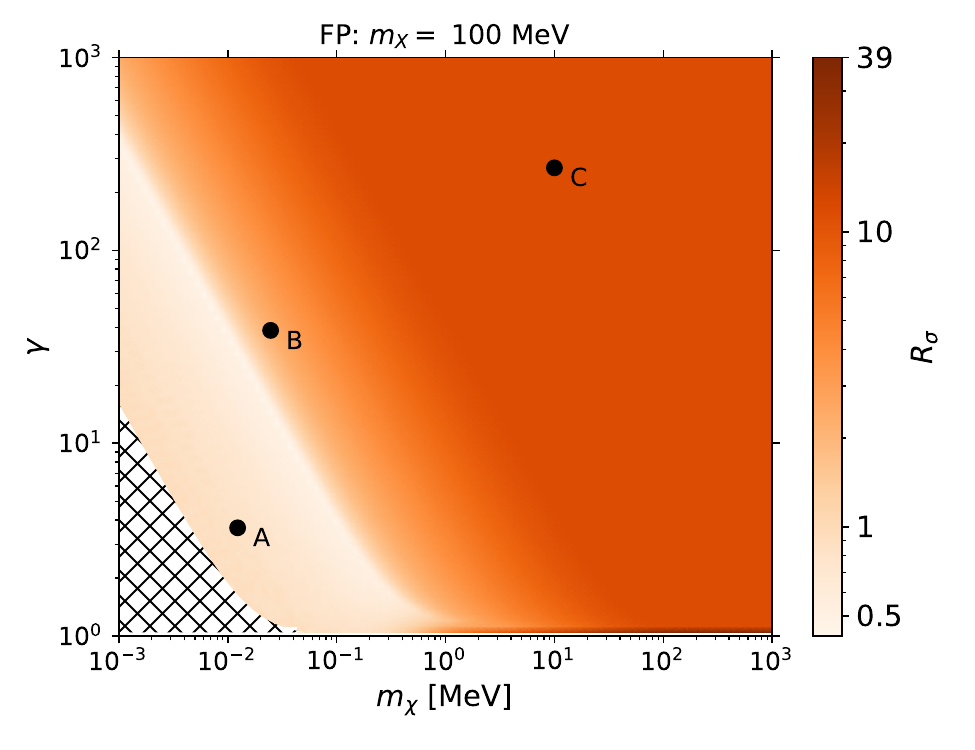}
\includegraphics[width=.325\textwidth]{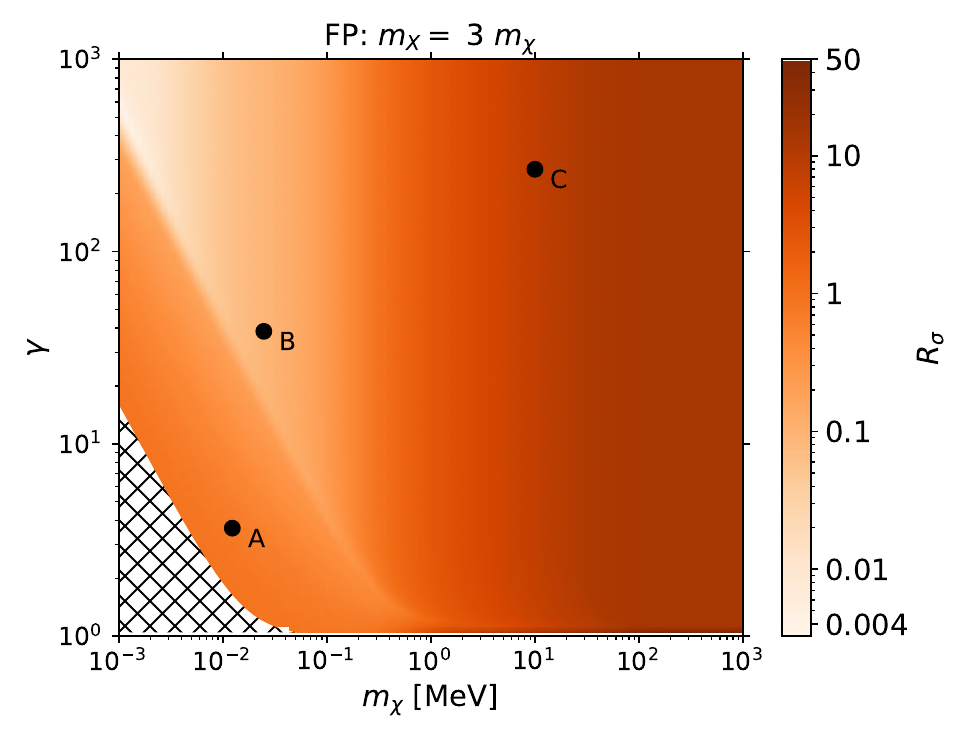}\\

\caption{$R_{\sigma}$ for the FP case representing different mediator mass regimes of $m_X =$ 50 keV, $m_X =$ 100 MeV, and $m_X = 3 \, m_{\chi}$ in the left, center, and right panels, respectively. The color scale in each panel is normalized to the maximum $R_\sigma$ value within that panel.}
\label{fig:csvar_ABC_FP}
\end{figure*}
\begin{figure*}[t!]
\centering

\includegraphics[width=.32\textwidth]{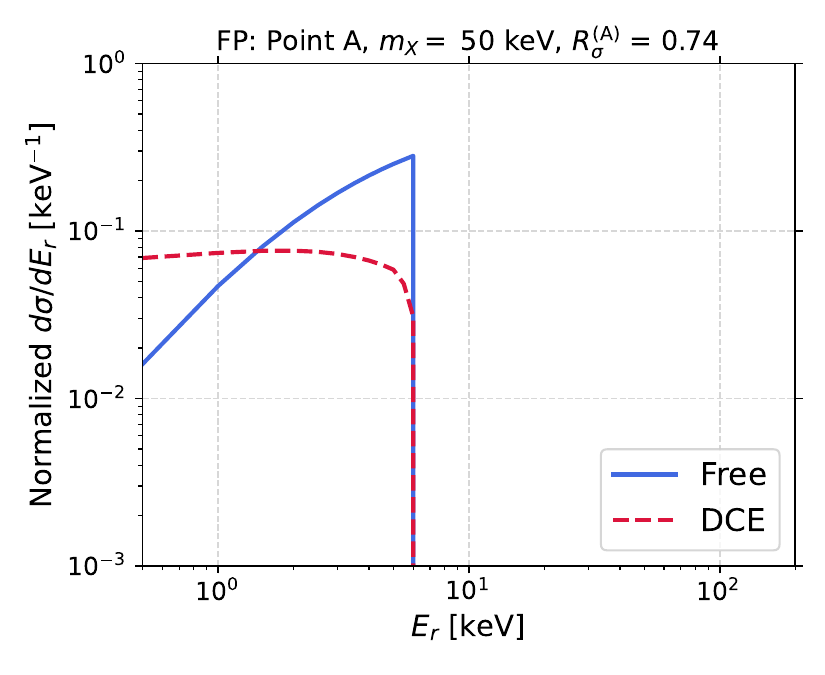}
\includegraphics[width=.32\textwidth]{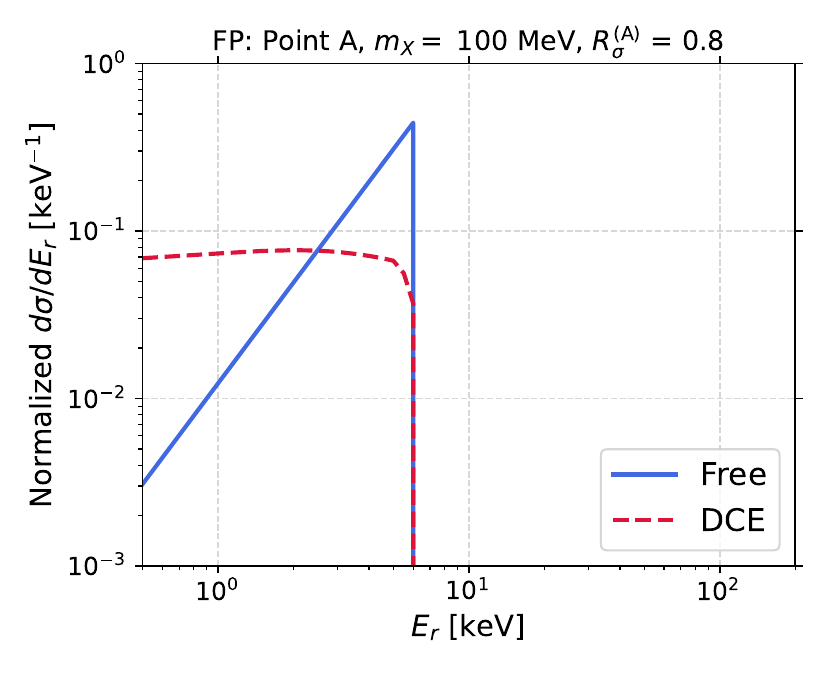}
\includegraphics[width=.32\textwidth]{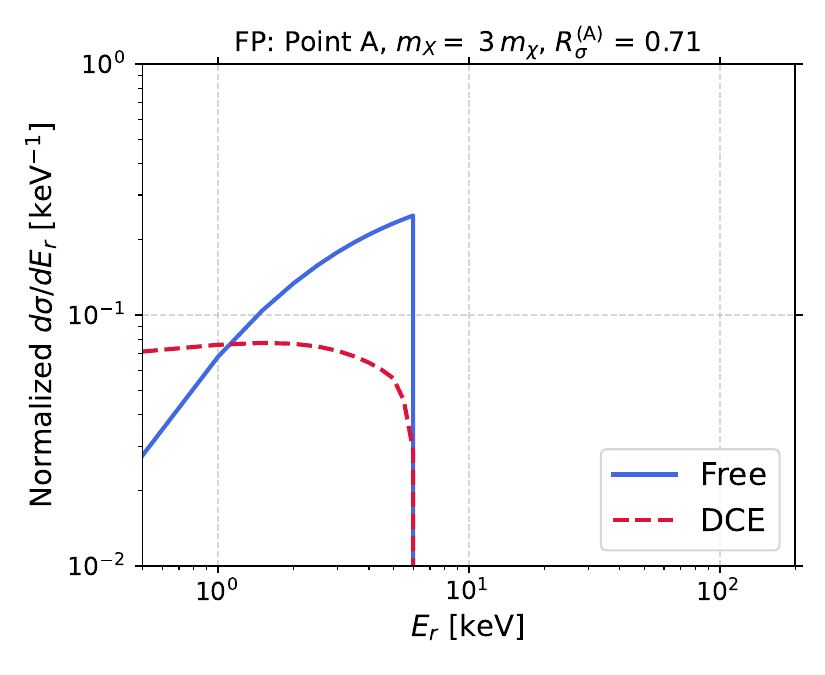}\\

\includegraphics[width=.32\textwidth]{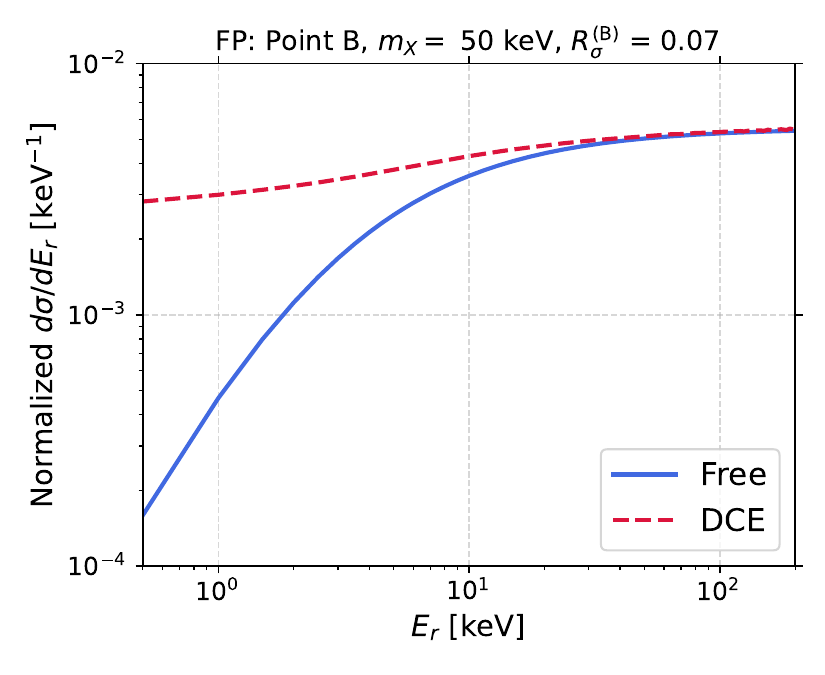}
\includegraphics[width=.32\textwidth]{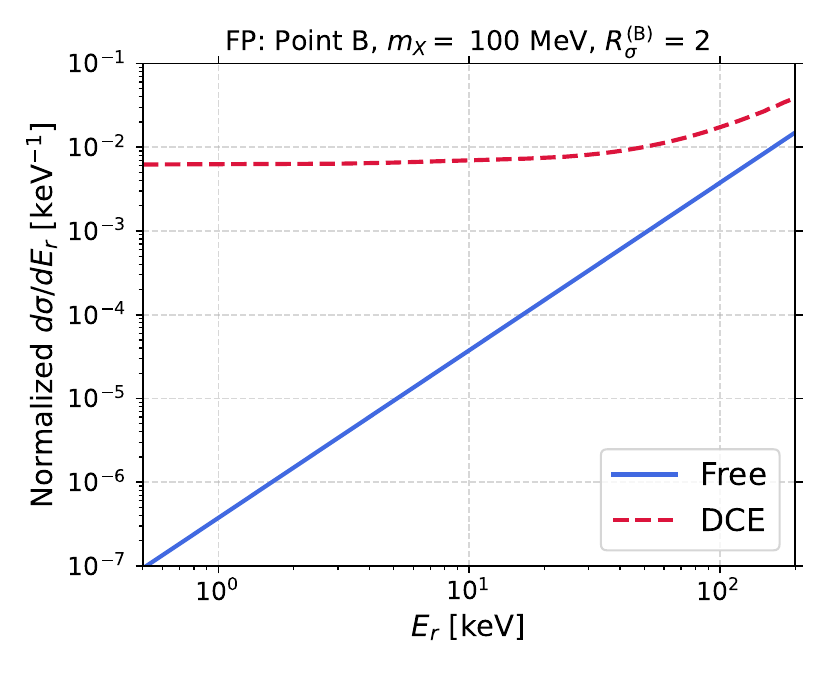}
\includegraphics[width=.32\textwidth]{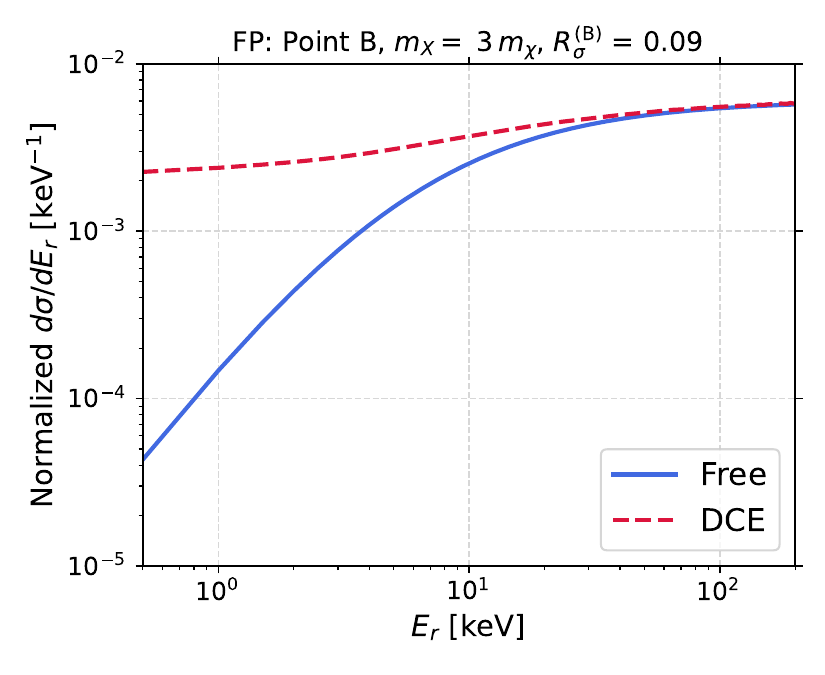}\\

\includegraphics[width=.32\textwidth]{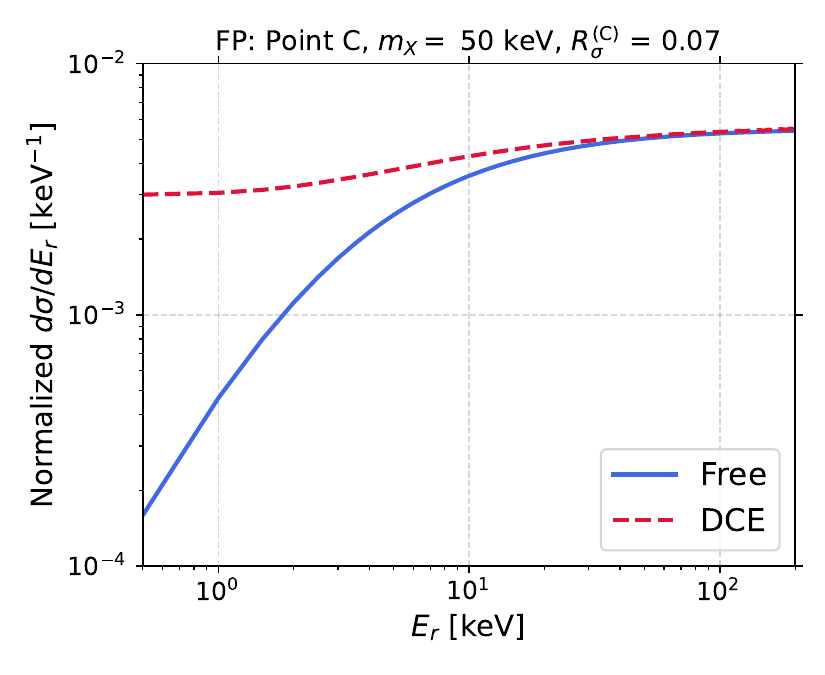}
\includegraphics[width=.32\textwidth]{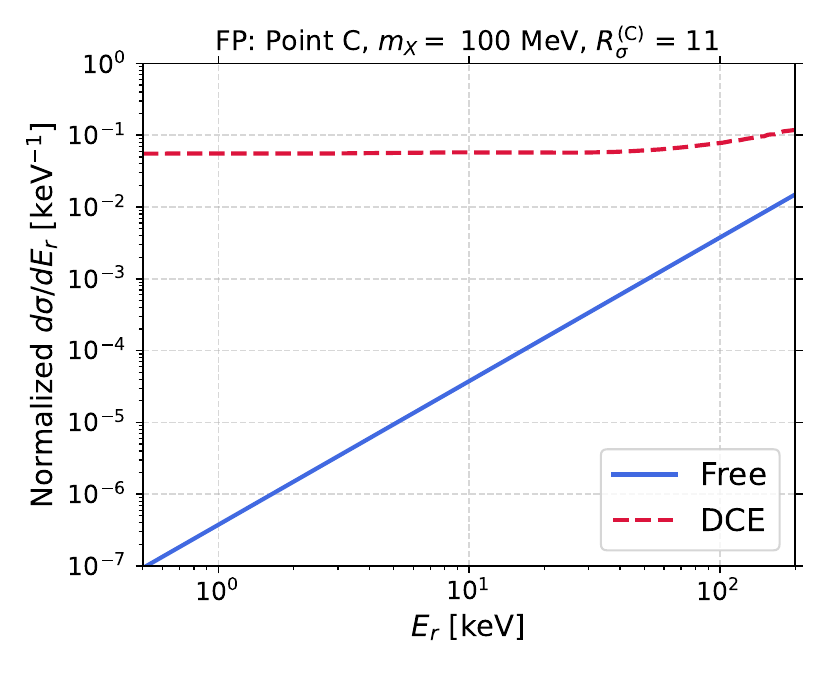}
\includegraphics[width=.32\textwidth]{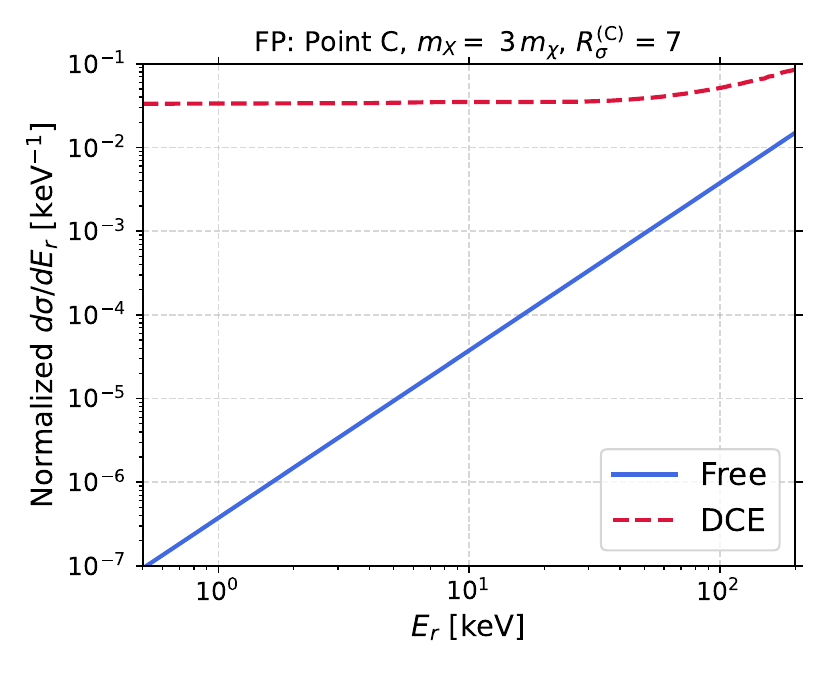}\\

\caption{Differential cross section for the FP case in terms of the electron recoil energy for the free electron treatment in solid blue and for the relativistic ionization form factor in dashed red. Left, center, and right panels respectively correspond to mediator masses $m_X =$ 50 keV, $m_X =$ 100 MeV, and $m_X = 3 \, m_{\chi}$. Top, center, and bottom panels respectively correspond to the kinematic points A, B, and C as indicated in Figure~\ref{fig:csvar_ABC_FP}.}
\label{fig:comparison_FP}
\end{figure*}
A question one may pose at this point is whether or not the observations above go through irrespective of the choice of the mediator type.
To address this question, we consider a contrasting scenario that is diametrically different from the FV case.
Figure~\ref{fig:csvar_ABC_FP} presents color maps of $R_\sigma$ for the FP case, using the same mediator mass choices as in the previous analysis.
Note that the matrix element squared in both $\beta \gg 1$ and $\beta \ll 1$ limits shows a rising behavior in the electron recoil energy.
In the left panel, we observe a gradual descent in the deviation between the atomic effects treatment and the free-electron approximation with increasing $\gamma m_{\chi}$.
In comparing the center panel of the FP case to that of the FV case we noticed a flipped behavior.
The value of the measure $R_{\sigma}$ ranges from small to extremely high as the energy $\gamma m_{\chi}$ increases.
The right panel shows a mixed behavior of the first two panels, featuring an intermediate region where the variation remains at or below the percent level.
Additionally, the variation is driven by the DM mass, owing to the choice scheme of the mediator mass.

The distributions at representative benchmark points A, B, and C, corresponding to those used earlier, are shown in Figure~\ref{fig:comparison_FP}.
When the mediator mass is small, as in the left panel, the distributions based on the free-electron treatment exhibit a rising trend and become noticeably flat with increasing electron recoil energy.
The center panel brings a new insight at a different trend.
Here the distributions are originally continuously increasing and the atomic effects tends to flatten out this behavior leading to extremely high variation.
In the FP case, we noticed that the value of the parameter $R_{\sigma}$ can be negligible when the mediator mass is light, whereas it remains non-negligible for a heavy mediator mass, in clear contrast to the FV case. 
This indicates that the deviation is minimized when the distribution is flat while it becomes significantly high if the distributions are either falling as in the FV case or rising as in here.
Finally the right panel includes a mixed behavior and allow for a small portion (near point B) of the kinematic parameter space to exhibit a flat trend and hence allows for negligible atomic effects.

\section{Detection Prospects at Detectors}
\label{sec:prospects}
Building on the formalism discussed thus far, we now examine the detection prospects at experimental facilities. 
For the time being, we defer the discussion of the origin of the DM and its properties to the next section. 
For a DM signal to be observable at a detector, the maximum electron recoil energy must be greater than the experimental threshold
\begin{eqnarray}
    E_r^{\rm max} > E_r^{\rm thr}.
\end{eqnarray}
In other words, there exists a nonzero probability of detecting DM with the given mass and energy as long as the above condition holds.
This energy threshold varies across experiments; for instance, certain direct detection experiments are sensitive to energy deposits below 1 keV, while large-volume neutrino experiments typically require energies above several tens of MeV.
\begin{figure}
    \centering
    \includegraphics[width=0.95\linewidth]{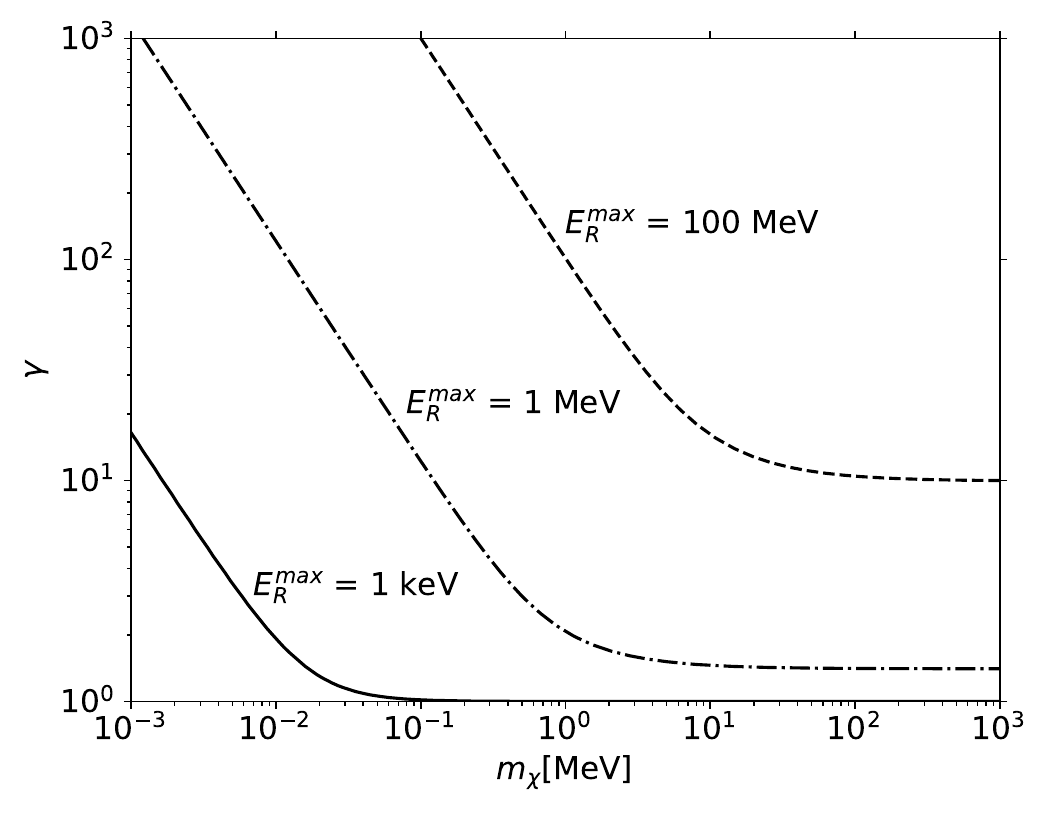}
    \caption{Maximum recoil energy based on the kinematics of the DM particle where solid, dash-dotted, dashed, and dotted contours represent 100 eV, 10 keV, 100 keV, and 1 MeV, respectively.}
    \label{fig:ermax_contours}
\end{figure}
Figure~\ref{fig:ermax_contours} illustrates the relevant parameter space in terms of DM speed and mass. The contour lines represent the maximum attainable electron recoil energy, with illustrative thresholds set at 1 keV, 1 MeV, and 100 MeV. For example, a DM particle sitting below the solid curve cannot produce electron recoils above 1 keV by any means, rendering this region of parameter space inaccessible to experiments even with low-energy detection thresholds of $\sim 1$~keV. 
In contrast, for large-volume neutrino experiments with MeV-scale energy thresholds, the region above the dot-dashed line becomes accessible.

By focusing on scenarios where DM moves at speeds closer to the speed of light, we observe that both low and high threshold experiments are theoretically capable of detecting fast-moving DM. 
In this scenario, we can take advantage of the fact that atomic electrons can be effectively treated as free particles at rest. Therefore, the total number of observable signal events is given by
\begin{equation}
    N^{\rm free} = \Phi \, \Delta T\, N_T\, \int^{E_r^+}_{E_r^-} \frac{d \sigma^{\rm free}}{d E_r} dE_r,
    \label{eq:master}
\end{equation}
where $E_r^\pm$ are defined as
\begin{eqnarray}
    E_r^+ &=&\min[E_r^{\max}, E_r^{\rm cut}], \\
    E_r^- &=& \max[E_r^{\min}, E_r^{\rm thr}],
\end{eqnarray}
and $E_r^{\rm cut}$ represents an upper limit imposed by detector saturation or other experimental constraints.
In addition to Eq.~\eqref{eq:master}, and for validation in low-threshold experiments, we also consider the full event rate including atomic effects. In this case, the total number of observable signal events is given by
\begin{equation}
    N = \Phi \, \Delta T\, N_T\, \int^{E_r^+}_{E_r^-} \frac{d \sigma}{d E_r} dE_r,
    \label{eq:master_full}
\end{equation}
where the integrand is given by Eq.~\eqref{eq:dsigma_4}.
The first factor in Eqs.~\eqref{eq:master} and \eqref{eq:master_full} represents the flux of the incoming DM, which depends on the source of the DM and its production mechanism. For the purposes of isolating the atomic effects, we fix the flux value to a reference value 
\begin{equation}
    \Phi^{\rm ref} = 10^{-7} \, {\rm cm}^{-1} \, {\rm s}^{-2}.
\end{equation}
Of course, as mentioned above, the flux is not constant in reality but varies with different model parameters. In the next section, we show how results can be readily obtained for realistic flux models through rescaling.
The observation time and the number of target electrons are denoted by $\Delta T$ and $N_T$, respectively. Our general approach involves four parameters: 
\begin{equation}
    {\rm free \,\, parameters} = \{m_{\chi}, \gamma, m_X, \overline{\sigma_{\chi _X}} \},
\end{equation}
where the coupling combination $g_{X} g_e$ can always be inferred from the reference cross section $\overline{\sigma_{\chi _X}}$. In the following, we illustrate the experimental sensitivity at two representative experiments:
\begin{itemize}
    \item {\textit{Super-Kamiokande (SK):}} SK is a water-based large-volume neutrino experiment with a volume of $22.4 \times 10^3 \, {\rm m}^3$. 
    The energy threshold of the experiment is chosen to be 100 MeV,
    and on average, the experiment detects an order of 726 sub-GeV and 197 multi-GeV electron recoil events annually~\cite{Dziomba2012}. SK has been collecting data for about 20 years which we set as an exposure time.

    \item {\textit{XENONnT Experiment:}} The primary objective of the multi-ton xenon-based experiment is to search for WIMP interactions with target nuclei; however, it also offers valuable sensitivity to electron recoil events. To facilitate a direct comparison between the predicted signal rate and the observed rate reported by the XENONnT Collaboration~\cite{XENON:2022ltv}, based on an exposure of 1.16 ton-years, we compute the signal rate while accounting for efficiency corrections, selection criteria, and binning. Our analysis focuses on the energy range ($1$ keV $ < E_R < 140$ keV)~\cite{XENON:2022ltv}.
\end{itemize}

We perform a signal-to-background log-likelihood analysis, with the exclusion significance defined as
\begin{equation}
    \sigma_{\rm exc} = \sqrt{2 \sum_{i=1} ^{N} \left[ B_i \ln \left( \dfrac{B_i}{B_i + S_i}\right) + S_i \right]},
\end{equation}
where $i$ runs over energy bins.
Here, $S$ denotes the signal predicted either by Eqs.~\eqref{eq:master} or~\eqref{eq:master_full}, depending on the case, while $B$ represents the background reported by the corresponding experiment.

\begin{figure*}[t!]
\centering

\includegraphics[width=.325\textwidth]{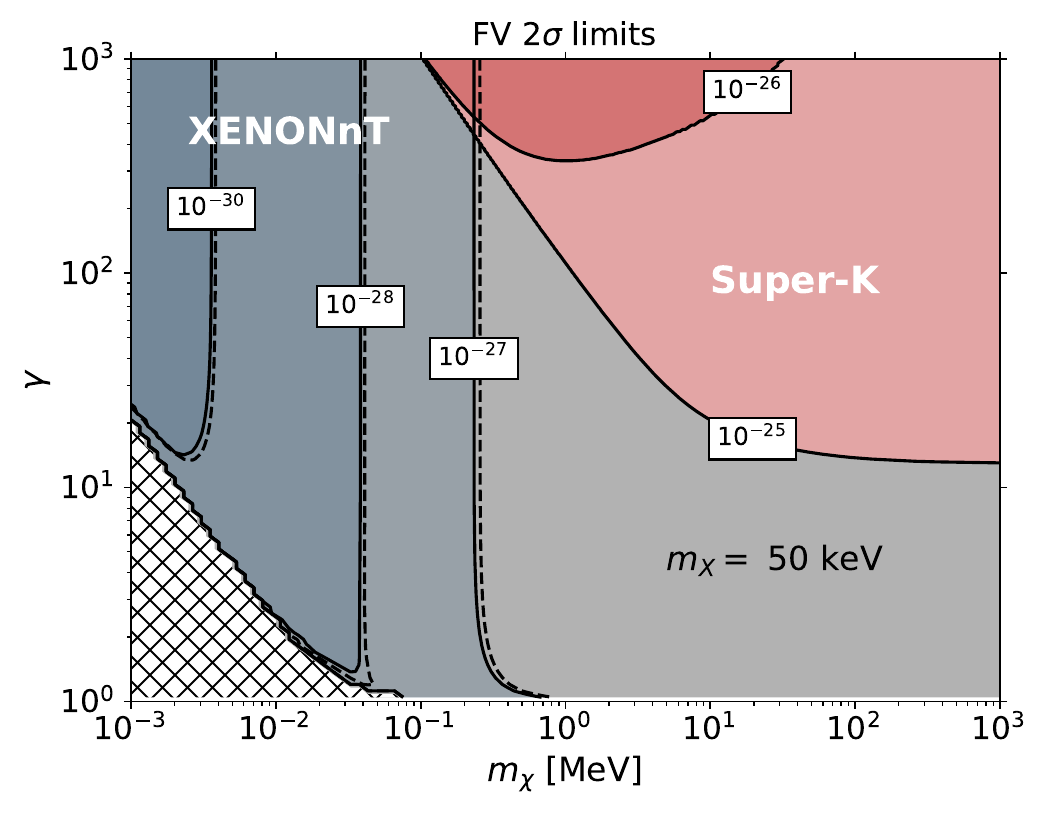}
\includegraphics[width=.325\textwidth]{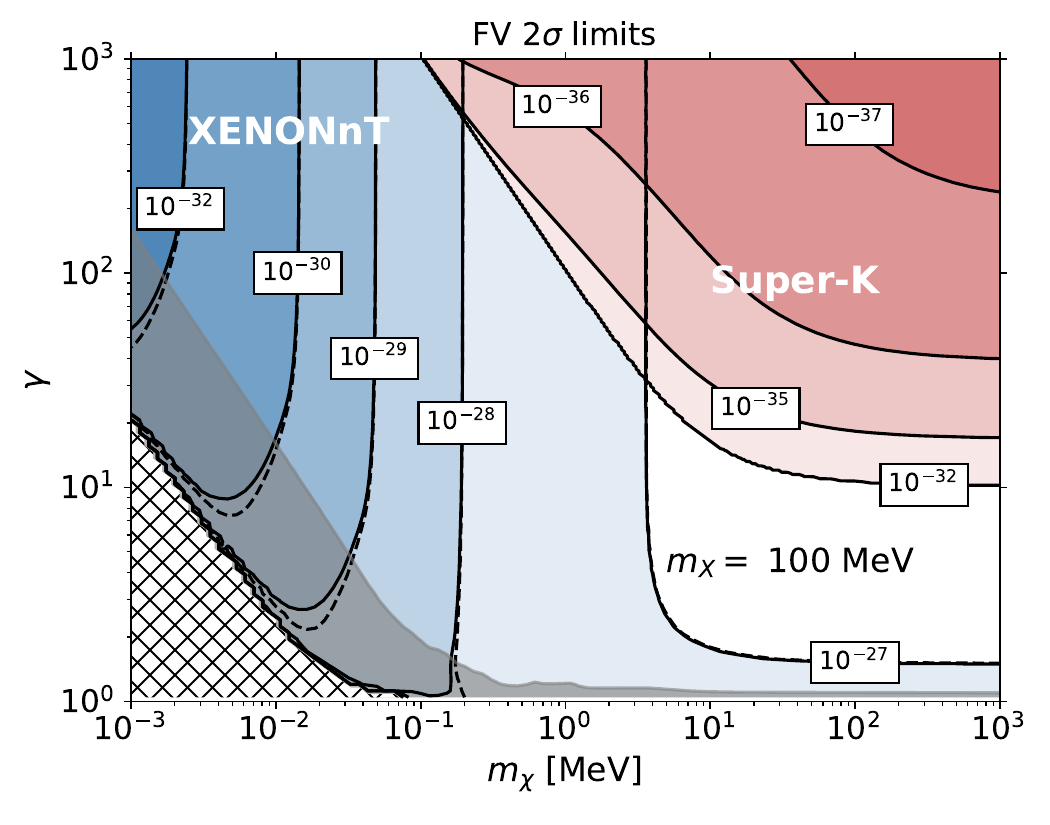}
\includegraphics[width=.325\textwidth]{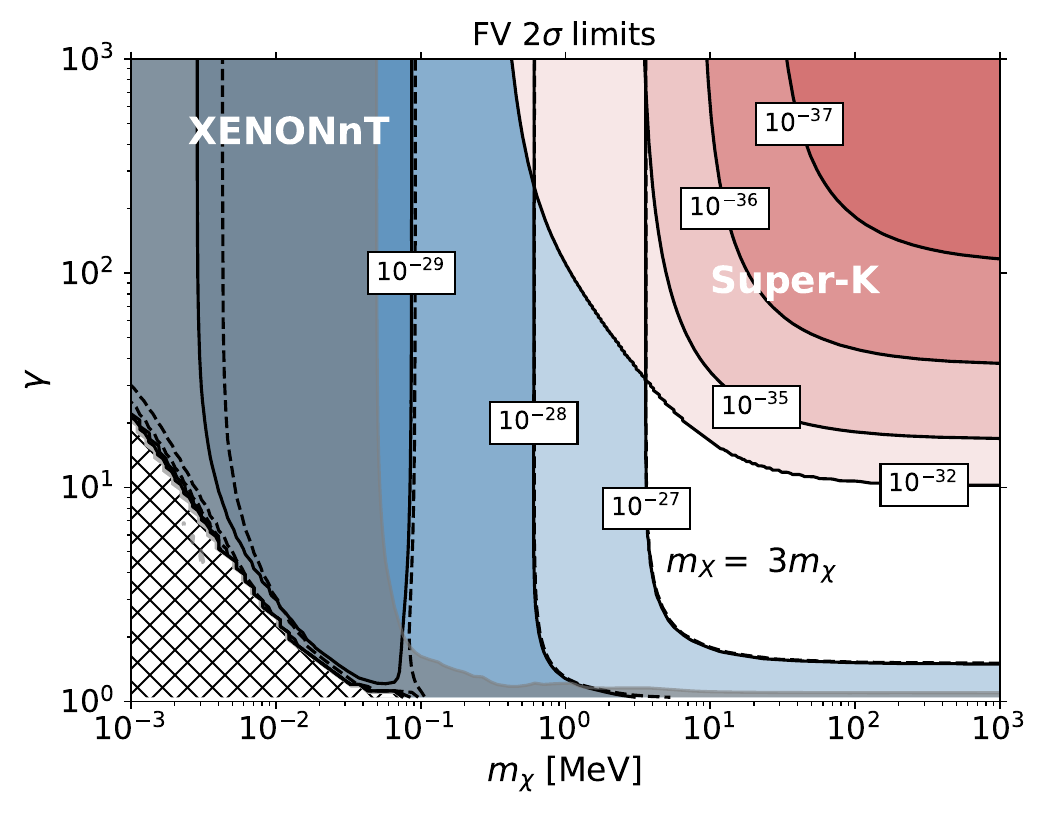}\\
\includegraphics[width=.325\textwidth]{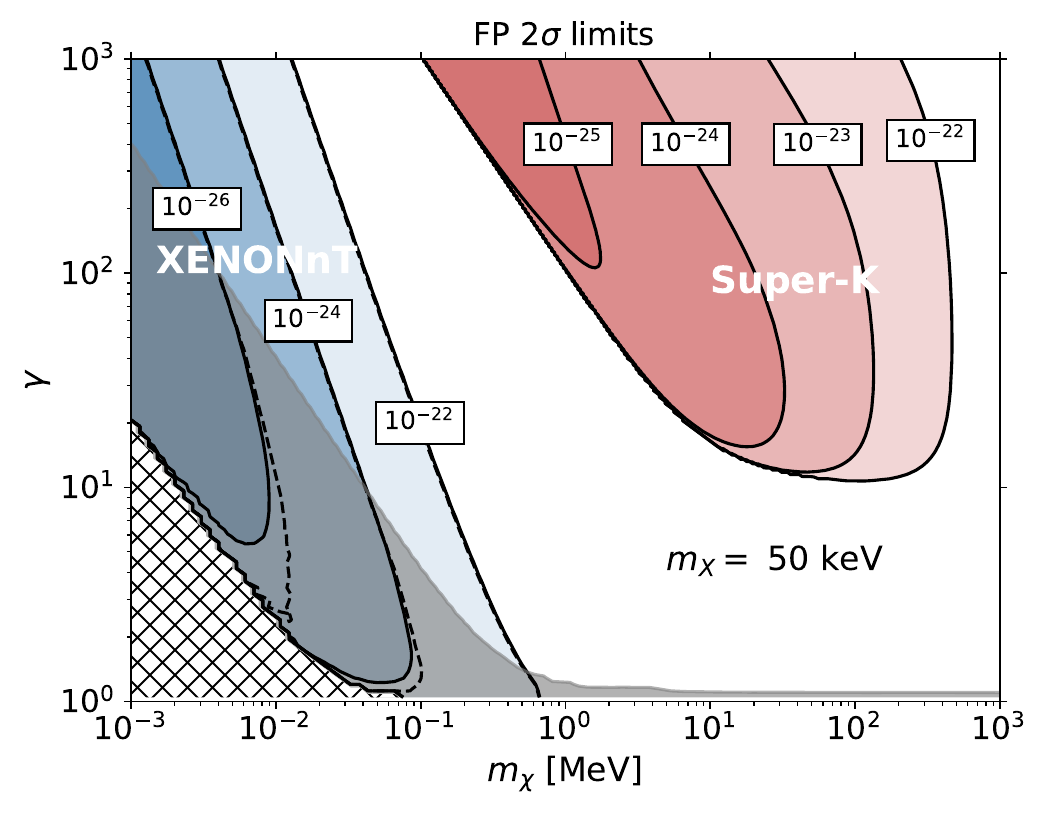}
\includegraphics[width=.325\textwidth]{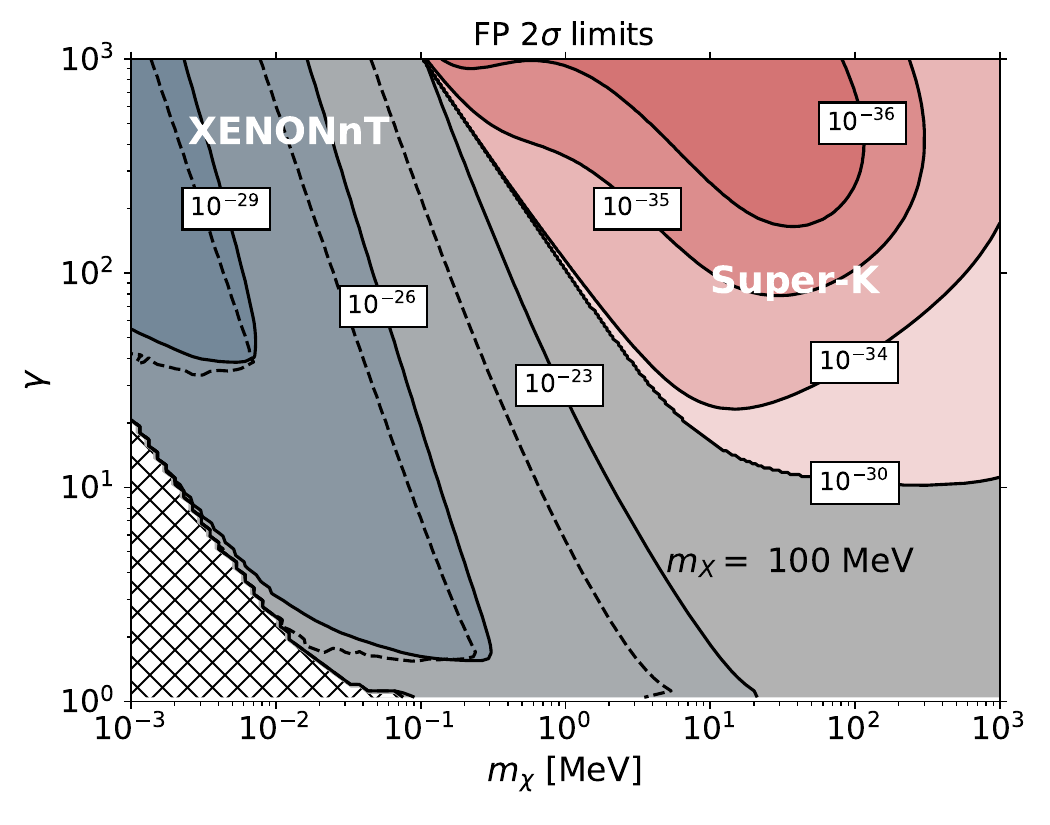}
\includegraphics[width=.325\textwidth]{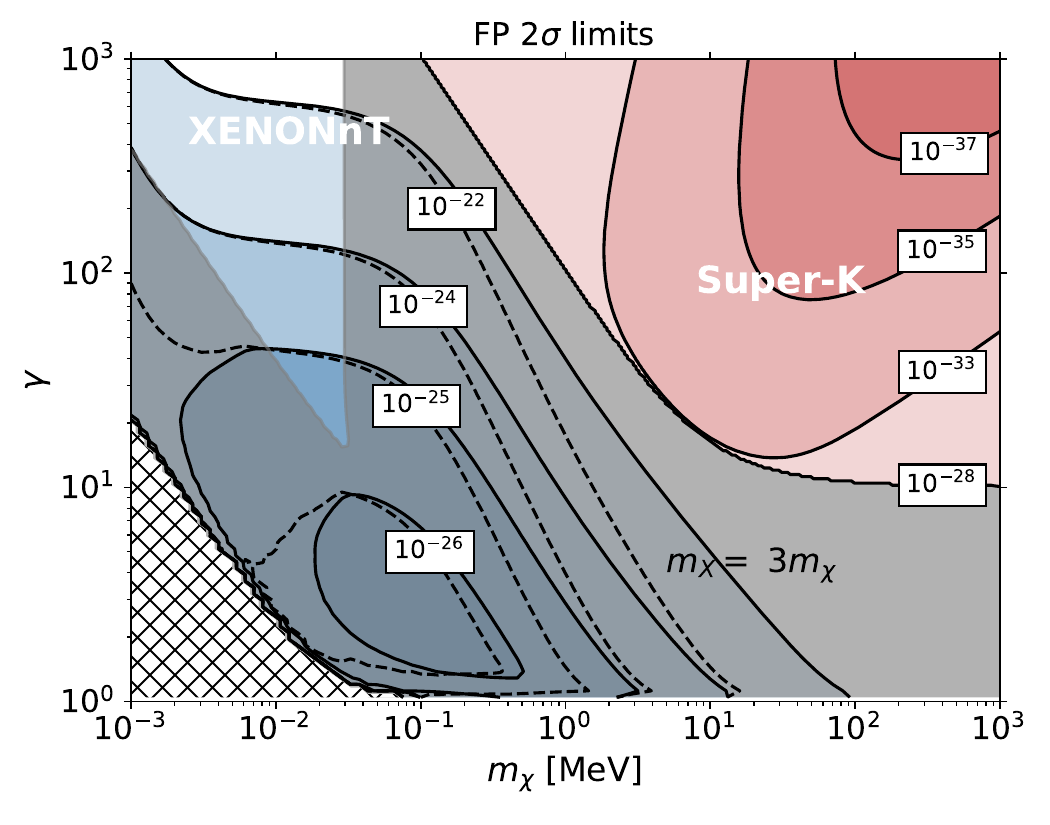}\\

\caption{$2\sigma$ exclusion limits for the FV case (top) and the FP case (bottom). Left, center, and right panels respectively correspond to mediator masses $m_X =$ 50 keV, $m_X =$ 100 MeV, and $m_X = 3 \, m_{\chi}$. The contour lines are indicated by the cross section in cm$^2$ where the blue (red) shaded regions correspond to the XENONnT (SK) limits. The gray shaded regions represent areas of the parameter space where $R_{\sigma} > 0.1$. The dashed contours in the XENONnT region indicate the corresponding limits derived from the signal defined in Eq.~\eqref{eq:master}.}
\label{fig:limits}
\end{figure*}
For illustrations, Figure~\ref{fig:limits} presents the $2\sigma$ exclusion limits for the FV case (top) and the FP case (bottom). 
The mediator mass choices are again $m_X = 50$ keV (left), $m_X = 100$ MeV (center), and $m_X = 3\, m_\chi$ (right).
Before delving into the model-specific results, we first highlight several general conventions that are consistently applied across all exclusion plots and analyses in this work. These key conventions are summarized as follows:
\begin{itemize} 
    \item[i.] Exclusion limits from the XENONnT experiment are uniformly represented by blue-shaded regions, while those from the SK experiment are represented in red. For XENONnT, the solid contours correspond to the limits derived from the signal defined by Eq.~\eqref{eq:master_full}, while the dashed contours correspond to those derived from the signal defined by Eq.~\eqref{eq:master}. For SK, all contours represent limits derived from the signal defined by Eq.~\eqref{eq:master}.
    
    \item[ii.] The intensity of the shading indicates the strength of the exclusion: lighter shades correspond to weaker constraints, whereas darker shades indicate stronger exclusions.

    \item[iii.] The number labeled on each contour line indicates the reference cross section, in cm$^2$, as defined in Eq.~\eqref{eq:refcsmod}, at which a $2\sigma$ exclusion is achieved.

    \item[iv.] The gray shaded region corresponds to $R_{\sigma} > 0.1$. This roughly translates to a theoretical uncertainty on the differential cross section greater than 10\%. 
    In other words, regions without gray shading correspond to areas in the $(m_{\chi}, \, \gamma)$ parameter space where the free-electron approximation remains approximately valid.
    This uncertainty applies only to the results associated with the XENONnT experiment, as XENONnT is a low-threshold experiment potentially sensitive to atomic effects, whereas SK operates at the MeV-scale threshold.

    \item[v.] Owing to its lower energy threshold, XENONnT generally probes a wider region of the kinematic parameter space, spanning both low and high values of $m_\chi$ and $\gamma$, depending on the specific model. 
    In contrast, SK is primarily sensitive to the upper-right region of the parameter space, where fast-moving DM particles are sufficiently energetic to induce detectable recoils. The hatched region in the bottom-left corner denotes the portion of parameter space where DM-induced electron recoil energies above 1 keV cannot be produced.
\end{itemize}

We now turn to model-specific observations. In the FV case shown in the top panel, the exclusion limits are stronger for the heavy mediator scenario compared to the light mediator, as expected. Notably, for XENONnT, the shape of the exclusion contour lines remains approximately consistent across mediator masses, with a gradual decrease in exclusion strength observed from left to right.
This observation admits a straightforward interpretation. At XENONnT, the electron recoil energies lie well below the electron mass, making the dimensionless parameter $x$ small. Consequently, the dominant term in the squared matrix element is proportional to $\alpha^2\gamma^2$ for both light and heavy mediator scenarios, as shown in Table~\ref{tab:RMES}.
In contrast to XENONnT, the exclusion limits at SK exhibit a different trend. For instance, in the light mediator case, the limits are stronger in the left region, whereas for a heavy mediator, the strongest limits appear in the upper-right region. This behavior can again be understood by examining the shape of the differential cross section, as given in Eq.~\eqref{eq:dcsdx}, and informed by the structure of the squared matrix element summarized in Table~\ref{tab:RMES}.
SK operates at high electron recoil energies, corresponding to the regime $x \gg 1$. In this limit, the differential cross section for the light mediator case becomes a decreasing function of the recoil energy. Moreover, not only does the shape exhibit a falling trend, but the total cross section also decreases with increasing $\alpha \gamma$, resulting in a higher event rate---and thus stronger limits---in the left region of the parameter space.
In contrast, for the heavy mediator case, the differential cross section remains relatively flat over a broad range of recoil energies, while the total cross section increases with $\alpha\gamma$. Consequently, the event rate---and hence the exclusion strength---is enhanced in the right region.
These trends in the differential cross section for the FV case, across light, heavy, and varying mediator masses, are illustrated in Figure~\ref{fig:comparison_FV}.

In the FP case shown in the bottom panel, the differential cross section exhibits an overall rising trend, as seen in Figure~\ref{fig:comparison_FP}. This rise is steeper for the heavy mediator scenario compared to the light mediator case, which tends to stabilize as the electron recoil energy increases.
At XENONnT, a comparison between the FP and FV cases reveals that the exclusion limits gradually weaken from left to right, although the contour lines exhibit a slight tilt.
This behavior arises from the fact that, unlike the FV case, the squared matrix element in the FP scenario is independent of both $\alpha$ and $\gamma$, as shown in Table~\ref{tab:RMES}.
At SK, where $x\gg1$, the matrix element squared for light mediator is always flat and the differential cross section is inversely proportional to $\alpha^2 \gamma^2$. As a result, the exclusion limit is stronger toward the left side of the SK region of interest.
In the heavy mediator case, the differential cross section scales as $(x/\alpha \gamma)^2$, leading to a stronger exclusion limit near the optimal region where the event rate is maximized, i.e., where $(x/\alpha \gamma)^2$ is maximized.

Across all panels of Figure~\ref{fig:limits}, for XENONnT we show, in addition to the limits derived from the full treatment including atomic effects (solid curves), the corresponding limits from the free-electron approximation (dashed curves). It is evident that the differences between the two treatments are large within the gray region and negligible outside it. This comparison delineates the parameter space in which the free-electron approximation is valid.

A summary of the limits for the various spin cases considered in this work is presented in Appendix~\ref{appendix:A}.

\section{Application to Benchmark DM Models}
\label{sec:interpretation}
In this section, we consider two representative mechanisms that can accelerate DM particles to high velocities. Specifically, we explore annihilation and cosmic-ray interactions as potential sources of fast-moving DM. 
Our focus is on characterizing the resulting DM flux, understanding the physical processes involved in its production, and discussing the implications for detection.
Here we briefly summarize the key ingredients needed to estimate the flux and refer readers to the sources cited in the Introduction and throughout this section for further details. Our focus is on interpretability and practical utility: specifically, we show how the limits derived for a fixed reference flux in the previous section can be straightforwardly reinterpreted for realistic model fluxes.

\subsection{Two-Component Boosted Dark Matter}
\begin{figure*}[t!]
    \centering
    \includegraphics[width=0.48\linewidth]{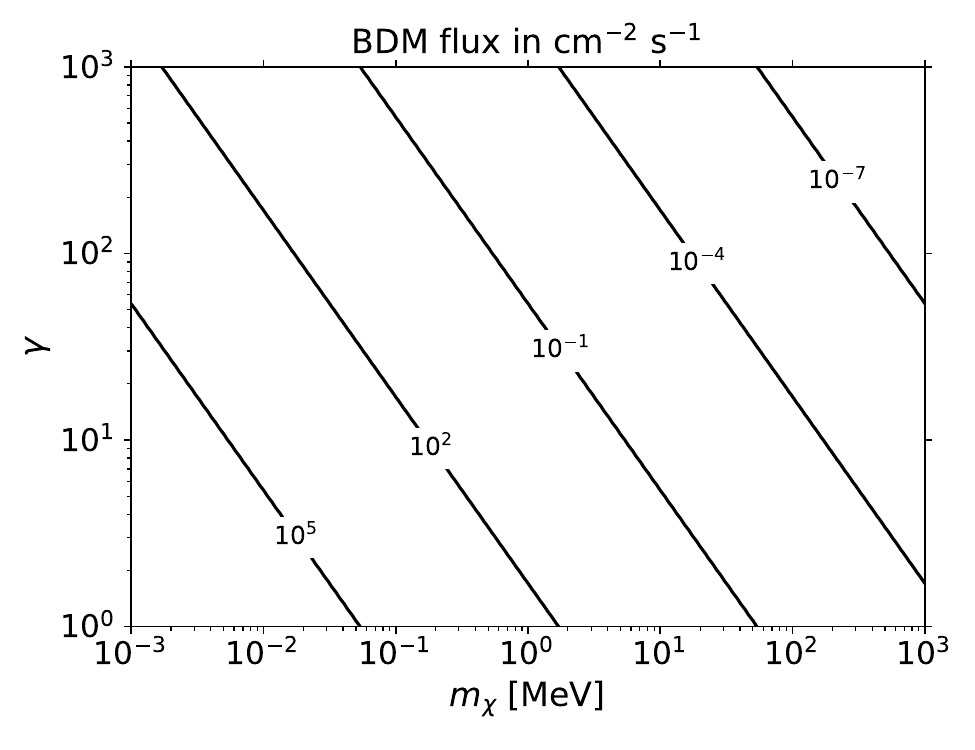}
    \includegraphics[width=0.48\linewidth]{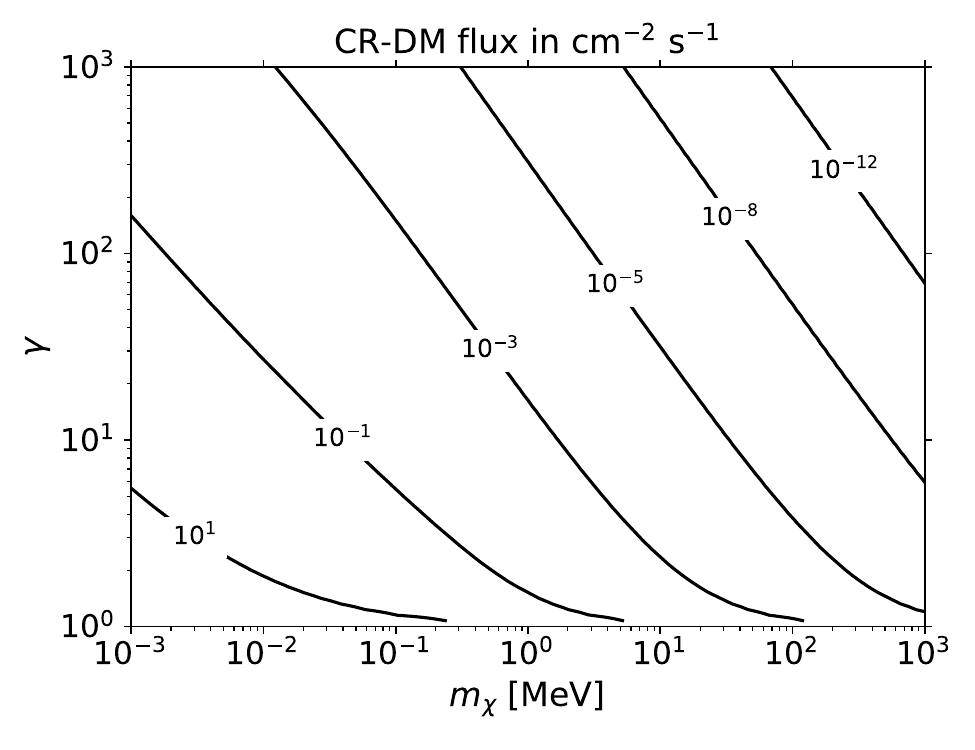}\\
    \includegraphics[width=0.48\linewidth]{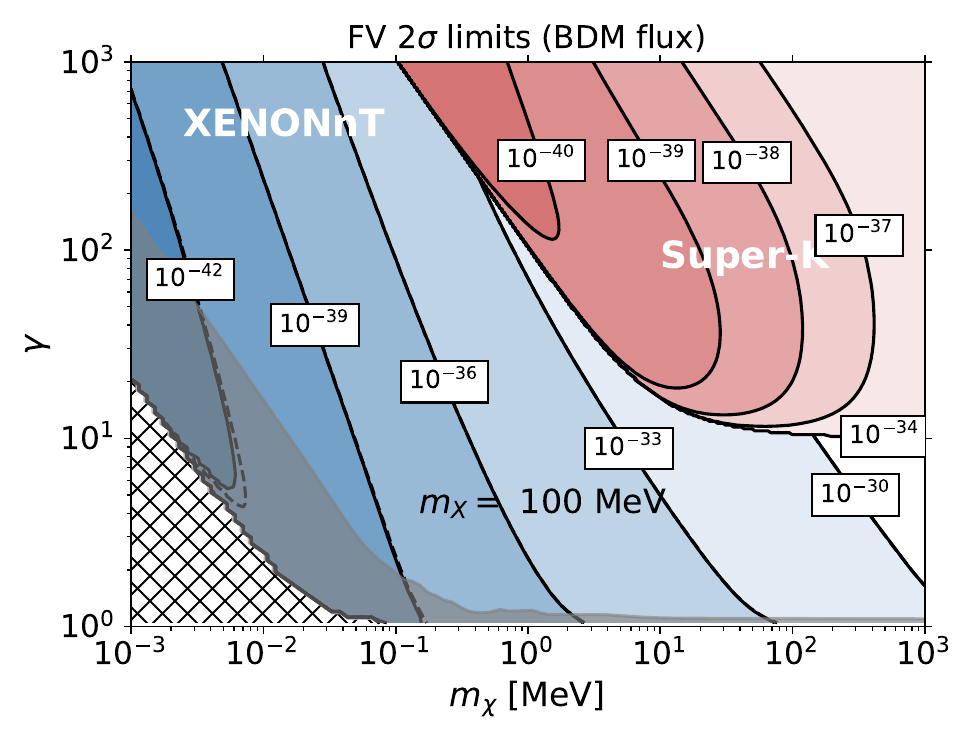}
    \includegraphics[width=0.48\linewidth]{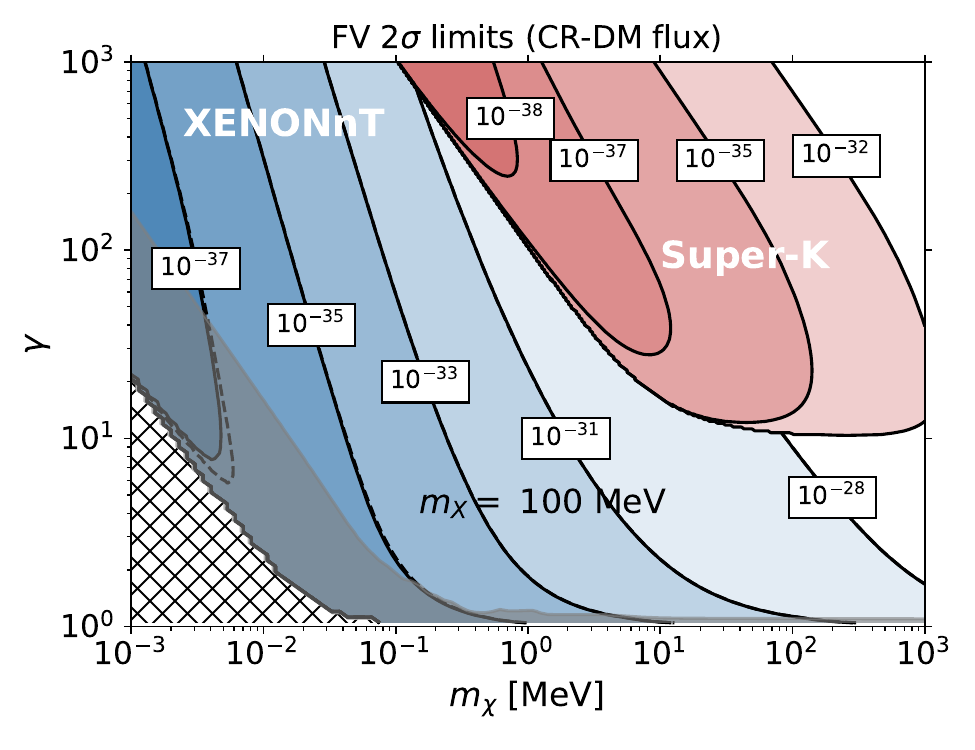}\\
    \caption{BDM and CR-DM fluxes (limits) in the top (bottom) panel shown in the two kinematic parameter space ($m_\chi$, $\gamma$).}
    \label{fig:fluxes}
\end{figure*}

To illustrate the flux of fast-moving or boosted DM (BDM) originating from the Galactic Center (GC), we consider a two-component dark matter model; one species is heavier than the other.
In this scenario, stable, relativistic particles in the dark sector are produced through the annihilation of heavier DM in the GC. 
This BDM produced in the present universe is subdominant, i.e., the expected flux is much smaller than that of the ambient (non-relativistic) DM. However, its large Lorentz boost factor allows it to overcome relatively high energy thresholds and be detected in large-volume terrestrial detectors through interactions with electrons, nucleons, or nuclei. 
This mechanism combines aspects of both direct and indirect detection: BDM is generated through halo DM annihilation (as in indirect searches) but detected through scattering in detectors (similar to direct detection experiments).

The example model of interest consists of two DM species, $\psi_A$ and $\psi_\chi$, with masses $m_A$ and $m_\chi$, respectively. For illustration, we assume that both $\psi_A$ and $\psi_\chi$ are Dirac fermions.  
The dominant component, $\psi_A$, does not interact directly with the SM, while $\psi_\chi$ interacts through a vector portal, e.g., dark photon.
The primary production mechanism of BDM is a pair annihilation, $\psi_A \bar{\psi}_A \to \psi_\chi \bar{\psi}_\chi$.
Since $\psi_\chi$ particles are created with a large boost factor, they can be detected in neutrino experiments such as SK or in ton-scale direct detection experiments such as XENONnT. For discussions of production mechanisms, the thermal evolution of the DM species, and constraints, see, for example, Refs.~\cite{Belanger:2011ww,Agashe:2014yua,Kong:2014mia}.

The flux of these BDM particles reaching Earth, $\Phi_{\rm GC}$, depends on the $\psi_A$ density, annihilation rate, and spatial distribution of $\psi_A$ in the GC. 
The differential flux per unit solid angle $\Omega$ per BDM energy $E_\chi$ is given by
\begin{equation}  
\frac{d\Phi_{\text{GC}}}{d\Omega dE_\chi} = \frac{1}{4} \frac{r_{\text{Sun}}}{4\pi} \left( \frac{\rho_{\text{local}}}{m_A} \right)^2 J \, \langle \sigma v \rangle_{AA\to\chi\chi} \, \frac{dN_\chi}{dE_\chi},  
\end{equation}  
where $r_{\text{Sun}}$ is the distance from the GC to the Sun, $\rho_{\text{local}}$ is the local $\psi_A$ density, taken as $0.3 \,{\rm GeV}/{\rm cm}^3$~\cite{Baxter:2021pqo}, and $J$ is the integral along the line of sight. 
The term $\langle \sigma v \rangle_{AA\to\chi\chi}$ represents the velocity-averaged annihilation cross section, and the BDM energy distribution is simply given by  
\begin{equation}  
\frac{dN_\chi}{dE_\chi} = 2 \delta(E_\chi - m_A),  
\end{equation} 
where the prefactor 2 describes the fact that two BDM particles are created per annihilation.
For a practical estimate, the total flux over the entire sky is parametrized in Ref.~\cite{Agashe:2014yua} as
\begin{eqnarray}  
\Phi^{4\pi}_{\text{GC}} &=& 4.0 \times 10^{-7} \, \text{cm}^{-2} \text{s}^{-1} \nonumber \\
&\times& \left( \frac{\langle \sigma v \rangle_{AA\to\chi\chi}}{5 \times 10^{-26} \, \text{cm}^3/\text{s}} \right) \left( \frac{20 \, \text{GeV}}{\gamma m_\chi} \right)^2.
\label{eq:BDMF}
\end{eqnarray}  

The upper-left panel of Figure~\ref{fig:fluxes} shows the BDM flux $\Phi^{\rm BDM}$ as a function of $m_\chi$ and $\gamma$. 
The flux increases as $\gamma m_\chi$ decreases, primarily due to the $(\gamma \, m_\chi)^{-2}$ dependence in Eq.~\eqref{eq:BDMF}. The contour lines (lower-left) maintain a similar shape across different masses and speeds, suggesting that lower mass and speed particles are more detectable compared to higher mass and speed particles. 

\subsection{DM Boosted by Cosmic Rays}
DM can also be accelerated to high energies through interactions with cosmic-ray (CR) electrons. 
While (a dominant fraction of) DM is non-relativistic, scattering with energetic CR electrons can impart significant energy in the present universe, potentially making (a certain fraction of) DM detectable in direct detection experiments.

The underlying process involves a CR electron transferring a portion of its energy to a DM particle during a collision. The resulting DM kinetic energy depends on both the initial energy of the electron and the scattering angle. The maximum kinetic energy transferable to the DM particle is determined by the electron's energy $T_e$ and mass. 
Given that CR electrons span a broad energy spectrum---from a few MeV to several GeV---they are capable of boosting DM particles across a wide range of kinetic energies.

The flux of CR-boosted DM depends on the local DM density $\rho_{\text{local}}$, the CR electron flux $d\Phi_e/dT_e$, and the interaction cross section between DM and electrons. The differential flux is given by~\cite{Cao:2020bwd, Herbermann:2024kcy}:
\begin{equation}
    \frac{d \Phi^{\rm CR}}{d T_\chi}=D_{\text {eff}} \, \rho_{\text {local}} \int_{T_e^{\min }}^{\infty} d T_e \frac{1}{m_\chi} \frac{d \Phi_e}{d T_e} \frac{d \sigma_{e \chi}}{d T_\chi}.
    \label{eq:CRF}
\end{equation}
Here, $D_{\text{eff}} = 10$ kpc represents the effective distance over which DM is considered. 
The CR electron flux $d\Phi_e/dT_e$ is parametrized in Ref.~\cite{Boschini:2018zdv}. 
If the original DM is non-relativistic, the differential cross section $d\sigma_{e\chi}/dT_\chi$ takes a simple form of $\overline{\sigma}/T_{\chi}^{\rm max}$ where the maximum kinetic energy is
\begin{eqnarray}
    T_\chi^{\max }=\frac{T_e^2+2 m_e T_e}{T_e+\left(m_\chi+m_e\right)^2 / 2 m_\chi}.
\end{eqnarray}
The lower integration limit is
\begin{eqnarray}
    T_e^{\min }=\left(\frac{T_{\chi}}{2}-m_e\right)\left(1 \pm \sqrt{1+\frac{2 T_{\chi}}{m} \frac{\left(m_e+m\right)^2}{\left(2 m_e-T_{\chi}\right)^2}}\right).
\end{eqnarray}
In terms of $\gamma$, we express the flux as
\begin{equation}
    \frac{d \Phi^{\rm CR}}{d \gamma} = m_{\chi} \, \frac{d \Phi^{\rm CR}}{d T_\chi}.
\end{equation}

The upper-right panel of Figure~\ref{fig:fluxes} illustrates the expected CR-boosted DM flux. 
Compared to BDM, $\Phi^{\rm CR}$ decreases more rapidly at high $\gamma m_\chi$, due to the rapid fall of the CR electron flux with energy. 

\bigskip 

We now illustrate how the limits can be obtained immediately for realistic fluxes. In this example, we choose the FV case with a mediator mass $m_X=100~\mathrm{MeV}$. Earlier, we performed the analysis for a fixed reference flux (top-middle panel of Figure~\ref{fig:limits}). Given the BDM and CR-DM fluxes at each DM mass and boost factor (top panel of Figure~\ref{fig:fluxes}), one can rescale the flux, which in turn rescales the derived cross-section limits, yielding the bottom panels of Figure~\ref{fig:fluxes}.

Focusing first on the BDM case (bottom-left), the shape and orientation of the flux contours reorient the resulting limits. For instance, at fixed flux, the XENONnT contours were primarily vertical, whereas with the BDM flux, they become diagonal. The sensitivity still weakens gradually with increasing DM mass. We find that XENONnT covers a broad region of parameter space, with cross-section limits ranging from $10^{-42}$ to $10^{-30}~\mathrm{cm}^2$.

The SK case similarly demonstrates how flux reorients the contours. With a constant flux, the limits were strongest in the upper-right portion of the SK region and weakened toward the lower-left. By contrast, the BDM flux increases toward the lower-left, so the SK limits become stronger on the left side of the region. Similar conclusions hold for the CR-DM flux case.

\section{Summary and outlook}
\label{sec:conclusion}

In this work, we have developed a comprehensive and mechanism-agnostic framework for interpreting electron-recoil signals induced by fast-moving dark matter (DM). Our analysis systematically incorporates relativistic kinematics, mediator-dependent matrix elements, and atomic-physics effects through a fully relativistic ionization form factor. This formalism provides a unified description of scattering processes for both fermionic and scalar DM across vector, scalar, pseudoscalar, and axial vector mediators.

We quantitatively demonstrated the asymptotic behavior of the relativistic ionization form factor and identified the energy regimes in which atomic effects are relevant. At high recoil energies, the form factor approaches the free-electron limit, validating the use of the free-electron approximation in those regions. Conversely, at lower energies, atomic effects can significantly modify the predicted event rates—sometimes enhancing rather than suppressing the signal—depending on the Lorentz structure of the interaction.

By comparing XENONnT and Super-Kamiokande, we highlighted the complementarity between low- and high-threshold experiments in probing fast-moving DM across wide kinematic ranges. We applied our formalism to two benchmark scenarios—two-component boosted DM and cosmic-ray-accelerated DM—and demonstrated how exclusion limits can be rescaled straightforwardly for realistic flux models. These results illustrate how the shape and orientation of the exclusion contours depend not only on detector thresholds but also on the underlying DM production mechanisms.

Overall, our framework bridges the gap between atomic-scale and relativistic treatments, enabling consistent interpretation of results from current and future direct-detection and neutrino-based experiments. It establishes a foundation for model-independent searches for relativistic DM, providing a quantitative map between flux models, mediator dynamics, and detector responses.

Looking ahead, several directions merit further exploration. A natural extension is to incorporate directional and time-dependent signatures, which may help discriminate between cosmic-ray-induced and annihilation-driven boosted DM. Future work may also include nuclear-recoil channels and composite or inelastic mediator models, as well as the integration of machine-learning-based analyses for non-standard recoil spectra. The formalism developed here can also inform studies of neutrino–DM interplay and quantum-sensor-based detection in emerging experiments \cite{Dong:2025mdk}.
Through these avenues, the mechanism-agnostic approach introduced in this paper paves the way for a deeper and more unified understanding of fast-moving dark matter across the next generation of terrestrial and astrophysical searches.


\section*{Acknowledgements}
We would like to thank the organizers of 2025 Particle Physics on the Plains conference during the completion of this project. KK is supported in part by the US DOE under Award No DE-SC0024407.

\appendix

\section{Additional Limit Plots}
\label{appendix:A}
\begin{figure*}[t!]
\begin{center}
\hspace{0.0cm} \includegraphics[width=.315\textwidth]{plots/SK_XEnT_flat_flux_FVL.pdf}
\hspace{0.0cm} \includegraphics[width=.315\textwidth]{plots/SK_XEnT_flat_flux_FVH.pdf}
\hspace{0.0cm} \includegraphics[width=.315\textwidth]{plots/SK_XEnT_flat_flux_FV3.pdf}\\

\hspace{0.0cm} \includegraphics[width=.315\textwidth]{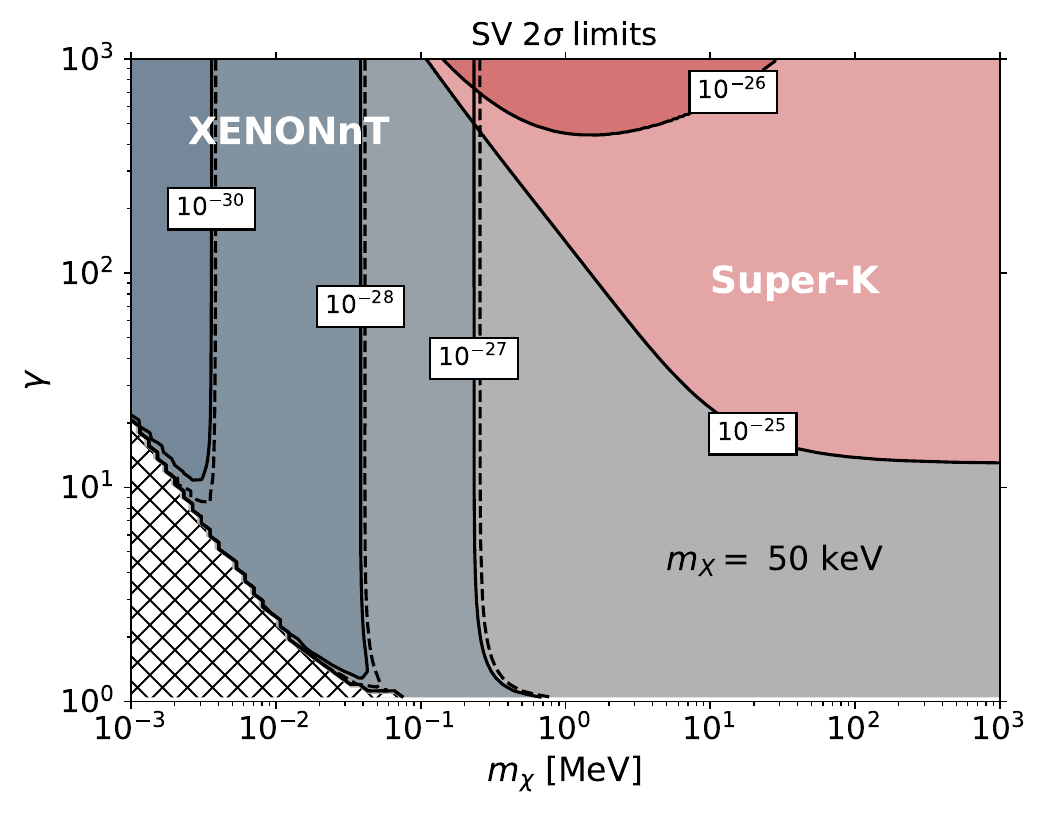}
\hspace{0.0cm} \includegraphics[width=.315\textwidth]{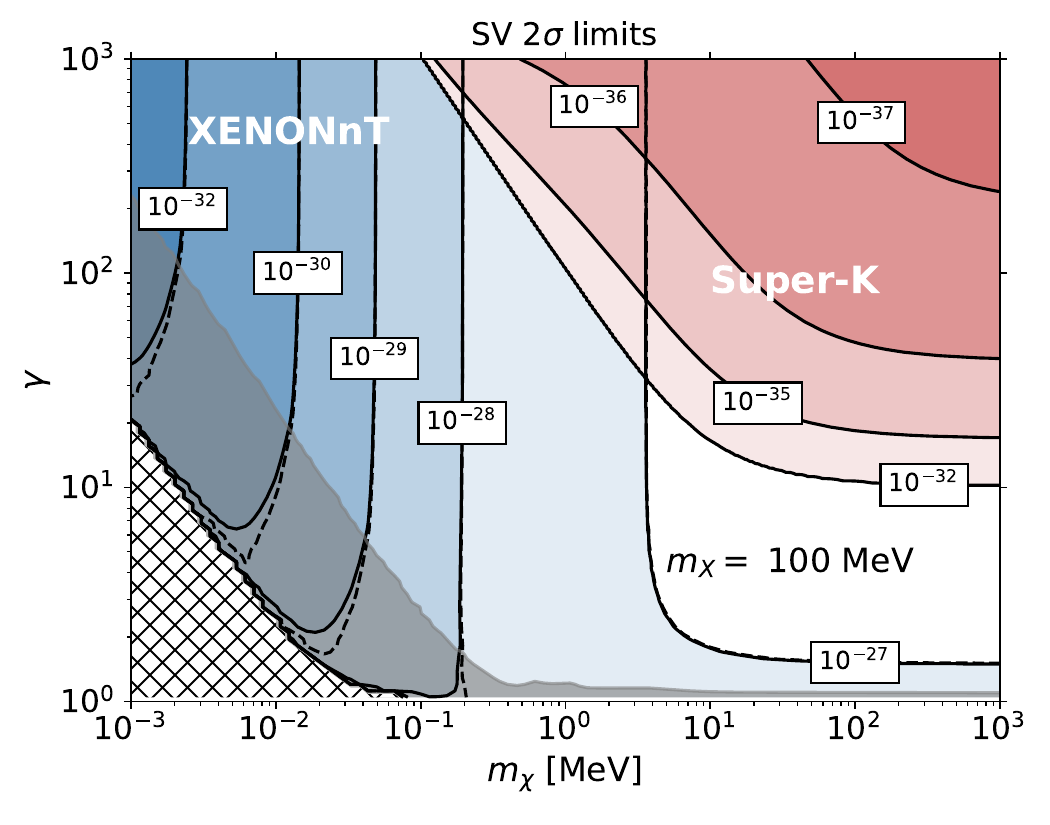}
\hspace{0.0cm} \includegraphics[width=.315\textwidth]{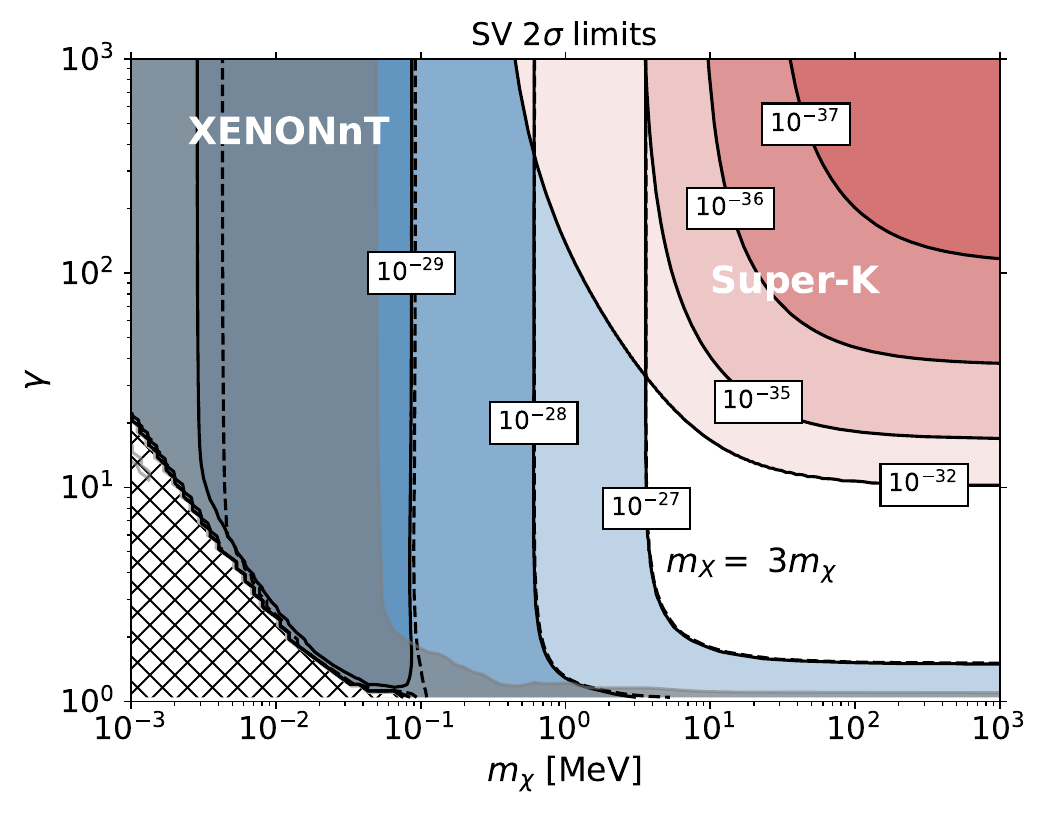}\\

\end{center}
\vspace*{-0.5cm}
\caption{Vector mediator $2\sigma$ exclusion limits for the FV case (top) and the SV case (bottom). Left, center and right panels respectively corresponds to mediator masses $m_X =$ 50 keV, $m_X =$ 100 MeV and $m_X = 3 \, m_{\chi}$. The contour lines are indicated by the cross section in cm$^2$ where the blue (red) shaded regions correspond to the XENONnT (SK) limits. The gray shaded regions represent areas of the parameter space where $R_{\sigma} > 0.1$. The dashed contours in the XENONnT region indicate the corresponding limits derived from the signal defined in Eq.~\eqref{eq:master}.}
\label{fig:limits_V}
\end{figure*}
In this appendix, we provide a list of limit plots corresponding to all renormalizable interactions studied in Table~\ref{tab:RMES}.
These plots follow the same conventions as that of Figure~\ref{fig:limits}.
We group interaction cases by the mediator types where vector mediator cases FV and SV are compared against each other first.
Then we move to the scalar mediator followed by the pseudoscalar mediator and finally the axial vector mediator.

In Figure~\ref{fig:limits_V}, we present the $2\sigma$ exclusion limits for the vector mediator scenario. 
Beyond the general features i-v outlined in Section~\ref{sec:prospects}, we highlight several model-specific observations. 
Notably, there are strong similarities between the exclusion limits for fermionic and scalar DM, both in the light and heavy mediator regimes, across the two experiments.
To better understand these similarities, we examine the structure of the differential cross sections for the two cases. 
In the context of the XENONnT experiment, where the electron recoil energy is relatively low, the dominant contributions to the differential cross section arise from the $x \ll 1$ regime (see Table~\ref{tab:RMES}). 
Conversely, for the SK experiment, which probes much higher electron recoil energies, the relevant terms come from the $x \gg 1$ limit.
For instance, in the light mediator case at XENONnT, the dominant contribution to the squared matrix element for both fermionic and scalar DM is given by the $4\alpha^2\gamma^2/x^2$ term, as shown in Table~\ref{tab:RMES}. 
Similarly, for the heavy mediator case at XENONnT, the leading term in both scenarios is $16\alpha^2\gamma^2/\beta^2$.
Overall, exclusion limits for the heavy mediator tend to be stronger than those for the light mediator. 
This is because the differential cross section is relatively flat in the heavy mediator regime, whereas it decreases with energy in the light mediator case, as illustrated in Figure~\ref{fig:comparison_FV}.
For the vector mediator, and especially in the case of the XENONnT experiment, the exclusion limits across all scenarios are primarily driven by terms proportional to $\alpha^2\gamma^2$. 
This common dependence leads to the observed uniformity in the exclusion contours across the panels in Figure~\ref{fig:limits_V}. 
The contour lines exhibit similar shapes, with the strongest exclusions occurring on the left side of the blue-shaded region, gradually weakening toward the right. Analogous features are observed in the SK experiment as well.

\begin{figure*}[t!]
\begin{center}
\hspace{0.0cm} \includegraphics[width=.315\textwidth]{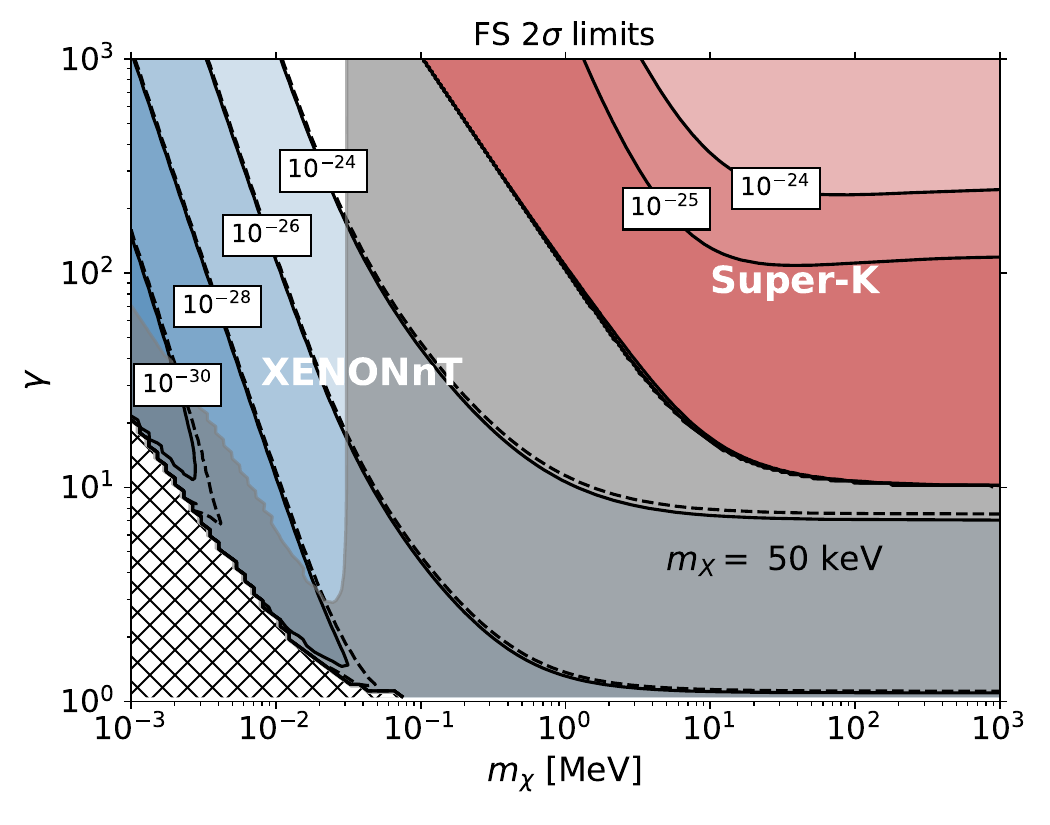}
\hspace{0.0cm} \includegraphics[width=.315\textwidth]{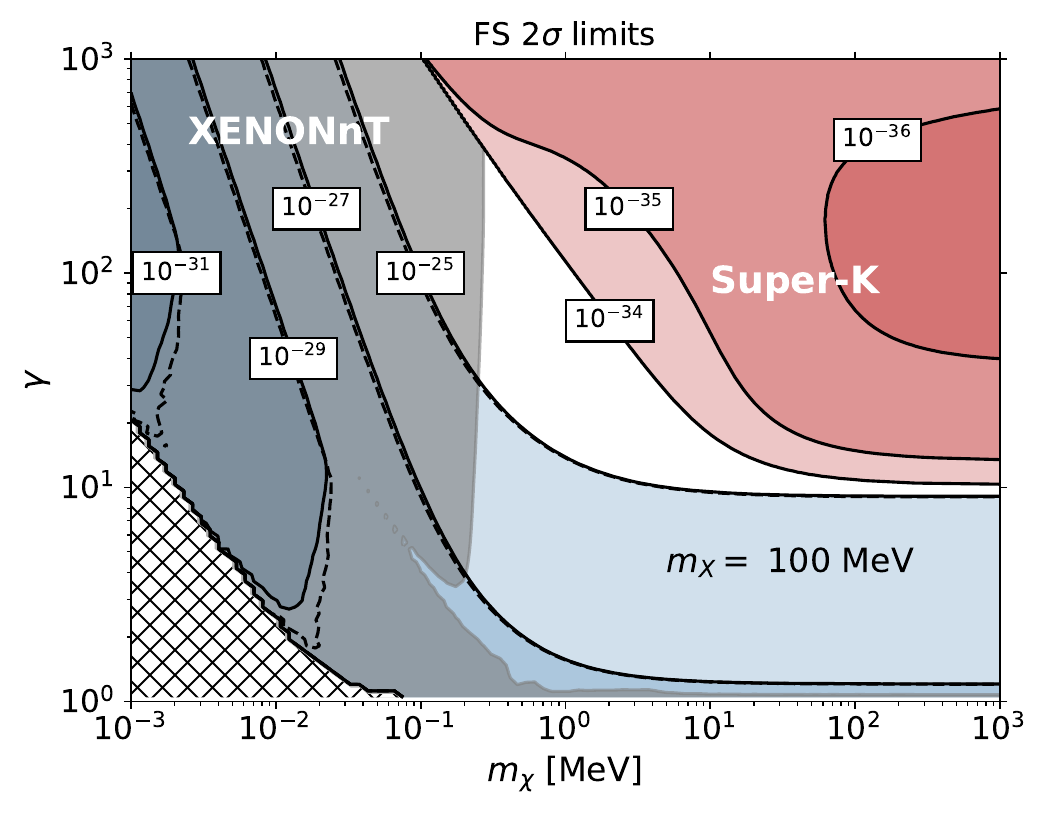}
\hspace{0.0cm} \includegraphics[width=.315\textwidth]{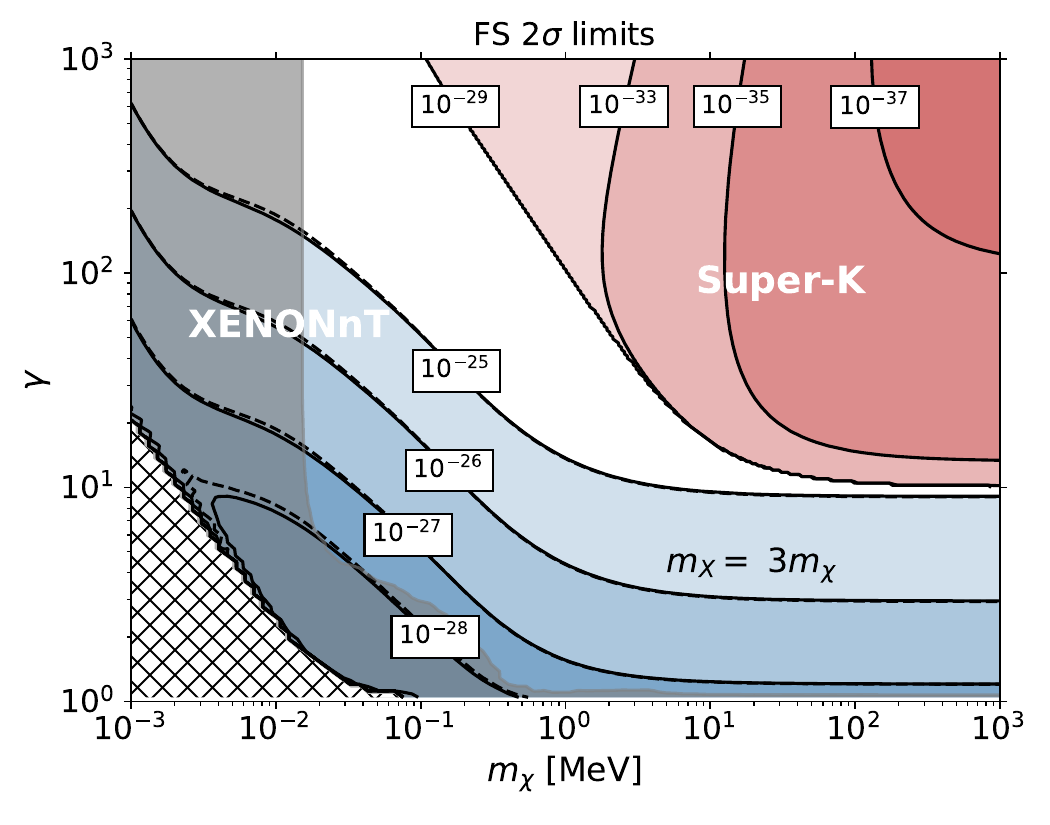}\\

\hspace{0.0cm} \includegraphics[width=.315\textwidth]{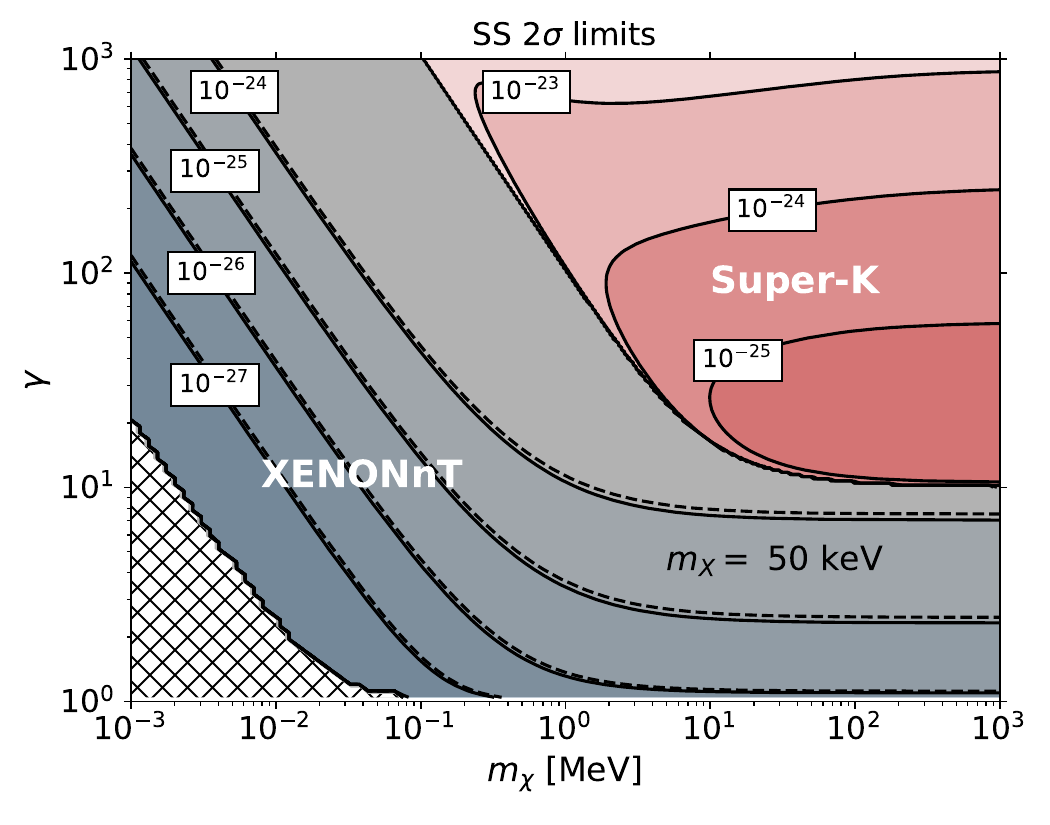}
\hspace{0.0cm} \includegraphics[width=.315\textwidth]{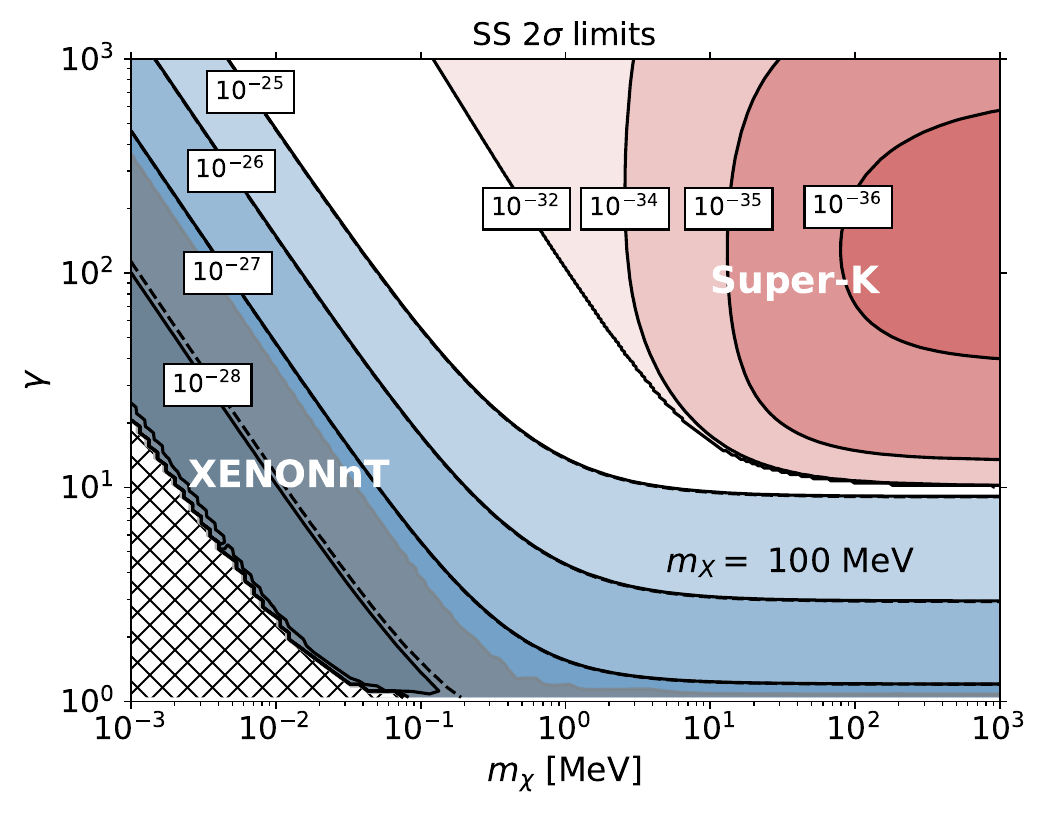}
\hspace{0.0cm} \includegraphics[width=.315\textwidth]{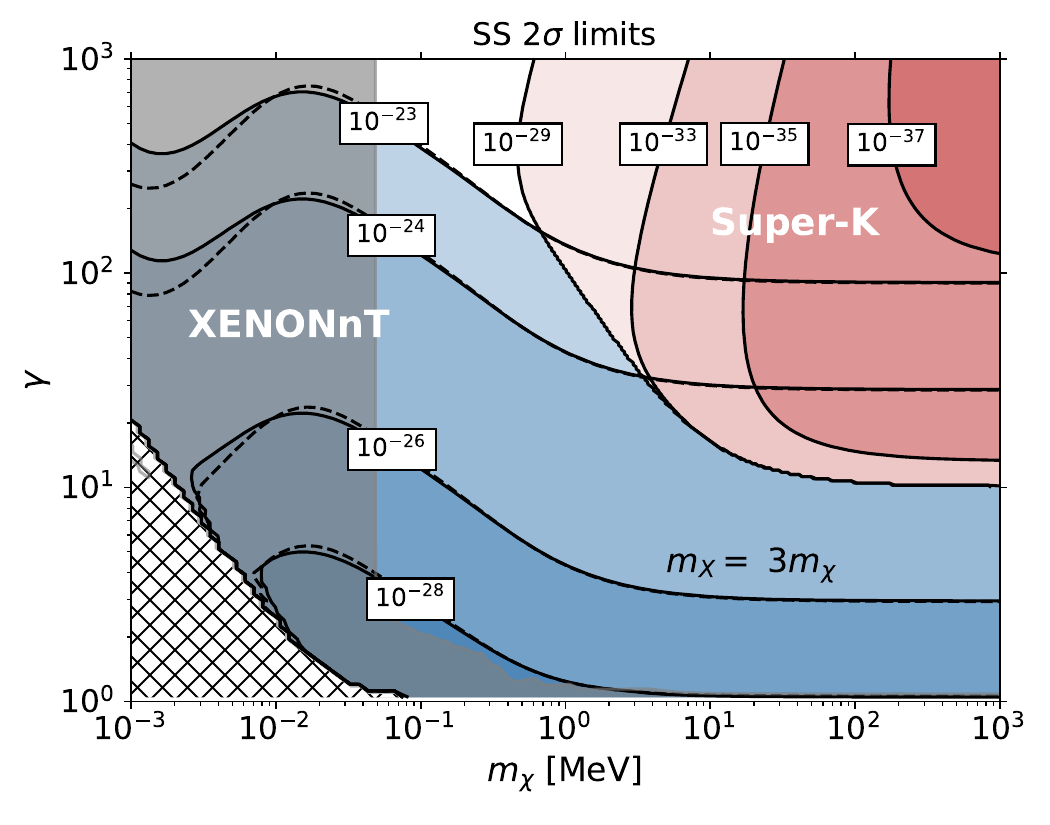}\\
\end{center}
\vspace*{-0.5cm}
\caption{Scalar mediator $2\sigma$ exclusion limits for the FS case (top) and the SS case (bottom). Left, center and right panels respectively corresponds to mediator masses $m_X =$ 50 keV, $m_X =$ 100 MeV and $m_X = 3 \, m_{\chi}$. The contour lines are indicated by the cross section in cm$^2$ where the blue (red) shaded regions correspond to the XENONnT (SK) limits. The gray shaded regions represent areas of the parameter space where $R_{\sigma} > 0.1$. The dashed contours in the XENONnT region show the limits derived from the signal defined in Eq.~\eqref{eq:master}.}
\label{fig:limits_S}
\end{figure*}
In Figure~\ref{fig:limits_S}, we examine the exclusion limits in the case where the interaction is mediated by a scalar particle. The top panel corresponds to fermionic DM, while the bottom panel shows results for scalar DM. 
The left, center and right panels depict scenarios with light, heavy and dynamic mediator mass choices, respectively.
The general considerations listed as items i–v in Section~\ref{sec:prospects} remain applicable here as well.
There is a noticeable degree of similarity in the exclusion contours across the different panels of the figure particularly in the case of XENONnT. 
In the $x \ll 1$ regime, relevant for low-energy recoils, the squared matrix element is dominated by $16\alpha^2/\beta^4$ for heavy mediators and by $4\alpha^2/x^2$ for light mediators, in both FS and SS cases.
For the SK experiment, where $x \gg 1$ dominates due to higher recoil energies, we observe some differences in the leading terms of the matrix elements. 
These distinctions can be seen in Table~\ref{tab:RMES}.
For instance, in the light mediator scenario, both FS and SS matrix elements fall as $1/x^2$, but with different behaviors: the SS case exhibits a continuous decline, while the FS case is stabilized by a constant contribution. 
Similarly, for the heavy mediator, the FS matrix element grows quadratically with $x$, whereas the SS case exhibits linear growth.
A notable feature in the scalar mediator case is the absence of dependence on the parameter $\gamma$ (see Table~\ref{tab:RMES}) in the squared matrix element. 
This omission significantly impacts the exclusion limits, especially when the matrix element is either constant or scales inversely with $x$, as is often the case for XENONnT limits in all panels and for the SK limits in the light mediator scenario.
In such cases, regions of parameter space with lower $\gamma$ values are more strongly constrained due to an enhanced cross section. 
This can be illustrated with a simple example from the XENONnT exclusion in the bottom-center panel of Figure~\ref{fig:limits_S}. 
According to Eq.~\eqref{eq:dcsdx}, the differential cross section behaves as
\begin{equation} 
\frac{d \sigma}{d x} \propto |\mathcal{A}|^2/(\alpha^2 \gamma^2). 
\end{equation}
Taking the SS case with a heavy mediator, we have $|\mathcal{A}|^2 = 16 \alpha^2/\beta^4$, leading to a differential cross section that scales as $16/(\beta^4 \gamma^2)$ and hence making the limit stronger towards lower $\gamma$ values. The right panel corresponds to the dynamical choice of the mediator mass ($m_X=3m_{\chi}$) where the mediator mass ranges from an order of a keV in the left part of the parameter space to an order of a GeV in the right part of the parameter space. This transition allows an admixture effect of both light and heavy mediators.

\begin{figure*}[t!]
\begin{center}
\hspace{0.0cm} \includegraphics[width=.315\textwidth]{plots/SK_XEnT_flat_flux_FPL.pdf}
\hspace{0.0cm} \includegraphics[width=.315\textwidth]{plots/SK_XEnT_flat_flux_FPH.pdf}
\hspace{0.0cm} \includegraphics[width=.315\textwidth]{plots/SK_XEnT_flat_flux_FP3.pdf}\\

\hspace{0.0cm} \includegraphics[width=.315\textwidth]{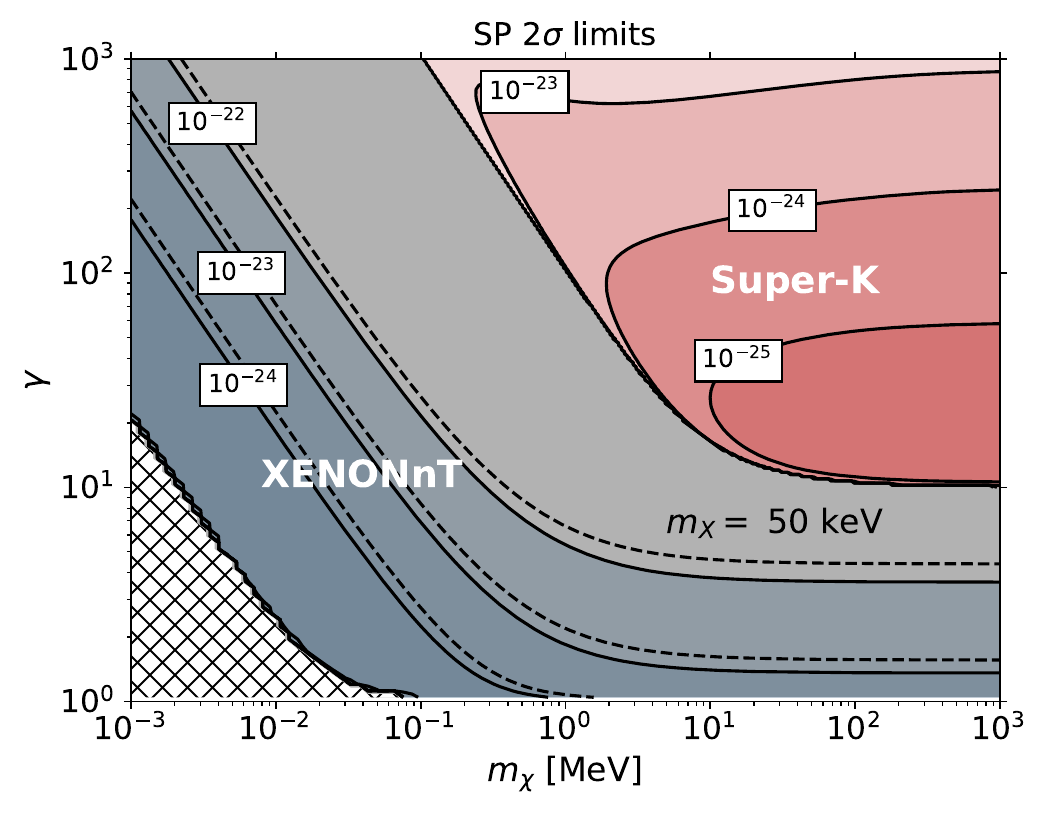}
\hspace{0.0cm} \includegraphics[width=.315\textwidth]{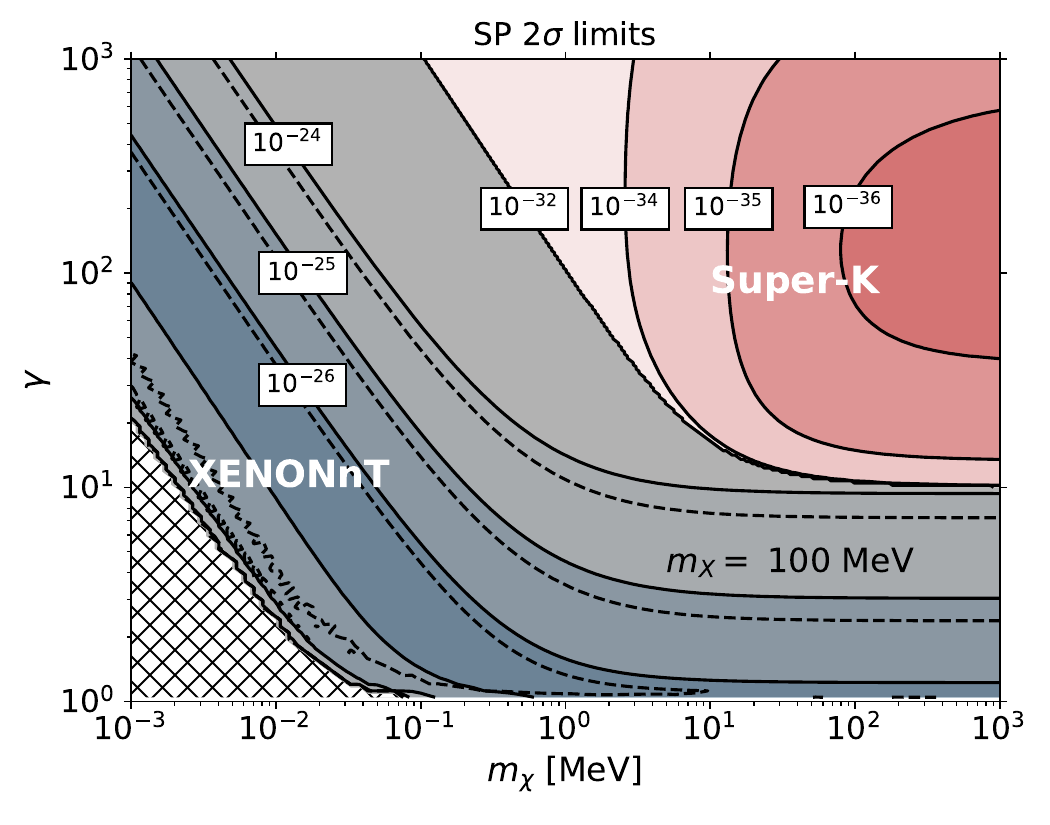}
\hspace{0.0cm} \includegraphics[width=.315\textwidth]{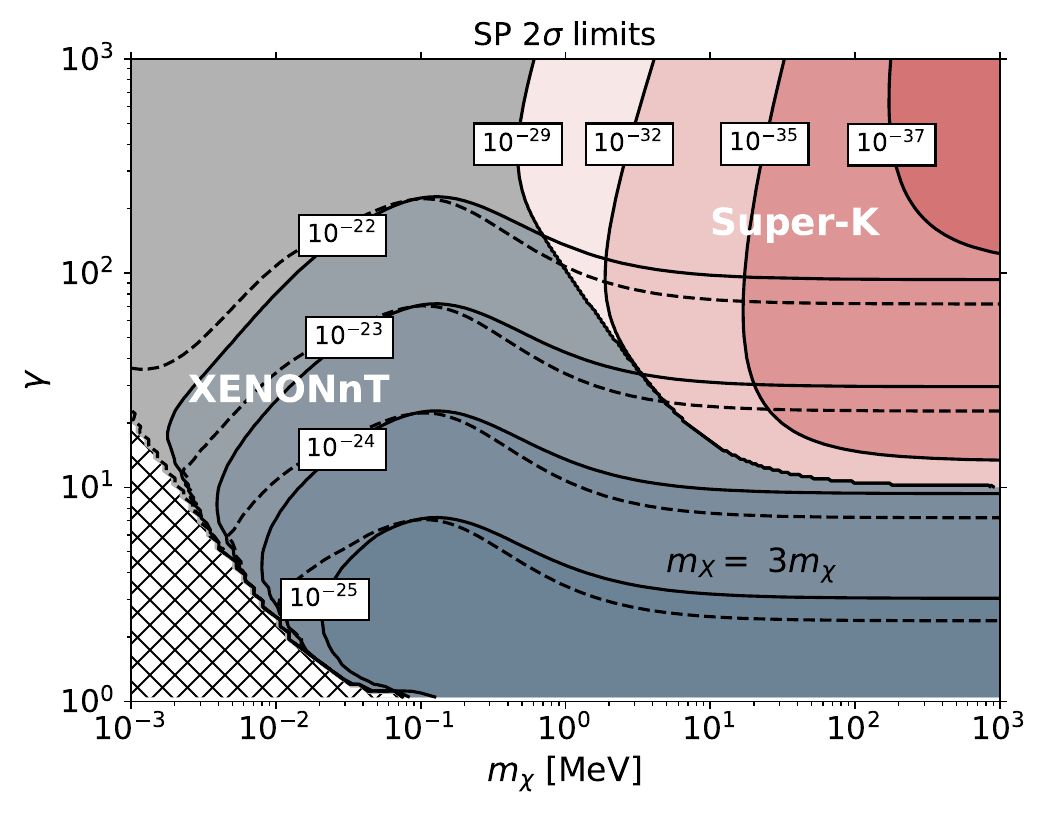}\\
\end{center}
\vspace*{-0.5cm}
\caption{Pseudoscalar mediator $2\sigma$ exclusion limits for the FP case (top) and the SP case (bottom). Left, center and right panels respectively corresponds to mediator masses $m_X =$ 50 keV, $m_X =$ 100 MeV and $m_X = 3 \, m_{\chi}$. The contour lines are indicated by the cross section in cm$^2$ where the blue (red) shaded regions correspond to the XENONnT (SK) limits. The gray shaded regions represent areas of the parameter space where $R_{\sigma} > 0.1$. The dashed contours in the XENONnT region denote the corresponding limits derived using the signal defined in Eq.~\eqref{eq:master}.}
\label{fig:limits_P}
\end{figure*}
In Figure~\ref{fig:limits_P}, we present the exclusion limits in the case of a pseudoscalar mediator. 
The top panel corresponds to fermionic DM, while the bottom panel shows the scalar DM case. The left, center and right panels illustrate the scenarios with $m_X = 50$ keV, $m_X = 100$ MeV and $m_X = 3\,m_\chi$.
As before, the general considerations listed in items i–v in Section~\ref{sec:prospects} also apply here.
To avoid redundancy, we note that the bottom panel of this figure closely resembles the corresponding panel in Figure~\ref{fig:limits_S}. 
This is expected due to the similar structure of the squared matrix element in both cases, as shown in Table~\ref{tab:RMES}.
In contrast, the top panel exhibits a different behavior. 
In the FP case with a light mediator, the squared matrix element is approximately unity\footnote{This approximation is true since the apparent $\beta^2/x$ term in table~\ref{tab:RMES} has a sub-leading contribution in the limit $\beta^2\ll 1$.}.
Consequently, the differential cross section is simply proportional to $1/(\alpha^2 \gamma^2)$, indicating that regions with smaller values of $\alpha$ and $\gamma$ are more strongly constrained. 
This trend is clearly visible in both the XENONnT and SK limits.
This particular case, where $|\mathcal{A}|^2 = 1$, provides a valuable reference point for isolating the effects of detector kinematics and specifications without complications from model-dependent dynamics. 
As such, it serves as a baseline to understand the general shape and orientation of the exclusion contours.
Once this baseline is established, deviations observed in other scenarios can be interpreted in terms of the additional contributions arising from the matrix element structure specific to each model.

\begin{figure*}[t!]
\begin{center}
\hspace{0.0cm} \includegraphics[width=.315\textwidth]{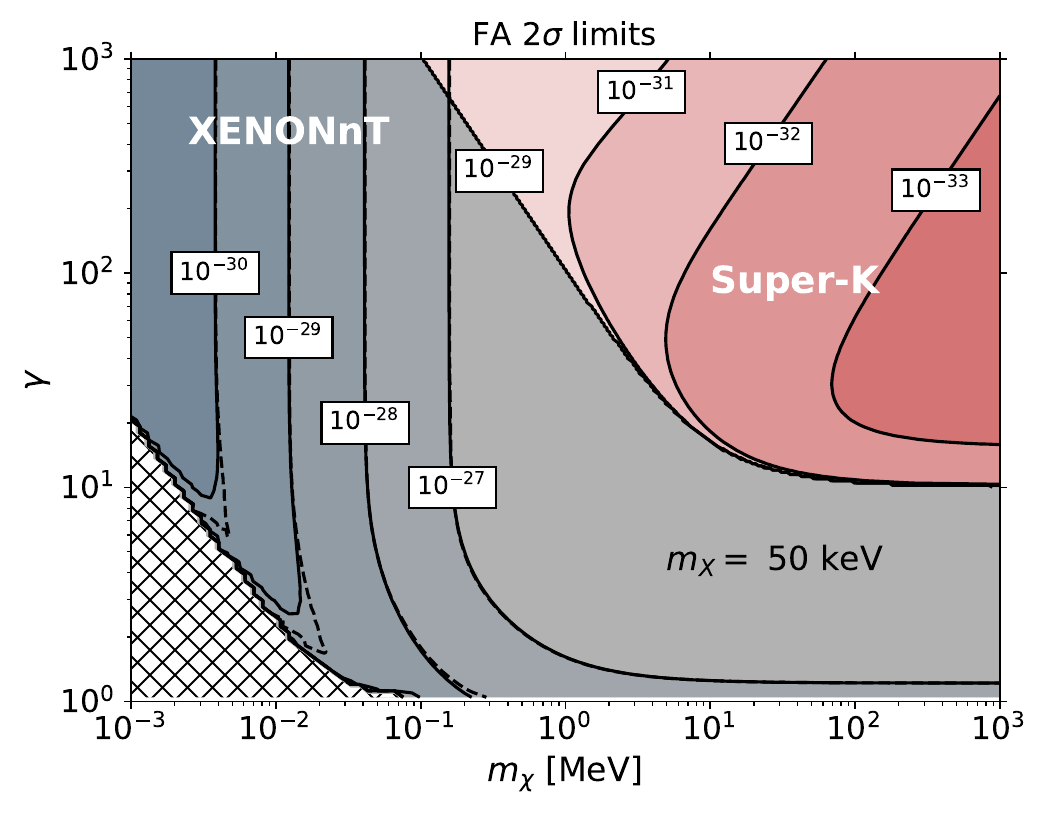}
\hspace{0.0cm} \includegraphics[width=.315\textwidth]{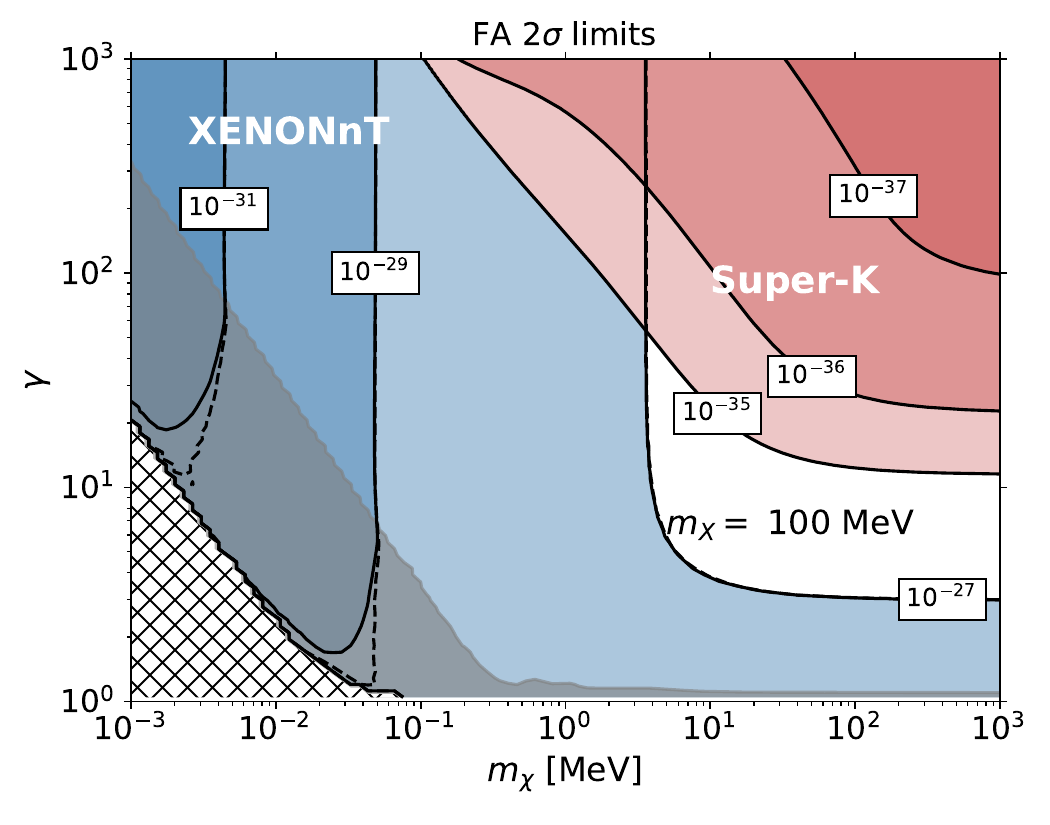}
\hspace{0.0cm} \includegraphics[width=.315\textwidth]{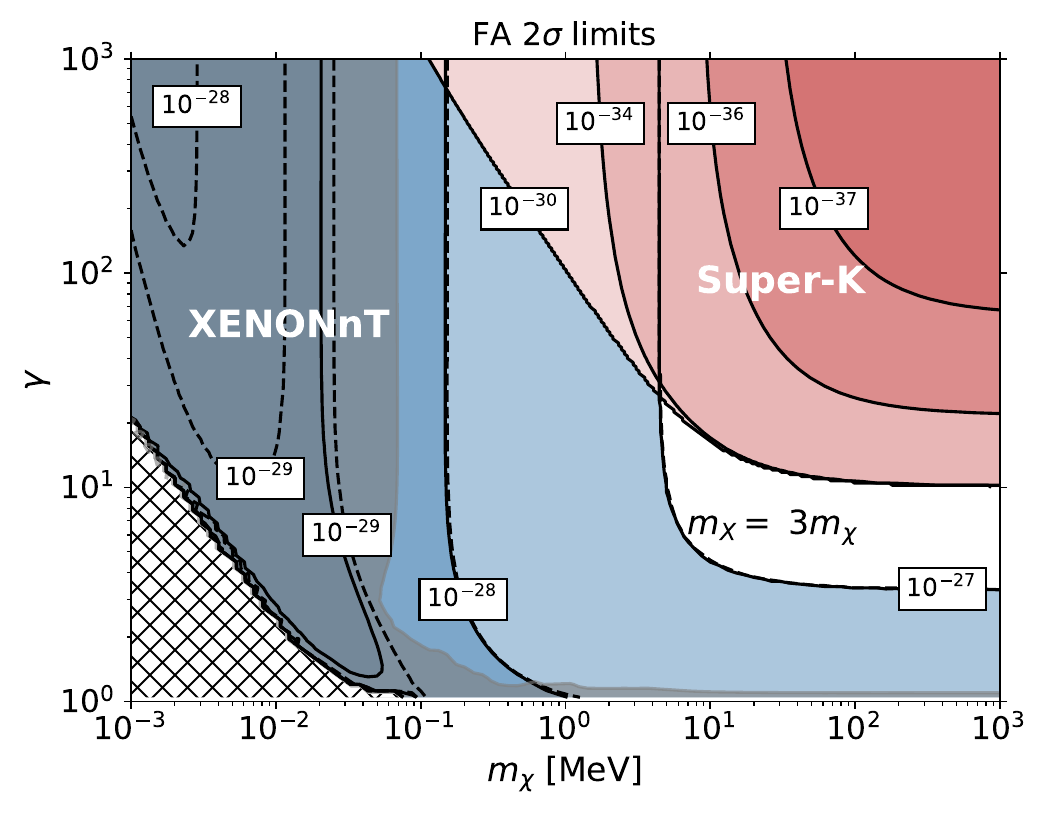}\\
\end{center}
\vspace*{-0.5cm}
\caption{Axial vector mediator $2\sigma$ exclusion limits. Left, center and right panels respectively corresponds to mediator masses $m_X =$ 50 keV, $m_X =$ 100 MeV and $m_X = 3 \, m_{\chi}$. The contour lines are indicated by the cross section in cm$^2$ where the blue (red) shaded regions correspond to the XENONnT (SK) limits. The gray shaded regions represent areas of the parameter space where $R_{\sigma} > 0.1$. In the XENONnT region, the dashed contours indicate limits derived from the signal defined in Eq.~\eqref{eq:master}.}
\label{fig:limits_A}
\end{figure*}
We now focus on a specific interaction scenario in which the mediator is an axial vector and the DM particle is fermionic. 
Figure~\ref{fig:limits_A} presents the $2\sigma$ exclusion limits, with the mediator mass choices follow the same as the previous cases
The general considerations outlined in items i–v for the vector mediator continue to apply here as well.
Here we observe that the differential cross sections in the FV and FA cases become very similar in the heavy mediator regime.
This similarity carries over to the exclusion limits, explaining the nearly identical constraints seen in both the XENONnT and SK experiments.
For the light mediator case, the FA scenario shares qualitative features with the FS case. 
In both, the cross-section decreases with increasing $x$ and tends to flatten out at large $x$. 
However, the rate of flattening differs: the FA case flattens more rapidly due to the additional $16 \alpha^2/\beta^4$ term in the squared matrix element.
As a result, in the FA case, the limits tend to strengthen with increasing $\alpha$, particularly noticeable in the SK experiment as shown in the left panel of Figure~\ref{fig:limits_A}.
This effect is less pronounced in the XENONnT experiment, as it primarily probes much smaller $x$ values where the influence of this term is comparatively weaker.

\bibliography{refs}

\end{document}